\documentclass[12pt,preprint]{aastex}

\usepackage{graphicx}
\usepackage{rotating}

\shorttitle{}
\shortauthors{Nesvorn\'y et al.}

\begin{document}
\baselineskip 19.pt

\title{NEOMOD: A New Orbital Distribution Model \\ for Near Earth Objects}

\author{David Nesvorn\'y$^1$, Rogerio Deienno$^1$, William F. Bottke$^1$, Robert Jedicke$^2$,
  Shantanu~Naidu$^3$, Steven~R.~Chesley$^4$, Paul W. Chodas$^4$, Mikael Granvik$^{5,6}$, David Vokrouhlick\'y$^7$,
  Miroslav Bro\v{z}$^7$, Alessandro~Morbidelli$^8$, Eric~Christensen$^9$, \\ Bryce T. Bolin$^{10,11,12}$} 
\affil{(1) Department of Space Studies, Southwest Research Institute, 1050 Walnut St., 
  Suite 300,  Boulder, CO 80302, USA}
\affil{(2) Institute for Astronomy, University of Hawaii, 2680 Woodlawn Drive, Honolulu, HI
  96822-1839, USA}
\affil{(3) Jet Propulsion Laboratory, California Institute of Technology, 4800 Oak Grove Dr.
Pasadena, CA 91109, USA}
\affil{(4) Department of Earth and Space Sciences, University of California,  
595 Charles Young Drive East, 5656 Geology Building, Los Angeles, CA 90095, USA}
\affil{(5) Department of Physics, University of Helsinki, P.O. Box 64, FI-00014, Finland} 
\affil{(6) Asteroid Engineering Laboratory, Lule\aa{} University of Technology, Box 848, 
SE-981 28, Kiruna, Sweden}
\affil{(7) Institute of Astronomy, Charles University, V Hole\v{s}ovi\v{c}k\'ach 2, CZ–18000 
Prague 8, Czech Republic}
\affil{(8) Laboratoire Lagrange, UMR7293, Universit\'e C\^ote d'Azur, CNRS, Observatoire de la 
C\^ote d'Azur, Bouldervard de l'Observatoire, 06304, Nice Cedex 4, France}
\affil{(9) Lunar and Planetary Laboratory, The University of Arizona, 1629 E. University Blvd. 
Tucson, AZ 85721-0092, United States}
\affil{(10) Division of Physics, Mathematics and Astronomy, California Institute of Technology, 
Pasadena, CA 91125, USA}
\affil{(11) Infrared Processing and Analysis Center, California Institute of Technology, 
Pasadena, CA 91125, USA}
\affil{(12) NASA Postdoctoral Program Fellow, Goddard Space Flight Center, 8800 Greenbelt Road, 
Greenbelt, MD 20771, USA}

\begin{abstract}
Near Earth Objects (NEOs) are a transient population of small bodies with orbits near or in the 
terrestrial planet region. They represent a mid-stage in the dynamical cycle of asteroids 
and comets, which starts with their removal from the respective source regions -- the main 
belt and trans-Neptunian scattered disk -- and ends as bodies impact planets, disintegrate 
near the Sun, or are ejected from the Solar System. Here we develop a new orbital model 
of NEOs by numerically integrating asteroid orbits from main belt sources and calibrating
the results on observations of the Catalina Sky Survey. The results imply a
size-dependent sampling of the main belt with the $\nu_6$ and 3:1 resonances producing
$\simeq 30$\% of NEOs with absolute magnitudes $H = 15$ and $\simeq 80$\% of NEOs with $H = 25$.
Hence, the large and small NEOs have different orbital distributions. The inferred flux of $H<18$ bodies 
into the 3:1 resonance can be sustained only if the main-belt asteroids near the resonance drift 
toward the resonance at the maximal Yarkovsky rate ($\simeq 2 \times 10^{-4}$ au Myr$^{-1}$ for 
diameter $D=1$ km and semimajor axis $a=2.5$~au). This implies obliquities $\theta \simeq 0^\circ$ 
for $a<2.5$~au and $\theta \simeq 180^\circ$ for $a>2.5$~au, both in the immediate neighborhood
of the resonance (the same applies to other resonances as well). We confirm the size-dependent
disruption of asteroids near the Sun found in previous studies. An interested researcher can use the 
publicly available NEOMOD Simulator to generate user-defined samples of NEOs from our model. 
\end{abstract}

\section{Introduction}

NEOs are asteroids and comets whose orbital perihelion distance is $q<1.3$ au. Asteroids, which 
represent the great majority of NEOs on short-period orbits, are the main focus here, but we also
include comets with $a<4.2$ au. The goal is to develop an accurate model of the orbital and absolute
magnitude distribution of NEOs that can be used to understand the observational incompleteness,
design search strategies, and evaluate the impact risk. We aim at setting up a
flexible scheme that can easily be updated when new observational data become available (from the
Vera C. Rubin Observatory, NEO Surveyor, etc.).

We closely follow the methodology developed in previous studies (Bottke et al. 2002, Granvik
et al. 2018; also see Greenstreet et al. 2012), and attempt to improve it whenever possible. This
does not always mean that a new level of realism/complexity is added to the model. For
example, Granvik et al. (2018) defined various NEO sources from numerical integrations where
main belt asteroids were \textit{drifted} into resonances (Granvik et al. 2017). This is arguably
a more realistic approach than simply placing test bodies into source resonances (Bottke et al.
2002). Here we opt for the latter method because it is conceptually simple and easy to modify.
We verify, when possible (e.g., Sect. 9), that the main results are not affected by this simplifying assumption.

We develop a new method to accurately calculate biases of NEO surveys and apply it to the Catalina
Sky Survey (CSS). The \texttt{MultiNest} code, a Bayesian
inference tool designed to efficiently search for solutions in high-dimensional parameter space
(Feroz \& Hobson 2008, Feroz et al. 2009), is used to optimize the model fit to CSS detections.
We adopt cubic splines to characterize the magnitude distribution of the NEO population. Cubic
splines are flexible and can be modified to consider a broader absolute-magnitude range and/or
improve the model accuracy. We use a large number of main-belt asteroids in each source ($10^5$),
which allows us to accurately estimate the impact fluxes on the terrestrial planets. Our model
self-consistently accounts for the NEO disruption at small perihelion distances (Granvik et al. 2016).

This article is structured as follows. We define NEO sources (Sect. 2), carry out $N$-body integrations 
to determine the orbital distribution of NEOs from each source (Sect. 3), and combine different sources 
together by calibrating their contributions from CSS (Sect. 4). The model optimization with {\tt MultiNest} 
is described in Sect. 5. The final model, hereafter NEOMOD, synthesizes our current knowledge of the orbital 
and absolute magnitude distribution of NEOs (Sect. 6). It can readily be upgraded as new NEO observations 
become available. We provide the NEOMOD 
Simulator\footnote{\url{https://www.boulder.swri.edu/\~{}davidn/NEOMOD\_Simulator}} -- 
an easy-to-operate code 
that can be used to generate user-defined samples of model NEOs. Planetary impacts are discussed in 
Sect. 7. Sect. 8 considers several modifications of our base model. In Sect. 9, we drift main-belt
asteroids toward source resonances to test whether the flux of bodies into resonances is consistent with 
the results inferred from the NEO modeling.  
 
\section{Source populations}

To set up the initial orbits of main belt asteroids in various NEO sources, we made use of the 
\texttt{astorb.dat} catalog from the Lowell observatory 
(Moskovitz et al. 2022).\footnote{\url{https://asteroid.lowell.edu/main/astorb/}} 
As of early 2022, the \texttt{astorb.dat} catalog contained nearly $1.2\times10^6$ entries, the 
great majority of which were main-belt asteroids. For each source, we inspected the known asteroid 
population near the source location to define the initial distribution of orbits for our numerical
integrations. We illustrate the method for the 3:1 resonance at $a=2.5$ au, which is a notable source of NEOs 
identified by many previous studies (e.g., Wisdom 1985, Gladman et al. 1997, Bottke et al. 2002, Morbidelli
\& Vokrouhlick\'y 2003,
Greenstreet et al. 2012, Granvik et al. 2018); other resonances are discussed later on.  

In Fig. \ref{res31}, the 3:1 resonance appears as a V-shaped gap -- this is the place 
where Jupiter's gravitational perturbations build up to boost object's orbital eccentricity 
(Wisdom 1982). The borders of the gap are approximately $a_1 = 2.5- (0.02/0.35)\, e$ au and 
$a_2=2.5 + (0.02/0.35)\, e$ au, where $e$ is the orbital eccentricity. The 3:1 source population 
is represented in this work by $10^5$ test bodies (not shown in Fig. \ref{res31}) placed 
within the gap borders. In reality, the main belt asteroids evolve into the resonance 
by the Yarkovsky thermal effect (Vokrouhlick\'y et al. 2015), but this is not considered
here. In Sect. 9, we drift asteroids into resonances and find that the orbital distribution of
NEOs is insensitive to how the resonant sources are populated (e.g., to the initial 
resonant amplitude distribution). One needs to be careful, however, with the eccentricity and
inclination distributions of source orbits (Bottke et al. 2002, Granvik et al. 2018).

We define two strips in $(a,e)$, one on the left and one on the right side of the 3:1 resonance
(Fig. \ref{res31}), and use the known asteroids in these strips to set up the eccentricity 
and inclination distributions for the 3:1 source. The idea is that bodies entering the 3:1 
resonance should have the $e$ and $i$ distributions similar to bodies in the strips. 
For 3:1, the left strip is defined as $a > 2.48 - (0.02/0.35)\,e$ au and $a <2.49 - (0.02/0.35)\, e$ au,
and the right strip is defined as $a > 2.51 + (0.02/0.35)\, e$ au and $a < 2.52 + (0.02/0.35)\, e$ au. 
{\it The Mars-crossing orbits are avoided.} Both strips have a fixed ($e$-independent) width to assure 
even sampling. To limit problems with the observational incompleteness, which may unevenly affect 
asteroid populations with different $e$/$i$, we only consider bodies with absolute magnitudes 
$H<18$ (cuts with $H<15$, $H<16$ or $H<17$ produce similar results). This means that the orbital distribution 
within a single source is size independent. The size dependence appears in our NEO model due to the 
size dependent weights of different sources (Sect. 5.1) and the size-dependent disruption (Sect. 5.4).

The orbital distribution of known asteroids in the strips is parameterized by analytic functions, which are 
then used to generate synthetic bodies. This two-step procedure is useful to leave the record of 
the adopted distributions (Table 1). Specifically, we experimented with the single Gaussian, double 
Gaussian, Rayleigh and Maxwell-Boltzmann distributions. For the 3:1 resonance, the eccentricity 
distribution is well approximated by a single Gaussian with the mean $\mu = 0.145$ and width 
$\sigma=0.067$, and the inclination distribution with a double Gaussian with $\mu_1=4.7^\circ$, 
$\sigma_1=2.7^\circ$, $\mu_2=13.5^\circ$, and $\sigma_2=2.5^\circ$, where the first Gaussian is 
given a 2.5 times greater weight than the second one (i.e., the weight ratio $w_1/w_2=2.5$; 
Fig.~\ref{ei}).\footnote{The double Gaussian distribution is given here by 
\begin{equation}
f(x)= 
w_1 \exp \left[ -{1 \over 2} \left({x-\mu_1 \over \sigma_1}\right)^2 \right] +
w_2 \exp \left[ -{1 \over 2} \left({x-\mu_2 \over \sigma_2}\right)^2 \right]
\ .
\end{equation}} 
Table 1 reports parameters of the adopted analytic distributions for all sources.  

For each draw of $e$ and $i$, the semimajor axis is assigned randomly between $a_1$ and $a_2$.
The perihelion ($\varpi$) and nodal ($\Omega$) longitudes are drawn from a uniformly random
distribution between 0 and
$2 \pi$ radians. The mean longitude $\lambda$ is chosen such that $\theta_{3:1} = 3 \lambda_J - \lambda -
2 \varpi = \pi$, where $\theta_{3:1}$ is the resonant angle of the 3:1 resonance and 
$\lambda_J$ is the mean longitude of Jupiter at the reference epoch ($\lambda_{\rm J}=343.68^\circ$ for 
MJD $=2459600.5$). With this choice, the initial resonant amplitude is simply 
$\Delta a=|a-2.5\, {\rm au}|$, and we can therefore easily check if different amplitudes would 
yield differing orbital distributions of NEOs (they do not; see Sect. 9 for additional tests).
This completes the description for the 3:1 resonance.

We followed the same procedure for the 5:2, 7:3, 8:3, 9:4, 11:5 and 2:1 resonances with Jupiter,
all of which can potentially be important sources of NEOs. In the preliminary tests, we also 
included the 7:2 resonance with Jupiter, and the 4:7 and 1:2 resonances with Mars. These 
resonances were tested to establish the importance of the `forest' of weak resonances in the inner 
main belt. Whereas these individual resonances are likely to be important for the NEO delivery, 
especially for large asteroids (Migliorini et al. 1998), we found that several trees cannot account 
for a forest. We thus followed the method described in Migliorini et al. (1998) to model \textit{all} 
weak resonances (also see Bottke et al. 2002). Specifically, we extracted all known asteroids from 
the \texttt{astorb.dat} catalog with $q>1.66$ au (i.e., no Mars crossers), $2.1<a<2.5$ au, 
$i<18^\circ$, and $H<18$ (163,971 bodies in total), and reduced that sample --
by random selection -- down to $10^5$ orbits that define our ``inner belt'' source. While it is 
not ideal to combine two different methods -- one that places synthetic bodies into strong 
resonances (see above for 3:1) and one based on real main-belt asteroids (here for the inner belt) 
-- we believe that this is the best practical approach to the problem at hand. The same method 
was used for the Hungaria ($q>1.66$ au, $a<2.05$ au, $i>15^\circ$) and Phocaea ($q>1.66$~au, 
$2.1<a<2.5$ au, $18^\circ<i<30^\circ$) asteroids. The known populations of Hungarias and Phocaeas were 
cloned 4 and 13 times, respectively, to obtain $10^5$ source orbits for each.\footnote{The cloning
consisted in applying a $10^{-6}$ relative change to the velocity vector of each object.}  

The $\nu_6$ resonance, which lies at the inner edge of the asteroid belt, requires a special
treatment. We place orbits in the strongly unstable part of the $\nu_6$ resonance where bodies 
are expected to evolve onto NEO orbits in $<10$ Myr (Morbidelli \& Gladman 1998). The left and 
right borders of the $\nu_6$ source region in $(a,i)$ are defined here as $a_1=2.062+0.00057\, 
i^{2.3}$~au and $a_2=a_1+0.04-0.002\, i$ au, with $i$ in degrees. To define 
the initial $e$ and $i$ distributions in the $\nu_6$ resonance, we consider the distribution of real
asteroids in the strip $a>2.12 + 0.00057\, i^{2.3}$ au and $a<2.18+0.00057\, i^{2.3}$ au, with $i$ 
in degrees. The eccentricity distribution of bodies in the strip can be approximated by a single 
Gaussian with the mean $\mu = 0.16$ and width $\sigma=0.067$, and the inclination distribution 
with a double Gaussian with $\mu_1=5.5^\circ$, $\sigma_1=2.3^\circ$, $\mu_2=15^\circ$, and 
$\sigma_2=3.0^\circ$, and $w_1/w_2=10$ (Table 1). The mean and nodal longitudes are uniformly 
distributed between 0 and $2\pi$ radians. 
We set $\varpi=\varpi_{\rm S}$, where $\varpi_{\rm S}$ is the perihelion longitude of Saturn 
at the reference epoch ($\varpi_{\rm S}=88.98^\circ$ for MJD $=2459600.5$). For each draw, the 
initial semimajor axis is randomly placed between $a_1$ and $a_2$ defined above. 

Nesvorn\'y et al. (2017) developed a dynamical model for Jupiter-family comets (JFCs). In brief,
the model accounted for galactic tides, passing stars, and 
different fading laws. They followed $10^6$ bodies from the primordial trans-Neptunian disk, 
included the effects of Neptune's early migration, and showed that the simulations reasonably well 
reproduced the observed structure of the Kuiper belt, including the trans-Neptunian scattered disk, 
which is the main source of JFCs. The orbital distribution and number of JFCs produced in the model were
calibrated on the known population of active comets. We refer the reader to Nesvorn\'y et al. (2017) 
for further details. 

Here we use the model from Nesvorn\'y et al. (2017) to set up the orbital distribution 
of comets in the NEO region. The comet production simulations from Nesvorn\'y et al. (2017) were 
repeated to have better statistics for $q<1.3$~au. Specifically, every body that evolved from the scattered 
disk to $q<23$ au was cloned 100 times, and the code recorded all orbits with $q<1.3$ au and 
$a<4.5$ au (with a 100-yr cadence). This data represents our model for cometary NEOs. The model includes 
the population of long-period comets but does not account for the long-period comet fading (Vokrouhlick\'y 
et al. 2019). Note that the current orbital distribution of JFCs is largely independent of details of 
the early evolution of the Solar System. We thus do not need to investigate different cases considered 
in Nesvorn\'y et al. (2017). 

In summary, we have 12 sources in total: eight resonances ($\nu_6$, 3:1, 5:2, 7:3, 8:3, 9:4, 11:5 and 2:1),
the forest of weak resonances in the inner belt, two high-inclination sources (Hungarias and Phocaeas),
and comets.

\section{Orbital integrations and binning}

The orbital elements of eight planets (Mercury to Neptune) were obtained from NASA/JPL Horizons for the 
reference epoch (MJD $=2459600.5$). We used the Swift \texttt{rmvs4} $N$-body integrator (Levison \& 
Duncan 1994) to follow the orbital evolution of planets and test bodies ($10^5$ per source).
The integrations were performed with a short time step (12 hours).\footnote{The optimal time step 
was determined by convergence studies. The results with 12 and 18 hour time steps, both in terms
of the orbital distribution produced from different sources and planet impact statistics, were found to 
be practically identical. Long time steps in excess of 1 day generate artifacts in the orbital distribution of 
NEOs with short orbital periods. Granvik et al. (2018) used a 12-hour time step as well.} For each source, 
we used 2000 Ivy Bridge cores of the NASA Pleiades Supercomputer, with each core following 8 
planets and 50 test bodies. The simulation set represented $\sim 10$ million CPU hours in total. 
A test body was removed from the integration when it impacted the Sun, 
one of the planets, or was ejected from the Solar System. All integrations were first run to 
$t=100$ Myr. The test bodies that had NEO orbits ($q<1.3$ au) at $t=100$~Myr were collected
and their integration was continued to $t=500$ Myr. We tested the contribution of long-lived 
NEOs for $t>500$ Myr and found it insignificant.

The orbits of model NEOs were recorded with a 1000-yr 
cadence. This is good enough -- with the large number of test bodies per source -- to faithfully
represent the orbital distribution from each source. For the $\nu_6$ and 3:1 resonances, we also
tested the high-cadence sampling, with the orbits being recorded  every 100 yr, and verified that the results
were practically the same. The high-cadence sampling, however, generated data files that were too 
large to be routinely manageable with our computer resources (hundreds of Gb per source).

The integration output was used to define the binned orbital distribution of NEOs from each source $j$,
${\rm d} p_j(a,e,i) = p_j(a,e,i)\ {\rm d}a\, {\rm d}e\, {\rm d}i$. We tested different bin sizes. 
On one hand, one wishes to represent the smooth orbital distribution as accurately as possible, 
without discontinuities. On the other hand, the {\tt MultiNest} fits become CPU expensive if too
many bins are considered. After experimenting with the bin size, we  
adopted the original binning from Granvik et al. (2018) for the {\tt MultiNest} runs and used
four times finer binning for plots (Fig. \ref{reside}). Table 2 reports the number of bins for the
\texttt{MultiNest} runs and the range of orbital parameters covered by binning.\footnote{Note that, 
as there are many more bins that known NEOs (Sect. 4), most bins do not contain a known NEO.} 

For each source, the orbital distribution was normalized to 1 NEO, 
\begin{equation}
\int_{a,e,i} p_j(a,e,i)\ {\rm d}a\, {\rm d}e\, {\rm d}i = 1\ ,      
\label{one}
\end{equation}   
effectively representing the binned orbital PDF (probability density function). We used the 
orbital range $a<4.2$ au, $q<1.3$~au, $e<1$ 
and $i<90^\circ$, hereafter the NEO model domain, because this is where all NEOs detected by CSS reside
(Sect. 4; except for (343158) Marsyas with $i=154^\circ$). The model can be easily 
extended to include retrograde orbits. As the binning is done only
in $a$, $e$, and $i$, the model ignores any possible correlations with the orbital angles (nodal, perihelion 
and mean longitudes). Some correlations would arise due to orbital resonances with planets 
(JeongAhn \& Malhotra 2014), but we do not investigate this issue here.

Given the vast number of bodies released from each source, the $N$-body integrator records a 
large number of planetary impacts. We record {\it all} impacts, including Mars impacts from 
impactors with $q>1.3$ au (not NEOs), and use this information to compute the impact flux from 
each source. When the source-specific impact fluxes are properly weighted by accounting for the 
size-dependent sampling of sources (Sect. 5.1), we obtain an accurate record of NEO impacts on 
planets (Mercury, Venus, Earth and Mars). These results are discussed in Sect. 7. 

\section{Catalina Sky Survey}

\subsection{Observations}

The Mt. Lemmon (IAU code G96) and Catalina (703) telescopes of the Catalina Sky Survey (CSS;
Christensen et al. 2012) produced nearly 22,000 NEO detections and redetections during the 
8-year long period from 2005 to 2012. The two surveys were complementary to each other, with 
the 1.5-m G96 telescope providing the narrow-field \& deep limiting magnitude observations and
the 0.7-m 703 telescope providing the wide field \& shallow limiting magnitude observations. The survey has a 
carefully recorded pointing history, amounting to well over 100,000 Fields of View (FoVs) for 
each site (for the 2005--2012 period), and a well characterized detection efficiency (Jedicke et al. 
2016). The orbital and magnitude distributions of NEOs detected by CSS were reported in Jedicke
et al. (2016). 

Here we use new detections and accidental redetections of NEOs by CSS -- 4510 individual NEOs in total.
We count each individual NEO only once (i.e., as detected) and do not consider multiple (accidental or
not) detections of the same object. With this setup, we mainly care about the detection
\textit{probability} of an object by CSS, and 
not about the number of images in which that same object was detected (hereafter the 
CSS detection {\it rate}). This has the advantage that we do not have to make decisions about
whether a particular detection was accidental or not.\footnote{Intentional redetections 
(e.g., the same object targeted multiple times) have no information content for our work. 
Accidental redetections of the same object typically happen when the object is relatively 
bright. The accidental redetections could thus improve the statistics for bright objects.} 
We consider the CSS detection rate only to compare our results with Granvik et al. (2018), where the  
accidental redetections were included.  

The detection probability (or bias for short) of an object in a CSS FoV\footnote{More accurately,
  this applies to a set of four FoV with the same pointing direction taken by CSS in short
  succession on the same night (FoV set or `frame'). The detection probability is the probability 
  that the CSS pipeline picks up an object in at least three of these four FoVs. Objects identified 
  in less then three FoVs by the CSS pipeline are not reported as detected.} can be split 
into three parts (Jedicke et al. 2016): (i) the geometric probability of the object to be located 
in the FoV, (ii) the photometric probability of detecting the NEO's tracklet, and (iii) the 
trailing loss.  

\subsection{Geometric probability}

To account for (i), we use the publicly available 
\texttt{objectsInField}\footnote{\url{https://github.com/AsteroidSurveySimulator/objectsInField}}
code ({\texttt{oIF}) from the Asteroid Survey Simulator (AstSim) package (Naidu et al. 2017).
The \texttt{oIF} code inputs several parameter sets: (1) the list 
of survey exposure times (MJD), (2) the pointing direction for each 
exposure, as defined by the right ascension (RA) and declination (DEC) of the field's 
center, (3) the sky orientation in the focal plane (the angle between sky north 
and the 'up' direction in the focal plane), (4) the FoV size and shape (rectangular or circular),
and (5) the observatory code as defined by the Minor Planet 
Center\footnote{\url{https://minorplanetcenter.net/iau/lists/ObsCodesF.html}}. 
The user needs to generate a database (.db)
file, for example with the help of the DB Browser for SQLite\footnote{\url{https://sqlitebrowser.org}},
containing all inputs. We refer the reader to the GitHub documentation of oIF for further 
details.

The oIF code inputs the orbital elements of a body at a reference epoch, propagates it 
over the duration of the survey (using the OpenOrb\footnote{\url{https://github.com/oorb/oorb}}
package, Granvik et al. (2009), and NASA/JPL's Navigation and Ancillary Information Facility (NAIF) 
utilities\footnote{\url{https://naif.jpl.nasa.gov/naif/utilities.html}}), and outputs the list 
of survey's FoVs in which the 
body would appear. To speed up the calculation, oIF uses a series of nested steps where the 
body's position relative to a specific FoV is progressively refined. The orbital propagation 
can use the Keplerian or $N$-body methods. 

\subsection{Photometric efficiency}

Once it is established that a body would geometrically appear in a given FoV, one has to account
for the photometric and trailing loss efficiencies in that FoV (items (ii) and (iii) above) to 
determine whether the object would actually be detected. To aid that, oIF  
reports the heliocentric distance, distance from the observer, and the phase angle of each body 
in each FoV. We can thus consider different absolute magnitudes $H$ of the body in question and 
compute its expected apparent magnitude $V$ in any FoV. This can be done by post-processing the 
oIF-generated output.     

The photometric probability of detection as a function of $V$ (Jedicke et al. 2016) can be given by
\begin{equation}
\epsilon(V) = \frac{\epsilon_0}{ 1 + e^{ ( { V - V_{\rm lim} } ) / V_{\rm width} } } \label{eq1}
\label{photo}
\end{equation} 
where $\epsilon_0$ is the detection probability for bright and unsaturated objects, $V_{\rm lim}$ is 
the (limiting) visual magnitude where the probability of detection drops to $0.5 \epsilon_0$, and 
$V_{\rm width}$ determines how sharply the detection probability drops near $V_{\rm lim}$. In addition,
we set $\epsilon(V)=0$ for $V>V_{\rm lim}+V_{\rm width}$ (Jedicke et al. 2016; no NEOs were detected 
for $V>V_{\rm lim}+V_{\rm width}$). The $\epsilon_0$, $V_{\rm lim}$, and $V_{\rm width}$ parameters were 
reported in Jedicke et al. (2016) for every night of CSS observations. This allows us to 
account for changing observational conditions and simulate CSS observations in detail.
The uncertainties of $\epsilon_0$, $V_{\rm lim}$, and $V_{\rm width}$ were not reported in Jedicke 
et al. (2016). We therefore cannot perform a detailed error analysis where these uncertainties 
would be propagated to the final results. For reference, the average values are $\epsilon_0 = 0.680$, 
$V_{\rm lim} = 19.42$ and $V_{\rm width}= 0.395$ for 703, and $\epsilon_0=0.853$, $V_{\rm lim} = 21.09$, 
and $V_{\rm width}=0.424$ for G96. 

\subsection{Trailing loss}

The trailing loss stands for a host of effects related to the difficulty
of detecting fast moving objects. If the apparent motion is high, the object's image (a streak) 
is smeared over many CCD pixels, which diminishes the maximum brightness and decreases S/N.
Long trails may be missed by the survey's pipeline (due to streaking), the object may not
be detected in enough images of an FoV set (as required for a detection), or the 
streaks in different images may not be linked together. 
The trailing loss is especially 
important for small NEOs; they can only be detected when they become bright,
and this typically happens when they are moving very fast relative to Earth during a close encounter. 
The oIF code provides the rate of motion ($w$ in deg/day) for each FoV where the test object 
was detected. We need to translate this rate into the trailing loss factor and estimate the 
fraction of objects not detected by the survey due to this effect.
 
The trailing loss of CSS was analyzed in Zavodny et al. (2008). It was deduced as a function of 
$V$ and $w$ from a series of CSS images where stars were 'trailed' by tracking at non-sideral 
rates of motion from 1.5 deg/day to 8 deg/day. The results are not available to us on a 
FoV-to-FoV basis -- we only have the 'average' trailing loss reported in Zavodny et al. (2008). 
This can be a source of important uncertainty because the trailing loss is known to vary with 
seeing (Vere\v{s} \& Chesley 2017), and should have varied over the course of CSS observations. 

An alternative method to estimating the trailing loss was proposed in Tricarico (2017), who compared 
the population of {\it known} NEOs that should have been detected by CSS to those actually detected, 
and looked into the overall variation of the detected fraction with $w$. The results were presented 
as the trailing loss average for G96 and 703 and should be representative for the bulk of detections 
($V=18$--20 for 703 and $V=20$--22 for G96). The detection efficiency was given as 
$\epsilon(w)=0.19+0.36/(w-0.06)$ for 703 and $\epsilon(w)=0.56+0.18/w$ for G96, with $0 \leq \epsilon(w) 
\leq 1$ and $w$ in deg/day. 

The CSS trailing loss inferred in Tricarico (2017) is very different -- in terms of the effect's 
overall importance -- from that obtained in Zavodny et al. (2008). For example, in Tricarico
(2017), the 703's detection efficiency drops to $\simeq 0.38$ for $w = 2$ deg/day, whereas 
Zavodny et al. (2008) found a practically negligible effect for $w<5$ deg/day and $V<22$ (for 
both CSS sites). The difference is puzzling. On one hand, Tricarico's method probably more 
closely mimics the actual detection of faint NEOs by CSS than the trailed-star method in 
Zavodny et al. (2008). On the other hand, Tricarico derived $\epsilon(w)$ as a function
of $w$, but not of $V$, while Zavodny et al. (2017) found that the trailing loss is 
sensitive to an object's apparent magnitude. 

Given that two different studies of the CSS trailing loss reported dissimilar results, we 
must make an uneasy choice on how to proceed. In Sect. 6, we first report the results of our 
base model, where we use the trailing loss from Zavodny et al. (2008). This allows us to 
directly compare the results with Granvik et al. (2018), where the same formulation of the 
trailing loss was used. 
Auxiliary NEO models, including those where we use the trailing loss from Tricarico (2017),
are discussed in Sect. 8. We point out that the trailing loss represents an important uncertainty 
in estimating the population of small NEOs, and we urge surveys to carefully characterize it.    

\subsection{CSS bias as a function of $a$, $e$, $i$ and $H$}

The detection probability of CSS, ${\cal P}(a,e,i,H)$, needs to be computed as a function of
$a$, $e$, $i$ and $H$. As we described in Sect. 3, the model NEO orbits are binned (Table 2). We therefore need
to compute ${\cal P}(a,e,i,H)$ in each bin. For each bin, we generated a large number 
($N_{\rm obj}=10,000$; the required number was determined by convergence tests) of test 
objects with a uniformly random distribution of $a$, $e$ and $i$ within the bin boundaries. The 
mean, perihelion and nodal longitudes were randomly chosen between 0 and 360$^\circ$. The \texttt{oIF} 
code was then used to determine the CSS geometric detection probability (or the detection 
rate). For each $H$ bin, we assigned the corresponding absolute magnitude to 10,000 
test NEOs and propagated the information to compute the photometric detection efficiency 
$\epsilon_{\rm P}(V)$ (Eq. \ref{photo}), individually for every FoV, and the trailing loss 
$\epsilon_{\rm T}(w,V)$. The geometric detection probability, $\epsilon_{\rm P}$ and $\epsilon_{\rm T}$ 
were combined to compute the detection probability of each test NEO in every FoV frame.
 
The rate of detection, ${\cal R}(a,e,i,H)$, is defined as the mean number of FoVs 
in which an object with $(a,e,i,H)$ is expected to be detected by the survey. We compute the 
mean detection rate as
\begin{equation}
{\cal R}(a,e,i,H) = \frac{1}{N_{\rm obj}}\sum_{j=1}^{N_{\rm obj}} \sum_{k=1}^{N_{\rm FoV}} \epsilon_{j,k}\ ,
\label{calr}
\end{equation} 
where $N_{\rm FoV}$ is the number of FoVs,
and $\epsilon_{j,k}$ is the detection probability of the body $j$ in the bin $(a,e,i,H)$ and 
FoV $k$.

The detection probability of CSS, ${\cal P}(a,e,i,H)$, is defined as the mean detection 
probability of an object with $(a,e,i,H)$ over the whole duration of the survey. We compute
the mean detection probability as 
\begin{equation}
{\cal P} = \frac{1}{N_{\rm obj}}\sum_{j=1}^{N_{\rm obj}} \bigg\{1-\prod_{k=1}^{N_{\rm FoV}} [1 - \epsilon_{j,k}]\bigg\}\ ,
\label{calp}
\end{equation}
where the product of $1 - \epsilon_{j,k}$ over FoVs stands for the probability of {\it non}-detection of 
the object $j$ in the whole survey. To combine 703 or G96, we have $1-{\cal P} = 
$(1$-{\cal P}_{703}$)$\times$(1$-{\cal P}_{G96}$).

Figures \ref{bias1}--\ref{bias3} illustrate the CSS bias in several examples. We find a good agreement
with the bias used in Granvik et al. (2018) when the CSS detection rate is averaged over the whole orbital
domain and plotted as a function of the absolute magnitude (Fig. \ref{bias1}). Some differences are noted
when the detection rate is plotted for different orbits. For example, our bias tends to vary more smoothly
with the orbital elements than the bias from Granvik et al. (2018). We attribute this to the large statistics
used here (e.g., 10,000 bodies per orbital bin).

The detection probability of CSS is $\gtrsim 0.7$ for large, $H \simeq 15$ NEOs, except for those on 
orbits with $a < 0.8$ au (Fig. \ref{bias2}). Fainter NEOs are detected with lower probability. 
Figure \ref{bias3} illustrates these trends in more detail. Interestingly, ${\cal P}$ shows dips and 
bumps as a function of NEO's semimajor axis (vertical strips in the top panels of Fig. \ref{bias2}). The dips, 
where the detection probability is lower, correspond to the orbital periods that
are integer multiplies of 1 year. This is where the synodic motion of NEOs allow them to hide and not
appear in the survey's FoVs. This effect has been reported before (e.g., Tricarico 2017). The average detection rate
is less sensitive to this effect because the hidden NEOs represent a relatively small fraction of the total
sample and have a small weight in the average when the detection rate is considered.    
          
\section{Parameter optimization with MultiNest}

We use {\tt MultiNest} to perform the model selection, parameter estimation and error analysis 
(Feroz \& Hobson 2008, Feroz et al. 2009).\footnote{\url{https://github.com/farhanferoz/MultiNest}} 
{\tt MultiNest} is a multi-modal nested sampling routine (Skilling et al. 2004) designed to compute 
the Bayesian evidence in a complex parameter space in an efficient manner. The parameter space 
may contain multiple posterior modes and degeneracies in high dimensions. For brevity, we direct 
those interested to the aforementioned works for further details. 

We use the following reasoning to define the log-likelihood in {\tt MultiNest}. Let $n_j$ be 
the number of objects detected by CSS in the bin $j$, and $\lambda_j$ the number of objects 
in the bin $j$ expected from the model. Here the index $j$ goes over all bins in $a$, $e$, $i$ 
and $H$. Assuming the 
Poisson distribution\footnote{More accurately, we should use the binomial distribution with 
the model-estimated probability of detection in the CSS FoV set given by $p=\lambda_j/N_{\rm img}$, 
where $N_{\rm img}=226,824$ is the total number of CSS FoVs. The 
Poisson distribution should be an adequate approximation of the binomial distribution as long as 
$N_{\rm img}$ is large enough {\it and} $\lambda_j$ is small enough. Both these conditions appear 
to be satisfied in the present case.}
with the expected number of events $\lambda_j$, the probability of drawing $n_j$ objects is
\begin{equation}
p_j(n_j) = {\lambda_j^{n_j} \exp(-\lambda_j) \over n_j!} \ . 
\end{equation}    
The joint probability over all bins is then 
\begin{equation}
P = \prod_j {\lambda_j^{n_j} \exp(-\lambda_j) \over n_j!} \ . 
\end{equation}   
The log-likelihood can therefore be defined as 
\begin{equation}
{\cal L} = \ln P = - \sum_j \lambda_j + \sum_j n_j \ln \lambda_j \ ,
\label{like}
\end{equation}
where we dropped the constant term $\sum_j \ln (n_j!)$. This definition is identical to that
used in Granvik et al. (2018), except that here work with the detection probability (not
efficiency) and first detection (i.e., no multiple redetections; Sect 4.1). The second term in Eq. 
(\ref{like}) is evaluated over all bins with 
detected objects. The first term penalizes models with large overall values of $\lambda_j$. For 
two or more surveys, $\cal{L}$ is simply the sum of individual survey's log-likelihoods.   

The models explored here range from simple ones with as few as 7 parameters (3 source weights 
and 4 magnitude distribution coefficients) to complex ones with as many as 30 parameters (12 sources 
with size dependent contributions, cubic spline representation of the magnitude distribution,
magnitude dependent disruption for bodies with low perihelion distance; Granvik et al. 2016). 
We first describe various issues that are common to these models and emphasize differences with 
respect to the previous work -- the tested models are discussed in Sects. 6 and 8. 

The model selection is based on the evidence term $\ln {\cal Z}$ computed by {\tt MultiNest}. The aim is 
to select one model from a set of competing models that represents most closely the underlying process 
that generated the observed data. The models are considered to be a priori equiprobable. To compare two 
models we compute the ratio of their posterior probabilities (the Bayes factor; $\Delta \ln {\cal Z}$) 
and use it to evaluate the statistical preference for the best one. Note that this procedure implicitly 
penalizes models with more parameters.   

There are three sets of priors: (1) coefficients $\alpha$ that determine the strength of different
sources, (2) parameters related to the absolute magnitude distribution, and (3) priors that define 
the disruption model. The motivation for (3) is 
explained in Sect. 5.3 (see Granvik et al. 2016). We limit our analysis to considerations 
based on the absolute magnitude distribution. The albedo and size distribution constraints from 
WISE (Mainzer et al. 2019) will be addressed in a forthcoming publication. 

\subsection{Strength of sources}

As for (1), the intrinsic orbital distribution of model NEOs is obtained by combining $n_{\rm s}$ sources:
$p(a,e,i)=\sum_{j=1}^{n_{\rm s}} \alpha_j\, p_j(a,e,i)$ with $\sum_{j=1}^{n_{\rm s}} \alpha_j = 1$. The
coefficients $\alpha_j$ represent the relative contribution of each source to the NEO population
(i.e., the fraction of NEOs from the source $j$). The binned distribution $p(a,e,i)$ is normalized
to 1 NEO and needs to be supplemented by the absolute magnitude distribution (Sect. 5.2).  

The main difficulty with implementing the $\alpha$ coefficients in \texttt{MultiNest} is that the
Bayesian tools typically work with independent priors. It is therefore not possible, for example, to 
choose each $\alpha_j$ randomly between 0 and 1, and rescale them later such that they sum to 1. 
Using a geometrical approach we found the following general algorithm for assuring that $\alpha_j$ 
have a multivariate, uniformly random distribution, and automatically sum to 1. We generate uniformly 
random deviates $0 \leq X_j \leq 1$ and compute
\begin{equation}
\alpha_j=\left [ 1 - (1-X_j)^{1 \over {n_{\rm s}-j}} \right ]  \left ( 1 - \sum_{k=1}^{j-1} \alpha_k  \right )
\label{alpha}
\end{equation}
for $1 \leq j \leq n_{\rm s}-1$, and
\begin{equation}
\alpha_{n_{\rm s}} = 1 -  \sum_{k=1}^{n_{\rm s}-1} \alpha_k \ .
\label{alpha2}
\end{equation}
The order in which different sources are linked to the index $j$ has no effect on the results. Kipping et al. 
(2013) derived an identical formula for $n_{\rm s} = 3$. The problem in question is related to the Dirichlet 
distribution with equal weights, but it is not immediately obvious to us how to construct an efficient 
algorithm based on that (as the inverse cumulative distribution, CDF, is needed in Eq. (\ref{alpha})).

The contribution of different sources to NEOs may be size dependent. 
This is because the weak orbital resonances in the inner belt are expected to produce an important 
share of large NEOs (Migliorini et al. 1998). Small main-belt asteroids instead drift across large radial 
distances by the Yarkovsky thermal effect (Vokrouhlick\'y et al. 2015), can pass over the weak resonances, and  
reach the strong $\nu_6$ source (Granvik et al. 2017). Granvik et al. (2018) accounted for 
the size dependency by adopting a separate size distribution for each source (see Sect. 5.2). 
Here we set $\alpha_j$ coefficients to be functions of the absolute magnitude. For simplicity, we adopt 
a linear relationship, $\alpha_j=\alpha_j^{(0)} + \alpha_j^{(1)}(H - H_\alpha)$, where $H_\alpha$ is some 
reference magnitude, and $\alpha_j^{(0)}$ and $\alpha_j^{(1)}$ are new model parameters. In practice, using 
Eqs. (\ref{alpha}) and (\ref{alpha2}), we set $\alpha_j(H_{\rm min})$ and 
$\alpha_j(H_{\rm max})$ for some minimum and maximum absolute magnitudes (e.g., $H_{\rm min}=15$ 
and $H_{\rm max}=25$), and linearly interpolate between them. This automatically assures that $\sum_j 
\alpha_j(H)=1$ for any $H_{\rm min}<H<H_{\rm max}$.      

\subsection{Absolute magnitude distribution} 

The differential absolute magnitude distribution is denoted by ${\rm d}n(H) = n(H) {\rm d} H$. Given that the 
magnitude distribution is not seen to wildly vary across the main belt (Heinze et al. 2019), and craters
on the main belt asteroids follow a common size distribution (Bottke et al. 2020), we use a similar setup
for different main-belt sources. Specifically, the magnitude distribution produced by source $j$ is 
set to be ${\rm d}n_j(H) = \alpha_j(H) n(H) {\rm d} H$. The magnitude distributions of different sources 
are similar, but change with $\alpha_j(H)$, which are assumed to linearly vary with $H$ (Sect.~5.1). 
For example, as the $\nu_6$ source is found to contribute more to faint NEOs than to bright NEOs 
(Sect.~6), the magnitude distribution of $\nu_6$ is slightly steeper than ${\rm d} n(H)$. When the 
contribution of different sources is combined, we find that $\sum \alpha_j(H) n(H) {\rm d} H = n(H) 
{\rm d} H$, which means that $n(H)$ stands for the absolute magnitude distribution of the whole 
NEO population. This is a convenient scheme.

Our choice of ${\rm d}n_j(H)$ greatly limits the number of model parameters. For the cubic spline 
representation of ${\rm d}n (H)$ (see below), and $n_{\rm s}$ sources, we have $2 n_{\rm s} + 5$ parameters
in total
($2 n_{\rm s}$ $\alpha$'s and 5 parameters defining ${\rm d}n (H)$). For comparison, Granvik et al. (2018) 
used different magnitude distributions for individual sources, in which each distribution was 
represented by the 3rd-order polynomial with 4 coefficients. This gives $4 n_{\rm s}$ parameters in total. 
The setup in Granvik et al. (2018) can account for large magnitude-distribution differences between 
different sources. With too many parameters, however, the model can be over-parameterized  and
not all the parameters can be constrained from the existing observations. 

Granvik et al. (2018) defined the magnitude distribution of each source using a smooth, second-degree 
variation of the differential slope. In terms of the log-cumulative magnitude distribution, 
$\log N(H)$, this is equivalent to a third-order polynomial representation: $\log_{10} N(H)= \log_{10} N_{\rm ref} 
+ \gamma (H-H_{\rm ref}) + \delta (H-H_{\rm c})^3$, where $N_{\rm ref}$ is the normalization constant,
$H_{\rm ref}$ is some constant reference magnitude ($H_{\rm ref}=17$ in Granvik et al.), $\gamma$ is 
the slope of the linear term, and the cubic term is centered at $H_{\rm c}$ and has the `twist' 
amplitude $\delta$. In this case, there are four free parameters for each source: 
$N_{\rm ref}$, $\gamma$, $\delta$ and $H_{\rm c}$.   

We tested this parameterization in our model and found that it has undesired limitations. First,
$\log_{10} N(H)$, as given above, is symmetric around $H_{\rm c}$, but the real magnitude distribution 
of NEOs is not symmetric; it is gently rounded just below $H=20$ but 
has a sharper dip leading to a steeper slope for $H>20$ (e.g., Harris \& D'Abramo 2015). It then 
becomes difficult to accurately 
fit observations in this model because the cubic polynomial representation is simply too rigid. In Granvik 
et al. (2018), the {\it asymmetric} magnitude distribution of NEOs was composed from many different 
sources each having a {\it symmetric} distribution (around a different $H_{\rm c}$ value). This should have 
produced some tension in the fit. Second, given the rigid nature of the cubic polynomial with a 
twist, the fit near $H=25$, where the magnitude distribution is steep, would influence the fit at $H=15$.
This is not desirable as the model should have enough flexibility to deal with the bright and faint bodies 
separately. Third, the cubic polynomial is difficult to generalize to a wider absolute magnitude range 
and/or higher accuracy. Higher-order polynomials, for example, have the inconvenient property that 
the polynomial coefficients sensitively depend on the order; they wildly change if the order is increased.       
 
Here we use cubic splines to represent $\log_{10} N(H)$. The magnitude interval of 
interest, $15<H<25$ for our base CSS model (Sect. 6), is divided into several segments. The more sections there are the 
more accurate the parameterization is, but we also have more parameters to deal with. After experimenting 
with different choices we opted for four segments and five parameters. There are four parameters defining
the average slope in each segment, $\gamma_j$, and one parameter that provides the overall calibration. 
We typically use $N_{\rm ref}=N(H_{\rm ref})$ with $H_{\rm ref}=17.75$ (diameter $D=1$ km for the reference 
albedo $p_{\rm V}=0.14$). The normalization constant and slope parameters are used to compute 
$\log_{10} N(H)$ at the boundaries between segments; cubic splines are constructed from that 
(Press et al. 1992). The splines assure that $N(H)$ smoothly varies with $H$. This representation 
has the desired properties: it is accurate, flexible, and can easily be generalized by adding 
more segments.\footnote{Initially, we sectioned the magnitude range $15<H<25$ evenly by having four 
intervals $H=15$--17.5, 17.5--20, 20--22.5, and 22.5--25, and found that the use of splines led to 
a substantial improvement of fits relative to those obtained with the third order polynomial. 
The results further improved when the division between the third and fourth intervals was set at 
$H=24$ (instead of 22.5). This is related to the asymmetry of the underlying distribution which is 
reproduced slightly better when the third and fourth segments have unequal lengths.}

Optionally, we can use additional constraints to inform the \texttt{MultiNest} fits. For example, the 
known sample of NEOs with $H<15$ is complete, and there are $\simeq50$ such objects in the JPL Small
Bodies Database.\footnote{\url{https://ssd.jpl.nasa.gov/tools/sbdb\_query.html}}
We can therefore fix $N(15)=50$ and compute the $\gamma_1$ slope such that this 
additional constraint is satisfied. With this, we only have four absolute magnitude distribution parameters 
in the \texttt{MultiNest} fit.

\subsection{Disruption model}

To account for the disruption of NEOs at small perihelion distances, following Granvik et al. (2016), we 
eliminate test bodies when they reach critical distance $q^*$. Granvik et al. (2016) found that
$q^*$ is a function of size with small NEOs disrupting at larger perihelion distances than the large ones. 
To demonstrate this, Granvik et al. (2016) divided the absolute magnitude range into three intervals, 
$H=17$--19, 20--22 and 23--25, and performed separate fits to CSS in these three cases. They found 
that $q^*(H)$ is roughly linear in $H$ with $q^* \simeq 0.06$~au for $H=17$--19, $q^*\simeq 0.12$ au 
for $H=20$--22, and $q^* \simeq 0.18$ au for $H=23$--25. We tested the same method here and found 
results consistent with Granvik et al. (2016). 

Performing separate fits in different magnitude ranges is somewhat awkward (because there are many 
other parameters to explore as well). 
Granvik et al. (2018) therefore used a different method where the effect of disruptions was 
approximated by a penalty function $P(a,e)=1-k[q_0-a(1-e)]$ for $q<q_0$ and $P(a,e)=1$ otherwise.
The two parameters of the penalty function, $k$ and $q_0$, which have some (unspecified)
relationship to $q^*$, were estimated from the CSS fit (Granvik et al. 2018). Given that 
the penalty function only depends on $a$ and $e$, this method cannot accurately reproduce the 
real effect of disruptions. This is because, when bodies are removed at $q^*$, it not only affects 
the $(a,e)$ distribution, but it also influences the inclination distribution (it becomes narrower
for shorter lifetimes) and absolute magnitude distribution (as $q^*$ is size dependent).
We find that this is not a minor issue (Fig. \ref{disrupt}). 
 
To circumvent these problems, here we assume that the $q^*$ dependence on $H$ is roughly linear, and parameterize 
it by $q^* = q_0^* + \delta q^* (H-H_q)$, where $H_q=20$. We use uniform priors for the two parameters, 
$q_0^*$ and $\delta q^*$. To construct the orbital distribution for any $q^*<0.3$ au, we first produce
the binned distributions (from each source) for $q^*=0$, 0.05, 0.1, 0.15, 0.2, 0.25, and 0.3 au. This 
is done by following the orbit of every simulated object and recording the time $t^*$ when the object 
reached $q<q^*$ for the first time. The binning is done for $t<t^*$. The object is assumed to disrupt 
at $t=t^*$ and not included for $t>t^*$. The fitting routine then linearly interpolates between  
distributions obtained with different $q^*$ to any intermediate value of $q^*(H)$. The resulting orbital 
distribution, $p_{q^*}$, which now also depends on the absolute 
magnitude, $p_{q^*}=p_{q^*}(a,e,i,H)$, is normalized to 1 ($\int p_{q^*}(a,e,i,H)\ {\rm d}a\, {\rm d}e\, 
{\rm d}i = 1$ for any $H$). 
         
The method described above assures that a single fit can be performed globally, for the full range 
of $H$, and at the same time we are using a physically-based approach to modeling the 
size/magnitude-dependent disruption distance. The linear dependence of $q^*$ on $H$ could be 
generalized to a more complex functional form when the need for that arises.

\subsection{Model summary} 

In summary, our biased NEO model is
\begin{equation}
{\cal M}_{\rm b}(a,e,i,H) =  n(H)\, {\cal P}(a,e,i,H) \sum_{j=1}^{n_s} \alpha_j(H)\, p_{q^*,j}(a,e,i,H)\, \ , 
\label{model}
\end{equation}
where $\alpha_j$ are the magnitude-dependent weights of different sources ($\sum_j \alpha_j(H) = 1$), 
$n_s$ is the number of sources, $p_{q^*,j}(a,e,i,H)$ is the PDF of the orbital distribution of NEOs 
from the source $j$, including the size-dependent disruption at the perihelion distance $q^*(H)$ (this
is the only $H$-dependence in the $p$ functions), 
$n(H)$ is the differential absolute-magnitude distribution of the NEO population (the log-cumulative
distribution is given by splines; Sect 5.2), and ${\cal P}(a,e,i,H)$ is the CSS detection probability
(Eq. \ref{calp}). For each \texttt{MultiNest} trial, Eq.~(\ref{model}) is constructed by the methods
described above. This defines the expected number of events $\lambda_j = {\cal M}_{\rm b}(a,e,i,H)$ 
in every bin of the model domain, and allows \texttt{MultiNest} to evaluate the log-likelihood 
from Eq. (\ref{like}).

The intrinsic (debiased) NEO model is simply
\begin{equation}
{\cal M}(a,e,i,H) =  n(H)\, \sum_{j=1}^{n_s} \alpha_j(H)\, p_{q^*,j}(a,e,i,H)\, \ . 
\label{model2}
\end{equation}  
By integrating Eq. (\ref{model2}) over the orbital domain, given that 
$\int p_{q^*,j}(a,e,i,H) \ {\rm d}a\, {\rm d}e\, {\rm d}i = 1$ and $\sum_j \alpha_j(H) = 1$, 
we verify that $n(H)$ stands for the (differential) magnitude distribution of the whole NEO population. 

\section{The base NEO model}

Our base NEO model accounts for $n_{\rm s}=12$ sources. Each source has a magnitude-dependent 
contribution (Sect. 5.1) and the source weights $\alpha_j(15)$ (for $H=15$) and $\alpha_j(25)$ (for $H=25$)
therefore represent $2(n_{\rm s}-1)$ model parameters (the last source's contribution is computed from
Eq. (\ref{alpha2})). There are four parameters related to the magnitude distribution,
$N_{\rm ref}$ and $\gamma_j$, $2 \leq j \leq 4$ ($15 \leq H \leq 25$).
The $\gamma_1$ parameter is fixed such that $N(15)=50$ (Sect. 5.2). In addition, the 
$q_0^*$ and $\delta q^*$ parameters define the disruption model. This adds to 28 model parameters in total. 
We used uniform priors for all parameters (see Sect. 5.1 for the multivariate uniform distribution
of $\alpha_j(15)$ and $\alpha_j(25)$). The CSS fits were executed with the \texttt{MultiNest} code
running on 2000 Ivy Bridge cores of the NASA Pleiades Supercomputer. Each fit required at least 
four Wall-clock hours to fully converge. 

The base model, as presented here, was identified by the Bayes factor analysis (Sect. 5). We generated 
a large number of rival models (about 50; Sect. 8) and computed their Bayes factors relative to the 
base model. These models tested the magnitude-{\it in}dependent $\alpha_j$, disregarded disruption of 
NEOs at small perihelion distances, adopted constant $q^*$ (independent of $H$), etc. The analysis showed 
an overwhelming statistical preference for the base model, ${\cal M}$. For example, the non-disruption 
and constant-$\alpha$ models are disfavored by $\Delta \ln {\cal Z}>20$ relative to ${\cal M}$. 
The models with fewer than 12 sources are disfavored by at least $5 \sigma$ relative ${\cal M}$,
except for the models without 7:3, 9:4, JFCs, or 11:5 (see below). There is a correlation between 
$\ln {\cal Z}$ and $n_{\rm s}$ with higher-$n_{\rm s}$ models generally giving higher Bayesian evidences. 
This probably means that the NEO population is supplied from a large number of sources and the 
CSS observations are sufficiently diagnostic to establish that.

Four rival models showed evidence terms comparable to the base model. The 11-source models without the 
7:3, 8:3 or JFC sources are favored by factors of 33, 18 and 3.7, respectively, relative to the
base model. The model without the 11:5 source is disfavored by a factor of 8.2 relative to the base
model. This means that the optimal model would be a 9-source model without 7:3, JFCs and 9:5 (but 
keeping 11:5). Here we prefer to report the results of the 12-source base model, because some of the Bayes 
factors reported above are relatively small. The base model also provides upper limits on the
contribution of these weak sources (see below).

\texttt{MultiNest} provides the posterior distribution of model parameters 
(Fig. \ref{triangle}).\footnote{Note that the posterior distribution does not account for uncertainties 
related to the photometric detection of NEOs by CSS (Sects. 4.3 and 4.4). The CSS photometric detection 
uncertainties are unavailable to us.} The posterior distribution is well behaved for most parameters (i.e., 
unimodal and Gaussian like). In some cases, the fit provides an upper bound on the contribution of a 
specific source. This most clearly happens for the 7:3 and 9:4 resonances, which are located in the 
sparsely populated region of the outer belt, and for JFCs. We use the posterior distribution to compute 
the median and standard 1$\sigma$ (68.3\% confidence interval) uncertainties of model parameters 
(Table 3). For parameters, for which the posterior distribution peaks near zero (e.g., 
the contribution of 7:3, 9:4 and JFCs), we also report the upper limit in Table 3.  For bright NEOs, 
for which the contribution of these weak sources was found to be slightly more substantial, we obtained 
$\alpha_{7:3}(15) < 0.012$, $\alpha_{9:4}(15)<0.020$, and $\alpha_{\rm JFC}(15)<0.017$ 
(68.3\% envelopes). The contribution of JFCs to the NEO population is inferred to be smaller than in 
previous works (e.g., $\simeq 6$\% contribution in Bottke et al. (2002), and 2--10\% $H$-dependent 
contribution in Granvik et al. (2018)). For faint NEOs ($H \simeq 25$), all middle and outer belt resonances, 
except for 5:2, have $\alpha(25) < 0.02$ (68.3\% envelopes). This implies that the contribution 
of the middle/outer belt to very small NEOs is minor.

We note several correlations between model parameters. A notable degeneracy is related to the 
contribution of the $\nu_6$ resonance and weak resonances in the inner main belt (Fig.~\ref{degenerate}). 
The orbital distributions produced by these sources are similar and \texttt{MultiNest} 
has difficulties to distinguish between them for $H=15$. Bottke et al. (2002) already discussed a related 
degeneracy between the $\nu_6$ resonance and their Intermediate Mars Crossers (IMC) source. 
There is a hint of correlation between $\nu_6$ and weak resonances even for $H=25$, where we only 
have an upper limit on the contribution of inner resonances. Faint NEAs detected by CSS are 
apparently more diagnostic for distinguishing these two sources.\footnote{Note that the residence time 
distribution from $\nu_6$ and inner resonances are similar but not equal; the degeneracy 
between these two sources is therefore not absolute. The inner resonances show the orbital distribution 
more peaked for $a>2$ au, whereas $\nu_6$ produces more evolved orbits with $a<2$ au.}
  
Additional correlations can be identified in Fig. \ref{triangle}. For example, $N_{\rm ref}$ and $\gamma_2$ 
are anti-correlated (labels 23 and 24 in Fig. \ref{triangle}), indicating that the models with lower 
$N_{\rm ref}$ require a steeper magnitude slope for $17.5<H<20$. Interestingly, the contributions of 
some individual sources, such as $\nu_6$, 3:1 and 5:2, to faint and bright NEOs are anti-correlated.
We speculate that this happens because the total contribution of a source to faint and bright NEOs 
is relatively well constrained from CSS. A smaller contribution for $H=15$ would then require a larger 
contribution for $H=25$ for things to balance. Other possibilities exist as well. 

The biased base model ${\cal M}_{\rm b}$ is compared to CSS NEO detections in Figs. \ref{bmodel} and 
\ref{dif}. The distributions in Fig. \ref{bmodel} are broadly similar. The 1D PDFs in Fig. \ref{dif} 
show the comparison in more detail. The model distribution in Fig. \ref{dif}(a) has the 
overall shape of CSS observations but the two semimajor-axis peaks at 1.5--2.4 au do not exactly 
align (they are shifted by 0.1--0.2 au). Statistical fluctuations may be responsible for 
this difference. We applied the Kolmogorov-Smirnov (K-S) test to CDFs corresponding to the 
distributions shown in Fig. \ref{dif} and found that the semimajor axis model distribution 
is not rejectable (K-S probability 9.7\%). The model $e$, $i$ and $H$ distributions match observations 
well (K-S probabilities 14\%, 32\% and 61\% for the eccentricity, inclination and absolute 
magnitude, respectively).

The base model correctly reproduces various orbital correlations with $H$. To demonstrate this,
we slice PDFs using different absolute magnitude ranges and show the results in Fig. \ref{bright}.
For example, the inclination distribution for $H=15$--20 is broader than the one 
for $H=20$--25. The eccentricity distribution is pyramidal in shape for $H=15$--20 and becomes 
more peaked for $H=20$--25. An interesting feature, which is not reproduced quite well in 
the model, is the population of faint NEOs with $H=20$--25, $a \simeq 1$--1.6 au and $e<0.4$ 
(K-S test probabilities $10^{-4}$ and 0.012 for $a$ and $e$, respectively). This population is 
not present in the CSS detections for $H<20$ and gradually appears for fainter NEOs.

The intrinsic (debiased) absolute magnitude distribution from our base model is shown in Fig.~\ref{harris}. 
It is practically identical ($<2$$\sigma$ difference for $17<H<25$) to that reported in Harris \& 
Chodas (2021). For $H<17$, the 3$\sigma$ envelope shown in Fig. \ref{bright} shrinks because we 
fixed $N(15)=50$ -- here the NEO population given in Harris \& Chodas (2021) is slightly higher.   
For reference, Harris \& Chodas (2021) obtained 4,625, 15,880 and $3.13\times10^5$ NEOs with $H<19.75$, 
$H<21.75$ and $H<24.75$, respectively (the magnitude cuts are given here to avoid problems with 
rounding of the magnitude values reported by JPL/MPC; Harris \& Chodas 2021). No error estimates 
were reported in Harris \& Chodas (2021). From our base model we find $4580\pm160$, 
$16020\pm550$ and $(2.89 \pm 0.15) \times10^5$ NEOs with $H<19.75$, $H<21.75$ and $H<24.75$, 
respectively, in very close agreement with Harris \& Chodas (2021). The relative 1$\sigma$ uncertainty 
of our estimates gradually increases from $\simeq3$\% for $H<20$ to $\simeq6$\% for $H<25$.
The uncertainty reported here was computed from the \texttt{MultiNest} posterior sample and does not 
account for various uncertainties related to the CSS detection efficiency (Sects. 4.3 and 4.4). As the 
CSS detection efficiency uncertainty likely increases with $H$ (e.g., due to issues related to 
the trailing loss; Sect. 4.4), our NEO-population estimates should become significantly more 
uncertain for faint magnitudes ($H \gtrsim 25$).\footnote{The absolute magnitude distribution given 
in Table 6 in Granvik et al. (2018) has the shape similar to ours but indicates a somewhat larger 
population of NEOs for $H > 20$ (Sect. 8).} The magnitude distribution in the extended magnitude 
range $15<H<28$ is discussed in Sects. 8 and 10. 

Heinze et al. (2019) estimated the slope of the absolute magnitude distribution for main belt asteroids.
They found $\gamma \simeq 0.22$ for $H=20$--23.5 and $\gamma \simeq 0.34$ for $H=23.5$--25.6.
Here our base NEO-population model suggests $\gamma \simeq 0.328 \pm 0.004$ for $H \simeq 20$ (Table 3) and 
a steeper slope for $H \simeq 25$ ($\gamma \simeq 0.566\pm 0.014$). This is roughly consistent with the 
results of Heinze et al. (2021), who found $\gamma=0.31$--0.34 for NEOs with $H \simeq 18$--22 and 
$\gamma=0.54$--0.57 for NEOs with $H \simeq 23$--28. The magnitude distribution of NEOs for $20 
\lesssim H \lesssim 25$ therefore appears to be significantly steeper ($>5$$\sigma$ difference) than 
that of main belt asteroids, but not much steeper ($\simeq 0.1$--0.2 difference in the slope index 
$\gamma$). This result is most likely related to the size-dependent delivery of main belt asteroids, 
via the Yarkovsky thermal force, to source resonances (e.g., Morbidelli \& Vokrouhlick\'y 2003). 


Various issues related to the photometric detection efficiency of CSS limit our ability to 
accurately predict the number of km-sized NEOs. The \texttt{MultiNest} fit gives $N(17.75)=931 \pm 30$
($H=17.75$ corresponds to $D=1$ km for $p_{\rm V}=0.14$), but the uncertainty given here does not account for 
the uncertainty in the CSS detection efficiency.\footnote{$N(17.75)$ reported here differs from $N_{\rm ref}$
given for $H_{\rm ref}=17.75$ in Table 3, because the $N_{\rm ref}$ parameter is defined by linear
interpolation (Sect. 5.2). $N(17.75)$, which stands for the number of NEOs with $H<17.75$, is obtained
from splines.} As we noted in Sect. 4, the uncertainties of parameters $\epsilon_0$, $V_{\rm lim}$, and 
$V_{\rm width}$ were not given in Jedicke et al. (2016). Ideally, we would need these uncertainties on a 
nightly basis. The changes of $\epsilon_0$ from night to night of CSS observations, which could be taken 
as a very conservative proxy for the uncertainty in the detection probability of bright NEOs, are 
$\sim 10$\% (Jedicke et al. 2016). The accurate characterization of survey's detection efficiency 
and its uncertainty is of the foremost importance for accurate population estimates.

We find that different main-belt sources have different contributions to small and large NEOs (Fig. 
\ref{alphas}). The models with the {\it size-independent} contribution of different sources are 
statistically disfavored ($\Delta \ln {\cal Z}>20$ relative to the base model) and can be 
ruled out. This relates back to Valsecchi \& Gronchi (2015) who pointed out that the orbital
distribution of bright NEOs ($H<16$) is significantly different from the model distribution in
Bottke et al. (2002). Granvik et al. (2018) already identified some complex size dependence in the NEO 
delivery process. Other works also speculated that the delivery process is size dependent 
(e.g., Nesvorn\'y et al. 2021). Here we find that the $\nu_6$ and 3:1 resonances jointly contribute 
to $\simeq 30$\% of $H=15$ NEOs and $\simeq 80$\% of $H = 25$ NEOs.\footnote{The $\nu_6$, inner weak 
and 3:1 resonances jointly contribute to $\simeq 54$\% of $H=15$ NEOs.}
This most likely happens because small main-belt asteroids radially drift by the Yarkovsky 
effect, pass through weak resonances, and reach the powerful $\nu_6$ and 3:1 resonances. Large main belt 
asteroids do not move much and are more likely to be removed from the asteroid belt by weaker resonances 
(Migliorini et al. 1998; Sect. 10). The $\nu_6$ resonance shows the strongest dependence on size 
with the $\simeq 10$\% contribution for $H =15$ and $\simeq 40$\% contribution for $H = 25$. The weak 
resonances in the inner main belt are found to produce over 20\% of NEOs with $H=15$, but their share 
drops to $<7$\% (1$\sigma$ limit) for $H = 25$ (Table~3). The contributions of $\nu_6$ and inner main-belt 
resonances show an anti-correlated dependence on size (Fig. \ref{alphas}).

We confirm the need for the size-dependent disruption of NEOs at small perihelion distances as originally 
pointed out in Granvik et al. (2016). The models without disruption are statistically disfavored 
($\Delta \ln {\cal Z}>20$ relative to the base model) and can be ruled out. Clearly, any model 
where the disruption is not taken into account produces a strong excess of low-$q$ (or high-$e$) orbits. 
The $q^*(H)$ dependence found here roughly matches the one inferred in Granvik et al. (2016), 
which is perhaps not that surprising given that we use the similar methodology and constraints
as Granvik et al. (2016). Figure \ref{qstar} shows the maximum likelihood base model with $q^*(18) 
\simeq 0.08$ au (compared to $q^*\simeq0.06$ au for $17<H<19$ in Granvik et al.) and $q^*(24) \simeq 0.2$ au
(compared to $q^*\simeq0.18$ au for $23<H<25$ in Granvik et al.). Based on this result we could tentatively 
suggest that the NEO disruption happens at a slightly larger perihelion distances than found in 
Granvik et al. (2016). However, given that there is some variability between different models (Sect. 8),
we believe that more work is needed to establish the $q^*(H)$ dependence with more confidence.     

The size-dependent contribution of main-belt sources and the size-dependent disruption of NEOs at 
small perihelion distances implies that the orbital distribution of NEOs must be size-dependent as well. 
Figure \ref{intr} compares the orbital distributions of large ($15<H<17.5$) and small ($22.5<H<25$)
NEOs. There are several differences. 
The eccentricity and inclination distributions of large NEOs are more extended than those of 
small NEOs. This is a direct consequence of the size-dependent disruption that favors removal 
of small NEOs with $e>0.6$. The inclination distribution of large NEOs is more extended because
large NEOs are less likely be disrupted; they tend to survive longer, thus allowing the inclination 
distribution to become increasingly wider over time. 

The results presented here can be used to estimate the completeness of the currently-known NEO
population. We illustrate the current completeness for $H<22$ in Fig. \ref{incom}. We find that the known 
population of $H<22$ NEOs is roughly a factor of $2$ incomplete ($\simeq 10,000$ known vs. 
$18,900\pm700$ estimated; Table 4). Undiscovered NEOs populate a wide range of orbits. The 
incompleteness rapidly increases 
toward fainter magnitudes. For example, for $H<25$, our model predicts that there are $\simeq20$ times 
more NEOs than currently known (i.e., the known population is only $\simeq 5\%$ complete). For $H 
\simeq 25$ we only know 1 in $\simeq 100$ NEOs (see the differential distribution in the bottom-right 
panel of Fig. \ref{incom}). We discuss the NEO population completeness in more detail in Sects. 8 and 10.

To aid similar estimates, and help to plan future observations, we developed the NEOMOD Simulator. 
The code inputs the base (or any other) model from \texttt{MultiNest}, provided as an ASCII table, 
and generates a user-defined sample of NEOs with the smooth orbital and absolute magnitude distributions. 
The NEOMOD Simulator will be available from \texttt{GitHub}.\footnote{See 
\url{https://www.boulder.swri.edu/\~{}davidn/NEOMOD\_Simulator} for a provisory distribution.} 
Fig. \ref{neomod} illustrates an example output from the NEOMOD 
Simulator, where the user requested to generate the full sample of $H<25$ NEOs from the base  
model described here. Statistically different NEO samples can be obtained by initializing the 
code with different random seeds. 

Potentially hazardous objects (PHOs) are defined as having a minimum orbit intersection distance (MOID) 
with Earth of less than 0.05 au (19.5 lunar distances) and $H \leq 22$ ($D \simeq 140$ m for 
$p_{\rm V}=0.14$). We used the code described in Wi{\'z}niowski \& Rickman (2013) to estimate the 
number of PHOs as a function of orbital elements. 10,000 objects were placed into each orbital bin in
$a$, $e$ and $i$, their nodal and perihelion longitudes were drawn from a uniformly random 
distribution, and MOID was computed for each orbit. We then evaluated the fraction of PHOs, following 
the definition above (MOID $< 0.05$ au), in each bin. The PHO fraction is the largest for orbits with 
$q \sim 1$ au, $Q=a(1+e)\sim 1$ au, $a \sim 1$ au, and/or $i<10^\circ$.    
Figure \ref{pho2} shows the completeness of the currently-known PHO population. The trends seen here 
are similar to those discussed for the whole NEO population above. The bulk of yet-to-be-discovered 
PHOs have orbits with $1.2<a<2.8$ au, moderate to large eccentricities, and $i \lesssim 40^\circ$. 
The PHO population completeness is $>90$\% for $a < 1.2$ au, $e < 0.3$ and $H<22$. This is because 
NEOs on these orbits have low MOID and can be more easily detected than NEOs in general. We find that there are $4000\pm150$ 
PHOs with $H<22$ in total, of which $\simeq 2300$ are known. The overall population completeness is slightly 
higher for PHOs ($\simeq 58$\%) than for $H<22$ NEOs in general ($\simeq 52$\%).   

\section{Planetary impacts}

Planetary impacts were recorded by the $N$-body integrator (Sect. 3). The record accounts for impacts of 
bodies with $q<1.3$ au (NEOs) and $q>1.3$ au (e.g., Mars-crossers). We thus have complete information 
to determine the impact flux on all terrestrial planets, including Mars. We followed $10^5$ test bodies from 
each source and have good statistics even from distant main belt sources (e.g., 9:4, 2:1). 
We find, in line with the results reported previously (e.g., Gladman et al. 1997, Bottke et al. 2006), that 
the impact probability per one body inserted in the source, $p_{\rm imp}$, strongly declines with the 
heliocentric distance of that source. For example, Hungarias, $\nu_6$ and weak inner-belt resonances have 
$p_{\rm imp} \simeq 0.01$--0.02 for impacts on the Earth, but $p_{\rm imp} \simeq 10^{-4}$ for the outer belt 
resonances such as 2:1 (Table 5). This happens for two reasons. First, the NEOs produced by distant 
sources typically end up having larger $a$ and $e$, and thus lower (intrinsic) impact probabilities with 
the Earth. Second, these NEOs have shorter dynamical lifetimes, $\tau$ (defined as the time interval 
spent with $q<1.3$ au) and are often removed before they can impact. For example, the $\nu_6$ source has 
$\tau \simeq 6.6$ Myr for a reference value $q^*=0.1$ au, and impacts from $\nu_6$'s NEOs on the Earth 
thus happen over this relatively long time interval. The 2:1 resonance produces much shorter lifetimes 
(e.g., $\tau \simeq~0.41$~Myr for $q^*=0.1$ au). 

Once the contribution of different sources to the NEO population is fixed,\footnote{Note that 
the NEO population is used here to calibrate the model but the impact statistics inferred from this
calibration accounts for impactors with $q>1.3$ au as well. This is because the $N$-body integrator 
recorded {\it all} planetary impacts, including those from $q>1.3$ au.} via the weights 
$\alpha_j$, we may ask how important each source is for planetary impacts. For that, we must fold in
both $p_{\rm imp}$ and $\tau$. The best way to accomplish this is to consider the {\it impact flux},
$f_{\rm imp}$, which is related to the impact probability and lifetime by $f_{\rm imp}=p_{\rm imp}/ \tau$.
Interestingly, the impact flux shows a much weaker dependence on the heliocentric distance of a source 
than the impact probability (Table 5). The low impact probabilities from 
more distant resonances are apparently compensated by shorter dynamical lifetimes. This suggests 
that the distant resonances could provide a surprisingly large share of impacts. For 
example, $f_{\rm imp}^{\nu_6} \simeq 0.003$ Myr$^{-1}$ per one body from the $\nu_6$ resonance and  
$f_{\rm imp}^{\rm 8:3} \simeq 0.0015$ Myr$^{-1}$ per one body from the 8:3 resonance (both given for 
the Earth and $q^*=0.1$ au). For large NEOs ($H=15$ corresponding to $D \simeq 3.5$~km for 
$p_{\rm V}=0.14$), we have $\alpha_{\nu_6}(15) \simeq 0.12$ and $\alpha_{\rm 8:3}(15) \simeq 0.09$ 
(Table 3). Combining these factors together we infer that the $\nu_6$ resonance contributes 
(only) $\sim 2.7$ times as many impacts as the 8:3 source for large impactors.

The situation dramatically changes when we consider impacts of small NEOs. For $H=25$ 
($D \simeq 35$ m for $p_{\rm V}=0.14$), we have $\alpha_{\nu_6}(25) \simeq 0.43$ and 
$\alpha_{8:3}(25) \simeq 0.010$ (Table~3), and the weighted impact flux ratio between the two resonances 
is thus $\simeq 90$. The low share of impacts from the 8:3 source is primarily the consequence 
of the size-dependent sampling of main-belt sources discussed in Sect. 6. The $\nu_6$ 
source is responsible for most impacts of small bodies on the terrestrial worlds. [An
impact is defined here when a body hits the top of planet's atmosphere. The atmospheric ablation
of small impactors and possible reduction of the impact flux on planet's surface is not 
considered.]      


    
To combine impacts from different sources, we compute the total impact flux, $F_{\rm imp}$, from  
\begin{equation}
F_{\rm imp} = n(H) \sum_{j=1}^n \alpha_j(H) {p_{{\rm imp},j}(q^*(H)) \over \tau_j (q^*(H))}\ ,  
\label{fimp}
\end{equation}  
where $n(H)$ is the absolute magnitude distribution of NEOs, $\alpha_j(H)$ are the 
magnitude-dependent source weights (Table 3), $p_{{\rm imp},j}$ is the probability of planetary 
impact for each body inserted in the source $j$, and $\tau_j$ is the mean lifetime of NEOs 
evolving from the source $j$. Parameters $p_{{\rm imp},j}$ and $\tau_j$ depend on $q^*$ and are 
therefore also a function of $H$ (via the linear relationship between $q^*$ and $H$, as defined 
in the base model; Fig. \ref{qstar}). We report them for a reference value $q^*=0.1$ au in
Table 5. 
 
Figure \ref{earth} shows $F_{\rm imp}(H)$ for the terrestrial planets. A rough approximation of the 
impact flux on the Earth was traditionally obtained when the magnitude distribution of NEOs, $n(H)$, 
was multiplied by constant collision probability ($1.5 \times 10^{-3}$ Myr$^{-1}$ for 
each NEO; Stuart 2001, Harris \& D'Abramo 2015). The $H$-dependent factors in Eq.
(\ref{fimp}), however, produce a more complex relationship between $n(H)$ and $F_{\rm imp}(H)$.
For example, Eq. (\ref{fimp}) gives  $\simeq 970$ impacts per Myr of $H<25$ NEOs on the Earth, whereas the 
approximate estimate from $n(H)$ would only give $\simeq 610$ impacts. It is therefore important
to carefully account for various size dependencies in Eq. (\ref{fimp}). For reference, we estimate 
one impact on the Earth from $H<17.75$ NEOs ($D>1$ km for $p_{\rm V}=0.14$) every $\simeq$ 630 kyr, 
in close agreement with the estimates given in Harris \& D'Abramo (2015) and Morbidelli et al. (2020). 

There are several sources of uncertainty in our impact flux estimates. The first one is related to 
the uncertainty of the NEO population estimate in Eq. (\ref{fimp}). As we discussed in Sect. 6, 
the relative 1$\sigma$ uncertainty of our base-model population estimate gradually increases from 
$\simeq 3$\% for $H<20$ to $\simeq 6$\% for $H<25$. The second source of uncertainty is the uncertainty 
of the impact fluxes $f_{{\rm imp},j}$ for bodies evolving from individual sources (Table 5). 
This uncertainty varies with source, target planet and $q^*$. In the best case, we record 
thousands of impacts on the Earth/Venus from the $\nu_6$ resonance for any $q^*$; this would imply a 
$\lesssim3$\% uncertainty. In the worst case, for the outer resonances, Mars/Mercury and large $q^*$, 
there are only a few impacts, but the outer resonances are not important for impacts anyway, so this 
should not be a major limitation of this work. The third and also the least understood source of 
uncertainty is related to the detection efficiency of CSS (photometric efficiency and trailing loss, 
Sect. 4; Jedicke et al. 2016). We are unable to quantify it here and leave this issue for future work.  
 
Table 6 reports the impact probabilities from different sources for Mercury, Venus and Mars. We used Eq. 
(\ref{fimp}) to compute the total impact flux on these planets as a function of impactor's absolute 
magnitude (Fig. \ref{earth}). The impact fluxes on the Earth and Venus are similar.
The impact flux on Mercury shows a shallower profile with $H$ mainly because small NEOs on orbits near 
Mercury are disrupted before they can impact. The size distribution of small ($<10$ km) craters on 
young Mercury terrains should be shallower than that found on the Moon and Mars, and this could have 
interesting applications to the Mercurian chronology as well.

\subsection{Impact ratios and the $R_{\rm b}$ parameter}
  
The Earth-to-Mars ratio in the number of impacts, E/Ma, is an important parameter often used to 
transfer the lunar crater chronology to Mars (e.g., Hartmann 2005, Marchi 2021). Here we find 
${\rm E/Ma} \simeq 2.8$ for $H = 15$ and ${\rm E/Ma} \simeq 4.3$ for $H = 25$ (Fig. \ref{earth}).
Adopting the standard Earth-to-Moon impact flux ratio, ${\rm E/Mo} = 20$, we estimate that
the Mars-to-Moon ratio in the number of impacts is ${\rm Ma/Mo} \simeq 7.1$ for $H = 15$ and 
${\rm Ma/Mo} \simeq 4.7$ for $H = 25$. In crater chronology studies this is often normalized 
to the unit surface area on these worlds (Mars has $\simeq 3.8$ larger surface area than the Moon),
giving the parameter $R_{\rm b}$, where the index b stands for bolides. We thus obtain $R_{\rm b}=2.0$ 
for $H = 15$ and $R_{\rm b}=1.2$ for $H = 25$. Both these values are significantly lower than
$R_{\rm b} \simeq 2.6$ used for NEO impacts in previous works (e.g., Hartmann 2005, Marchi 2021).
 
The ratio of impact fluxes is size-dependent as a consequence of the size-dependent contribution 
of different main belt sources (small and large bodies have different orbital distributions). For 
small bodies, the $\nu_6$ and 3:1 sources dominate, and these resonances have -- on their own -- 
$R_{\rm b} \simeq 0.8$ (they quickly move asteroids into the NEO zone where they can impact 
Earth/Moon rather than Mars). For large bodies, the weak resonances are important; they have -- 
on their own -- $R_{\rm b} \simeq 2.8$ (because asteroids in the weak resonances spend a long time 
on Mars-crossing orbits and have a greater chance of Mars impact). We emphasize that these
estimates accurately account for {\it all} impacts, including the ones from $q>1.3$ au. The $R_{\rm b}$
parameter is relatively low for small impactors {\it not} because we would be missing any Mars 
impacts from $q>1.3$ au. It is low simply because the $\nu_6$ and 3:1 resonances give fewer 
impacts on Mars relative to the Moon.   

The method we use here to estimate the $R_{\rm b}$ parameter is the best we can think of. First, by calibrating 
the model on NEO observations, we infer the flux of asteroids from different main-belt sources. 
We already know the impact probability per one body evolving from each source (all planetary impacts 
were recorded by the $N$-body integrator; Sect. 3), and this allows us to accurately estimate 
the impact flux on the terrestrial worlds. Still, there are some approximations. The main caveat 
of this method is that we assume that Mars impactors are on unstable orbits (e.g., in strong or 
diffusive resonances, scattered by Mars) that typically evolve, over long timescales, to $q<1.3$ au, 
and we can therefore calibrate them from NEOs. Our estimates would be inaccurate if many Mars 
impactors remain on semi-stable orbits with $q>1.3$ au over very long time intervals (longer than our 
integration timespan, 500 Myr). We plan on verifying this assumption in forthcoming work.          
We also did not account for weak resonances with $a>2.5$ au (higher order than 11:5). If these weak 
resonances were included as an additional source in the model, and the model was recalibrated, 
then perhaps the weight might (slightly) shift from stronger resonances, such as 5:2 and 8:3, to weaker 
resonances, and this could influence $R_{\rm b}$. In any case, this effect could only change $R_{\rm b}$ for 
large, $H\lesssim18$ asteroids, not the small ones. Full resolution of this problem is left for 
future work.   

Previous estimates of $R_{\rm b}$ for asteroids were inferred from the Mars-crossing population 
of large asteroids. For example, Bottke et al. (2002) estimated $R_{\rm b}=2.8$ for $H<18$, which
is significantly larger than our $R_{\rm b}\simeq1.8$ for $H < 18$. Bottke et al. (2002) inferred 
$R_{\rm b}$ from the Mars-crossing population known in 2002. They accounted for secular variations
of Mars's orbit, computed the impact probabilities on a grid in $(a,e,i)$ space, and approximately 
compensated for the observational incompleteness. This allowed them to estimate that the mean  
interval between impacts of $H<18$ asteroids on Mars is $\tau_{\rm Mars}(18) \simeq 1$ Myr, thus 
giving $R_{\rm b}=2.8$. Here we employed the same method with the asteroid catalog available in 
2022. If the cataloged asteroids with $q<1.8$ au and $H<18$ are assumed to be a complete sample, 
we find $\tau_{\rm Mars}(18) \simeq 2$ Myr, some two times longer than reported in Bottke et al. 
(2002). This would indicate $R_{\rm b}\simeq1.4$. A $\simeq 70$\% completeness for $q<1.8$ au and $H<18$ 
would give $R_{\rm b} \simeq 1.8$ in agreement with the estimate inferred from our NEO-based method. 
To obtain $R_{\rm b}\simeq 2.8$ from Bottke et al. (2002) the current population of $q<1.8$ au and $H<18$ 
asteroids would have to be only $\simeq 50$\% complete.
  
\section{Auxiliary models}

To this point we only presented the results of the 28-parameter base model. We now discuss several 
model modifications to explain some of our choices that we made to assemble the base model. We also 
explore the model validity beyond the range of parameters considered in the base model. 
       
In the first modification, the base model domain was extended to fainter magnitudes, $15<H<28$. The 
modified model produced a reasonable fit to the CSS observations (i.e., relatively large evidence;
Fig. \ref{aux1}). The extension to fainter magnitudes, however, revealed an intriguing difference 
relative to the intrinsic (debiased) 
magnitude distribution given in Harris \& Chodas (2021) (Fig. \ref{aux2}). Our distribution is slightly 
shallower for $H>25$ and leads to a smaller population of NEOs with $H<28$. Specifically, Harris 
\& Chodas (2021) estimated $\simeq 3.6 \times 10^7$ NEOs with $H<28$, whereas we have $(1.4 \pm 0.2) 
\times 10^7$ NEOs with $H<28$, a value that is roughly 2.6 times lower (see Sect. 10 for a discussion 
of constraints from bolide observations). The difference could be explained if we overestimated the CSS 
detection efficiency for $H > 25$, perhaps because of some issue with the trailing loss (Sect. 4.4). 
Alternatively, some of the assumptions in Harris \& Chodas (2021) may not be quite right. Harris \& Chodas 
(2021) did not derive any formal uncertainty of their population estimates but suggested that their 
extrapolation to the faintest magnitudes may be up to a factor of $\sim 5$ uncertain. 
We discuss this issue in Sect. 10.   

The absolute magnitude distribution given in Table 6 in Granvik et al. (2018) has the shape similar to 
ours but indicates a somewhat larger population of NEOs (Fig. \ref{aux2}). The difference is 
statistically significant. For example, Granvik et al. (2018) estimated $(8.02 \pm 0.45)\times 10^5$
NEOs with $H<24.875$, while we only have $(3.4 \pm 0.2)\times 10^5$ NEOs with $H<24.875$ 
(1$\sigma$ uncertainties quoted here). Our estimates for $H<25$ closely agree with those given 
in Heinze et al. (2021). Heinze et al. estimated  $(3.72 \pm 0.49) \times 10^5$ NEOs with $H<25$ 
while we only have $(3.6 \pm 0.2)\times 10^5$ NEOs with $H<25$ (Table 4). For $25<H<28$,
the magnitude distribution given in Heinze et al. (2021) is similar to that of Harris \& Chodas 
(2021) but steeper than ours (Fig. \ref{aux2}). This leads to $(2.64 \pm 0.88) \times 10^7$ 
NEOs with $H<28$ in Heinze et al. (2021) and $(1.4 \pm 0.1) \times 10^7$ NEOs with $H<28$ 
here. Given the relatively large uncertainty in Heinze et al. (2021), however, this difference
is not statistically significant (only $\simeq 1.4$$\sigma$).   

In the second modification, we kept the extended magnitude range ($15<H<28$), and used 
the trailing loss from Tricarico (2017) (Sect. 4.4). In this case, the intrinsic magnitude 
distribution of model NEOs again closely follows Harris \& Chodas (2021) to about $H = 25$. 
A relatively large difference then appears for fainter magnitudes, where the modified model 
gives a very shallow slope and only $(8.9 \pm 0.9) \times 10^6$ NEOs with $H<28$. This is a factor 
of $\simeq 4$ below the estimate of Harris \& Chodas (2021), and a factor of $\simeq 1.6$ below the 
estimate obtained above with the trailing loss from Zavodny et al. (2008). This may indicate
that the trailing loss from Tricarico (2017) overestimates the CSS detection efficiency for 
very faint NEOs. In broader sense, this highlights the dependence of the population estimates 
obtained here for very faint NEOs on the adopted trailing loss model.  

In the third modification, we used the trailing loss from Zavodny et al. (2008), but did not fix 
$N(15)=50$. Instead, we let the \texttt{MultiNest} fit decide what the population of $H<15$ should 
be based on the CSS data for $H>15$. If $N(15)$ is not fixed, the model overestimates, by roughly 
a factor of two, the number of NEOs brighter than $H=15$. This most likely happens because the 
statistical power of the brightest CSS NEOs in the \texttt{MultiNest} fit is not large enough to 
properly fix $N(15)$. We therefore impose $N(15)$ in the base model as an external constraint (Sect. 6).

We also tested a slight modification of the fitting procedure, where G96 and 703 were treated as 
separate surveys. The log-likelihood in Eq. (\ref{like}) was computed separately for them, and 
was subsequently combined to evaluate the total log-likelihood. Strictly speaking, combining the 
surveys at the level of log-likelihoods must be better than combining their detection 
efficiencies and object detections. This is because the detection bias of the G96 survey only applies 
to NEO detections in the G96 survey (and not 703), and vice versa. Note that this method is 
different from testing the two surveys separately; it makes use of the full statistical power of them 
combined. 

The results of this test were similar to those obtained with the standard method but we also 
noted several differences. The contribution of the $\nu_6$ resonance to NEOs with $H=15$ is  
smaller than reported in Table 3 (here $\alpha_{\nu_6}(15)=0.036\pm0.023$). This can indicate 
that -- at least for some parameters -- the systematics in the model fitting may be the dominant 
source of uncertainties and some source weights may be more uncertain than indicated in Table 3. 
The differences for all other source weights and other parameters are smaller than 30\%.
For some reason, the new fitting procedure also gives $N(17.75)=1010 \pm 19$ -- a $\sim 8$\% larger population 
than that obtained in the base model (and smaller uncertainty). This indicates that an accurate 
population estimate (also) depends on the details of the fitting algorithm. A detailed 
investigation of this approach is left for future work. Our preliminary results from 2013-2021 CSS 
observations, for which we derived the CSS detection efficiency from scratch (Nesvorn\'y et al., 
in preparation), favor combining surveys at the level of log-likelihoods and indicate 
$N(17.75)=931 \pm 21$. It thus appears that the detection efficiency of the original CSS for 
$H<17.75$ was slightly underestimated (Jedicke et al. 2016) and this was compensated by combining 
the detection efficiences of G96 and 703 (Granvik et al. 2018).


Following Granvik et al. (2016), our base model accounted for the size-dependent disruption of NEOs 
at low perihelion distances. We extensively tested various NEO models where the disruption module 
in \texttt{MultiNest} was switched off. All modified models without disruption showed a strong excess 
of high--$e$ and low--$q$ orbits, and Bayes factors that strongly disfavored them ($\Delta \ln {\cal Z}>20$
in favor of the base model). We also tested several models where the disruption module in \texttt{MultiNest} 
was switched on, but the dependence of $q^*$ on $H$ was ignored (i.e., fixed $q^*$ for all sizes). Again, 
the evidence term showed a strong preference for the models with the size-dependent disruption 
($\Delta \ln {\cal Z}>20$ in favor of the base model). This confirms the results of Granvik et al. (2016).

\section{Models with the Yarkovsky drift}

The methodology described above, where the contribution of different main-belt sources is inferred
from the NEO population, is agnostic as to whether the main-belt sources can actually provide that
contribution. This depends on the influx of main-belt asteroids into resonant sources and 
complex interaction of drifting orbits with weak resonances in the inner belt and for the Hungarias and Phocaeas.
To test this, we performed new numerical integrations in which bodies were {\it not} placed onto 
unstable orbits in the resonances. Instead, we collected real main-belt asteroids near a resonance, 
accounted for the Yarkovsky effect (Vokrouhlick\'y et al. 2015), and followed bodies as they drifted 
into the resonance and became NEOs. Two cases were considered: one with the {\it maximum} (theoretically 
possible) Yarkovsky drift and one where the drift was set to the {\it mean} (theoretically 
estimated) Yarkovsky rate. In either case, asteroids were assumed to drift \textit{toward} the 
resonance. The first case maximizes the asteroid flux into the source. The second case would correspond 
to a situation where asteroids drift toward the resonance with random obliquities. 

Adopting thermal parameters appropriate for the S and C type asteroids (Vokrouhlick\'y
et al. 2015), we estimate that the maximum Yarkovsky drift of a reference $D=1$ km body
at $a=2.5$ au is ${\rm d}a/{\rm d}t = 1.61^{+1.67}_{-0.82}\times10^{-4}$ au Myr$^{-1}$
for S, and ${\rm d}a/{\rm d}t = 2.35^{+2.74}_{-1.20}\times10^{-4}$ au Myr$^{-1}$ for C.
For comparison, if the measured Yarkovsky drifts for Golevka (S type) and Bennu (C type) are
rescaled to the same size and orbital radius, we obtain $2.25 \times 10^{-4}$ au Myr$^{-1}$ and
$1.82 \times 10^{-4}$ au Myr$^{-1}$, respectively (Greenberg et al. 2020). 
Given these results, we decided to make 
no distinction between S and C type asteroids, and adopted the drift rate 
\begin{equation}
{{\rm d}a \over {\rm d}t} = 2 \times 10^{-4}\, {\rm au \over Myr} \times \cos \theta\, \left( {1\,{\rm km} \over D } \right )  
\left( {2.5\,{\rm au} \over a } \right )^2\ ,
\label{dadt}
\end{equation}
where $\theta$ is the asteroid obliquity. 

We considered all main-belt asteroids near the 3:1 resonance that could potentially drift into
the resonance in 100 Myr. Near the 3:1 resonance, the maximum accumulated drift of a $D=1$-km body 
over 100 Myr is  $\simeq 0.02$ au. We therefore set, with a generous safety margin, 
$a_1(e)=2.46-(0.02/0.35)\, e$ au and $a_2(e)=2.54+(0.02/0.35)\, e$ au, and collected all known 
main-belt asteroids with $H<17.75$, $q>1.66$ au and $a_1(e)<a<a_2(e)$ (31,121 in total). The numerical 
integrations were performed with the modified {\it Swift} integrator, where artificial force terms
were added to account for ${\rm d}a/{\rm d}t$ from Eq. (\ref{dadt}). The diameters in Eq. (\ref{dadt})
were estimated from the absolute magnitudes of selected asteroids and the reference albedo
$p_{\rm V}=0.14$.  The orbits of eight planets and all selected asteroids were integrated with a 12-day 
time step for 100 Myr.

In the case with the maximum 
drift rate, we set $\theta=0$ for $a<2.5$ au and $\theta=180^\circ$ for $a>2.5$ au.
We found that $\eta=11$,107 asteroids reached the NEO region in 100 Myr. The number of NEOs expected from 
this influx in a steady state is $\eta \tau/(100\, {\rm Myr})$ where $\tau$ is the mean NEO lifetime for 
objects evolving from the 3:1 source (Table 5). 
For $q^*=0$--0.1 au, which should be appropriate for $H<17.75$ (Fig. \ref{qstar}), we 
have $\tau = 1.4$--2.5 Myr. We can thus estimate that the 3:1 resonance should contribute 
$\simeq 155$--277 NEOs with $H<17.75$. This is roughly consistent with the result of Morbidelli
\& Vokrouhlick\'y (2003), who found, in the case where the effects of YORP and collisions were 
suppressed, $\simeq 161$ $H<18$ NEOs from 3:1. For comparison, we inferred from the base model in Sect. 6 
that the 3:1 source should produce $\simeq 24 \pm 4$\% of NEOs with $H<17.75$ (Fig. \ref{alphas}).
This gives $\simeq 180$--268 NEOs for $N(17.75)=931 \pm 30$ (Sect. 6). We conclude that the model
with the maximum Yarkovsky drift of large main-belt asteroids toward the 3:1 resonance is consistent
with what is needed from the NEO-population modeling (Sect. 6).

The same simulations were repeated with the mean Yarkovsky drift toward the 3:1
resonance (the mean rate is $1/2$ of the maximum rate for random orientation of the spin axes;
Vokrouhlick\'y et al. 2015), and found $\eta \simeq 5$,000. This case can be ruled out because 
it only gives $\simeq70$--124 NEOs with $H<17.75$. Given these results, the case with fully random 
obliquities, where the main-belt asteroids would drift toward or away from the 3:1 resonance, was not 
investigated in detail -- we roughly estimate that this case would only give $<70$ NEOs with 
$H<17.75$.

We conclude that asteroids near the 3:1 resonance must be drifting toward the resonance with the 
(near) maximum Yarkovsky drift rates ($\simeq 2 \times 10^{-4}$ au Myr$^{-1}$ for $D=1$ km). This most 
likely happens because large, slow-drifting asteroids cannot cross the 3:1 resonance and this produces 
a dynamical bias, where all asteroids currently in the immediate neighborhood of the 3:1 resonance 
must be drifting toward it. In addition, the YORP effect must have driven their obliquities to 
$\theta \simeq 0$ or $\theta \simeq 180^\circ$, and this maximized the Yarkovsky drift and resonance 
feeding rate.
 
This result has several interesting consequences. First, in the immediate neighborhood of the 3:1 resonance, 
$\sim$km-class asteroids should have $\theta \simeq 0$ for $a<2.5$ au and $\theta \simeq 180^\circ$ 
for $a>2.5$ au. This prediction is testable by lightcurve observations (see the note at the end of 
the main text). Second, the spin re-orientation timescale of $\sim$km-class main-belt 
asteroids via collisions or YORP (Vokrouhlick\'y et al. 2015) should be relatively long. As bodies keep their
drift directions, they must be drifting toward the resonance and not away from it; the bodies currently 
drifting away from the resonance would have to have a relatively recent ($<100$ Myr) reorientation 
event. Third, the YORP effect must have driven obliquities of km-class bodies to either $\theta \simeq 0$ or 
$\theta \simeq 180^\circ$. This rules out, on the population level, the YORP models/shapes that lead to 
$\theta \sim 90^\circ$ and sets limits on the importance of spin-orbit resonances (Vokrouhlick\'y 
et al. 2003, 2006).   

Similar tests were performed for the $\nu_6$ and 5:2 resonances. For the 5:2 resonance, we found 
$\eta=10$,169 -- the influx in the 5:2 resonance is thus similar to the influx in the 3:1 
resonance. We estimate $\simeq 31$--46 NEOs with $H<17.75$ from 5:2 in the steady state.  
For comparison, our NEO model nominally implies $\simeq 56$ NEOs from the 5:2 source, but this value has a 
relatively large uncertainty (Table 3) and is consistent within 1$\sigma$ with the drift-inferred 
values. In addition, the population of $H<17.75$ main-belt asteroids near the 5:2 resonance is 
probably incomplete and that may account for some of the difference as well.
For the $\nu_6$ resonance, where ${\rm d}a / {\rm d}t<0$ was assumed for all orbits, we found a 
lower influx, $\eta=4$,040, because the region adjacent to the $\nu_6$ resonance is sparsely populated. 
With the relatively long lifetimes of orbits evolving from $\nu_6$ (Table 5), this implies 
$\simeq 237$--318 NEOs with $H<17.75$, to be compared with $\simeq 186$ inferred in Sect. 6 for the 
$\nu_6$ source. An accurate comparison is somewhat complicated in this case because many asteroids 
in the drift simulations reached the NEO orbits via weak resonances, and not from $\nu_6$.  

The simulations presented here offer an opportunity to test whether the NEO orbital distributions
obtained from different sources sensitively depend on the initial conditions. The model described 
in Sect. 6 was based on the orbital distributions obtained from the simulations where test bodies 
were inserted onto unstable orbits in resonances. Here we instead drifted real main-belt asteroids 
into resonances. We can therefore compare the orbital distributions of NEOs obtained from the two 
methods to see if there are any important differences. We find that the distributions obtained from 
the two methods are practically identical ($<1\%$ differences for $\nu_6$, 3:1 and 5:2). This justifies 
our preferred approach to this problem described in Sect. 5.1.

\section{Discussion}

\subsection{Magnitude distribution of NEOs}

Harris \& Chodas (2021) determined the absolute magnitude distribution of NEOs by comparing  detections 
of new NEOs with redetections of previously known NEOs (also see Harris \& D'Abramo 2015). If all objects 
were equally detectable, the ratio of new detections to redetections in a survey is proportional to the 
number of not yet discovered NEOs, thus giving clues about the observational incompleteness. Given that 
the observational bias is the same for both the new detections and redetections, by using the ratio of 
the two, the method is relatively insensitive to the observational bias and different surveys can be 
clumped together to improve the statistics. Harris \& Chodas (2021) approximately accounted for observational 
biases of different surveys to correct the estimates for unequal detectability of NEOs on different orbits. 
They also corrected a small error in Harris \& D'Abramo (2015) related to a rounding problem. 

The method based on redetections is limited to a magnitude range where the numbers of new-detections and 
redetections are statistically large, which, according to Harris \& D'Abramo (2015), corresponds to the magnitude 
range $17.5<H<23.5$. To extrapolate the results to fainter magnitudes, where there are no or too few 
redetections, Harris \& D'Abramo (2015) and  Harris \& Chodas (2021) assumed that a survey detects an 
increasingly smaller fraction of the NEO population and estimated -- from the statistics of close encounters 
of faint NEOs to the Earth -- that this fraction is proportional to $10^{-0.8 H}$. Finally,
anchoring the results to the re-detection ratio approach at $H \simeq 23.5$, they produced the absolute 
magnitude distribution of NEOs for $25\leq H \leq 31$.

To demonstrate the applicability of their estimate, Harris \& Chodas (2021) used the fixed impact flux 
probability with the Earth, $f_{\rm imp}=1.5\times10^{-3}$ Myr$^{-1}$ for each NEO, and compared their impact 
statistics with the one inferred from observations of bolides (Brown et al. 2002). Brown et 
al. (2002) analyzed satellite records of bolide detonations in the Earth atmosphere to estimate the impact 
flux of $\sim1$--10 m bodies. For $D \simeq 10$ m, roughly equivalent to $H=28$ for our reference albedo 
$p_{\rm V}=0.14$, the average interval between impacts was found $\simeq 10$ yr (with a factor of $\simeq 2$ 
uncertainty). The infrasound data from Silber et al. (2009), as reported by Brown et al. (2013), indicate
a somewhat shorter interval but the error bars of these estimates overlap with the bolide data. For 
comparison, Harris \& Chodas (2021) estimated the average interval between impacts of $H<28$ bodies to 
be $\simeq 18$ yr (Fig. \ref{impacts}).  
 
Here we find that the magnitude distribution is relatively shallow for $H>25$ and estimate a 
somewhat smaller population of faint NEOs (Sect. 8). We also find, however, that the 
Earth-impact probability of faint NEOs is relatively large (because they evolve onto NEO 
orbits via the $\nu_6$ resonance), and that this larger impact probability at least partially 
compensates for the smaller population. For example, we have $f_{\rm imp} \simeq 1.5\times10^{-3}$ Myr$^{-1}$
for $H=15$ and $f_{\rm imp}\simeq2.6\times10^{-3}$ Myr$^{-1}$ for $H=28$. The mean interval between impacts 
for $H<28$ is estimated here to be $\simeq 30$ yr (Fig. \ref{impacts}). This is a factor of 1.6 and 3 
longer than the nominal intervals from Harris \& Chodas (2021) and Brown et al. (2002, 2013), 
respectively. Adopting our estimate, the probability of having four impacts in the last 30 years 
from $D>10$ m projectiles would only be 1.5\%. Brown et al. (2013) suggested that the current 
impactor flux for near-Earth asteroids that are 10--50 m in diameter may be higher than the long 
term average.

Note that all estimates quoted above have significant uncertainties. Brown et al. (2002, 2013) reported 
a factor of $\simeq 2$ uncertainty in their estimates from bolide and infrasound observations, but the
fact that these two estimates agree means that the combined uncertainty would be smaller. Harris \& Chodas 
(2021) suggested a factor of few uncertainty in their estimate. Our impact flux estimate is at 
least $\simeq 10$\% uncertain (1$\sigma$ from the magnitude distribution uncertainty for $H=28$; Fig. 
\ref{aux2}) and probably much more given that we were not able to characterize the uncertainty of 
the CSS detection efficiency (Sect. 4). It is possible, for example, that the CSS detection efficiency 
is overestimated by a factor of $\simeq2$--3 for $H\simeq28$. If so, this would bring our impact flux 
up by the same factor. It is also possible that the difference between our estimates and bolide/infrasound 
data has some interesting physical explanation. We are testing different possibilities and will report 
on the results in forthcoming publications.
 
\subsection{PM/AM ratio}

There has been some debate about the PM/AM ratio of meteorites/bolides (Morbidelli \& Gladman 1998; Wisdom 2017, 
2020). The PM/AM ratio measures the relative frequency of meteorite falls before (6--12 h) and after (12--18 h) 
noon. It is usually reported as the number of afternoon falls (12--18 h) over the number of day-time
falls (6--18 h), to express the observed excess of afternoon falls, here denoted as ${\cal E}$. Ordinary 
chondrites (OCs), for example, have ${\cal E}=0.63 \pm 0.02$ (Wisdom 2017). Morbidelli \& Gladman (1998) obtained 
${\cal E}=0.52$ and 0.48 for impactors from the $\nu_6$ and 3:1 resonances, respectively (no 
cutoff on collisional lifetime or entry velocity imposed here), and suggested that the PM excess 
of reported OC falls should be a consequence of the collisional removal of meteoroids (young NEOs 
tend to have a stronger PM excess). Wisdom (2020), as an update 
on Wisdom (2017), estimated ${\cal E}=0.533 \pm 0.002$ and $0.604 \pm 0.007$ from the $\nu_6$ and 3:1 
resonance. He argued that the previous (lower) estimates of Morbidelli \& Gladman (1998) were wrong because 
-- to calculate ${\cal E}$ -- Morbidelli \& Gladman (1998) incorrectly assumed orbits with a uniformly 
random distribution of the argument of perihelion, $\omega$.

Here we take the opportunity to rectify this issue. Our $N$-body integration recorded a large number
of Earth impacts from bodies started in the $\nu_6$ (2527 in total) and 3:1 resonances (398 in total).
For each impact, we propagated the impactor to the Earth's surface and determined the geocentric coordinates 
of the impact. This allowed us to estimate ${\cal E}$ without any uncertainty related to the $\omega$ 
distribution. The night-time impacts were ignored. To be consistent with the previous work 
(Morbidelli \& Gladman 1998; Wisdom 2017, 2020), the Earth obliquity was neglected in this test. 

We obtained ${\cal E}=0.47 \pm 0.02$ and $0.50 \pm 0.05$ for the $\nu_6$ and 3:1 resonances, respectively.
These values are better aligned with Morbidelli \& Gladman (1998) than with Wisdom (2017, 2020). The 
PM excess reported in Wisdom (2017) for the 3:1 resonance is roughly $2\sigma$ above 
our value. The reasons behind this are uncertain. Part of the difference may be explained by the relatively
short integration timespan (20 Myr in Wisdom 2017, 2020). When the integrations were extended to 40 Myr,
Wisdom (2020) found ${\cal E}=0.587 \pm 0.007$ for the 3:1 resonance. Here we find the same trend: the 
early impacts show higher PM excess than the late ones (e.g., ${\cal E} \simeq 0.56$ from the 3:1 resonance 
and $t<10$ Myr). 

For reference, we also computed the PM excess with the disruption model (Sect. 5.3).   
For example, for $q^*=0.3$ au, which should be appropriate for 1-10 m meteoroids, we find ${\cal E}=0.51 \pm 0.05$ 
and $0.58 \pm 0.08$ for the $\nu_6$ and 3:1 resonances. Meteoroid disruption close to the Sun can thus
significantly influence the PM excess. The observed statistics of PM/AM falls shows higher excess 
(${\cal E} = 0.63 \pm 0.02$) than the values derived here for the $\nu_6$ and 3:1 resonances. Morbidelli
\& Gladman (1998) suggested that the excess increases in the model when it is accounted for the collisional
lifetime of meteoroids. A possible solution to this problem could thus be that the PM excess 
is influenced by the physical lifetime of meteoroids (collisional disruption, disruption at low perihelia,
YORP spin-up, etc.)  
 
\subsection{Size dependencies}

The size-dependent sampling of main-belt sources found here, both in terms of their contribution 
to the NEO population and Earth impacts, helps to resolve the following scientific problem.
Granvik et al. (2018) estimated that the outer-belt contribution to NEOs is 
practically negligible ($\simeq3.5$\% for the 2:1 resonance complex). They suggested that
$\simeq 80$\% of impactors on the terrestrial worlds are produced from the $\nu_6$ resonance,
and over 10\% of impactors are produced from the 3:1 resonance, Hungarias and Phocaeas,
leaving only $<10$\% for the middle/outer belt. Based on this, Granvik et al. (2018) proposed
that the majority of \textit{primitive} NEOs/impactors must come from the $\nu_6$ resonance.
Nesvorn\'y et al. (2021) instead found that the middle/outer belt can supply nearly
50\% of {\it large} NEOs, $\simeq70$\% of large primitive NEOs, and $\simeq35$--40\% of large
impactors ($D \gtrsim 5$ km).

Nesvorn\'y et al. (2021) speculated that these differences may be a consequence
of the size-dependent delivery process. On one hand, small main-belt asteroids can drift
over a considerable radial distance by the Yarkovsky effect and reach NEO space
from the powerful $\nu_6$ resonance at the inner edge of the asteroid belt 
(e.g., Granvik et al., 2017). The $\nu_6$ resonance is known to produce highly evolved NEO orbits and 
high impact probabilities on the Earth (Table 6; Gladman et al. 1997). On the other hand, large
main-belt asteroids often reach NEO orbits via slow orbital evolution in weak resonances (Migliorini
et al. 1998, Morbidelli \& Nesvorn\'y, 1999, Farinella \& Vokrouhlick\'y 1999). Whereas each of
these resonances adds only a little, their total contribution to the population of large NEOs can
be significant.

Here we find supporting evidence for this thesis. For $H = 15$ ($D=3.5$ km for the reference albedo
$p_{\rm V}=0.14$), we find that the middle/outer main belt produce $\simeq 40$\% of NEOs. When
extrapolated to $D>5$ km, this should be consistent with the similarly large contribution reported
in Nesvorn\'y et al. (2021). For $H = 25$ ($D=35$ m for a reference albedo $p_{\rm V}=0.14$), however,
the contribution is only $\simeq10$\% (the 3:1 source is excluded here). The $\nu_6$ and 3:1 resonances produce
$\simeq 80$\% of small NEOs for $H = 25$ (and $\simeq 90$\% of small Earth impactors; Sect. 7),
which is in line with the findings reported in Granvik et al. (2018).    

\section{Summary}

The main results of this work are summarized as follows.

\begin{description}
\item (1) We developed a new NEO model (NEOMOD). The model is based on numerical integrations of 
  bodies from 12 sources (11 main-belt sources and comets). A flexible method to accurately calculate
  biases of NEO surveys was applied to the Catalina Sky Survey (CSS) observations from 2005 to 2012
  (Christensen et al. 2012). The \texttt{MultiNest} code (Feroz \& Hobson 2008, Feroz et al. 2009)
  was used to calibrate the model on CSS detections. The algorithms developed here can be readily
  adapted to any current or future NEO survey.
\item (2) The methodology used in Granvik et al. (2018) was improved. We adopted the cubic splines
  to characterize the magnitude distribution of the NEO population. The cubic splines are flexible and 
  can be modified
  to consider a broader absolute-magnitude range and/or improve the model accuracy. We used a large 
  number of main-belt asteroids in each source ($10^5$), which allowed us to accurately estimate the
  impact fluxes on the terrestrial planets. Our model self-consistently accounts for the NEO disruption 
  at small perihelion distances (Granvik et al. 2016).

\item (3)  We used 10,000 test objects per orbital bin, 18,480 orbital bins, 56 
  absolute magnitude bins, and nearly 250,000 FoVs to compute the CSS detection probability as 
  a function of NEO's $a$, $e$, $i$ and $H$. We considered different approaches to modeling the trailing
  loss of CSS. The trailing loss represents an important uncertainty in estimating the population of small
  NEOs, and we urge surveys to carefully characterize it. 
  
\item (4) Our base model is available via the NEOMOD Simulator (Sect. 6), a code that
  can be used to generate a user-defined sample of model NEOs. Researchers interested in the probability
  that a specific NEO evolved from a particular source can obtain this information from the ASCII table
  that is available along with the Simulator. Optionally, the NEOMOD Simulator can output the information 
  about the impact probability of model-generated NEOs with the Earth.         
\item (5) We found that the sampling of main-belt sources by NEOs is {\it size-dependent} with the 
  $\nu_6$ and 3:1 resonances contributing $\simeq 30$\% of NEOs with $H = 15$, and $\simeq 80$\% of 
  NEOs with $H = 25$. This trend most likely arises from how the small and large main-belt asteroids 
  reach the source regions. The size-dependent sampling suggests that small terrestrial impactors 
  preferentially arrive from the $\nu_6$ source, whereas the large impactors can commonly come from
  the middle/outer belt (Nesvorn\'y et al. 2021).
\item (6) We confirm the size-dependent disruption of NEOs reported in Granvik et al. (2016), and find a 
  similar dependence of the disruption distance on the absolute magnitude. As a consequence of the 
  size-dependent disruption and item (5), small and large NEOs have different orbital distributions.
\item (7) Although the base NEOMOD fit only applies to $H<25$, the fit in the extended magnitude 
  range shows a shallower absolute magnitude distribution for $25<H<28$ and smaller number of NEOs 
  with $H<28$ than Harris \& Chodas (2021). The average time between terrestrial impacts of $D \simeq 
  10$ m bolides is found to be $\simeq 30$ yr -- $\simeq 3$ times longer than the nominal estimate 
  from Brown et al. (2002, 2013). These differences may point to some problem with the detection efficiency
  of CSS for $25<H<28$. Alternatively, they may have some interesting physical explanation.
\item (8) We compute the PM excess of meteorite falls for meteoroids evolving from the $\nu_6$ and 3:1 
  resonances to find ${\cal E}=0.47 \pm 0.02$ and $0.50 \pm 0.05$, respectively. These values are better 
  aligned with Morbidelli \& Gladman (1998) than with Wisdom (2017, 2020). The observed statistics of 
  PM/AM falls shows higher excess (${\cal E} = 0.63 \pm 0.02$) than the values derived here for the 
  $\nu_6$ and 3:1 resonances. The PM excess can be influenced by the physical lifetime of meteoroids.
\item (9) The model-inferred contribution of the 3:1 source to large NEOs ($H \lesssim 18$) implies
  that the main-belt asteroids should drift toward the 3:1 resonance at the maximum Yarkovsky drift
  rates ($\simeq 2 \times 10^{-4}$ au Myr$^{-1}$ for a $\simeq 1$-km diameter body at 2.5 au). This suggests
  that the main-belt asteroids on the sunward side of the 3:1 resonance ($a<2.5$ au) have obliquities
  $\theta \simeq 0^\circ$; the ones with $a>2.5$ au should have $\theta \simeq 180^\circ$ (in the immediate
  neighborhood of the resonance). A similar inference applies to the $\nu_6$ and 5:2 resonances (it should 
  apply to other resonances as well). These predictions are testable from lightcurve observations.
\item (10) The contribution of inactive comets to the NEO population is inferred to be smaller than in previous 
  works ($\alpha_{\rm JFC}<0.017$; 68.3\% envelope). For comparison, Bottke et al. (2002) found a 
  $\simeq 6$\% contribution of JFCs and Granvik et al. (2018) suggested a 2--10\% $H$-dependent contribution.
  As the Bayes factor slightly favors a model without any comet contribution, the evidence for cometary
  NEOs can be thus hard to extract from the CSS observations alone. This may imply that JFCs disrupt 
  rather than becoming dormant (see Sect. 6 in Nesvorn\'y et al. 2010).
\item (11) We estimate that the Mars-to-Moon ratio in the number of impacts is ${\rm Ma/Mo} \simeq 7.1$ 
  for $H = 15$ and ${\rm Ma/Mo} \simeq 4.7$ for $H = 25$. In crater chronology studies this is often normalized 
  to the unit surface area on these world giving the parameter $R_{\rm b}$, where the index b stands for 
  bolides. We obtain $R_{\rm b}=2.0$ for $H = 15$ and $R_{\rm b}=1.2$ for $H = 25$. Both these values 
  are significantly lower than $R_{\rm b} \simeq 2.6$ used for NEO impacts in previous works.
\end{description}
\textit{Note added as this manuscript was being completed.} J. \v{D}urech and J. Hanu\v{s} (2022, 
private communication) recently analyzed the Gaia data release 3
(DR3) to determine the obliquities for $\simeq 9500$ asteroids. The distribution of obliquities confirms
the trend predicted in Sect. 8 and item (9) above. The obliquities of main belt asteroids immediately
sunward of strong orbital resonances are $\theta<90^\circ$, with a concentration for $\theta<45^\circ$;
capture in spin-orbit resonances is probably important here. The obliquities on the opposite side of
resonances are $\theta \simeq 180^\circ$, exactly as predicted here to produce sufficiently large feeding
rates.
 

\acknowledgements
The simulations were performed on the NASA Pleiades Supercomputer. We thank the NASA NAS computing division 
for continued support. The work of DN, RD, and WFB was supported by the NASA Planetary Defense Coordination 
Office project ``Constructing a New Model of the Near-Earth Object Population''. The work of SN, SRC and PWC 
was conducted at the Jet Propulsion Laboratory, California Institute of Technology, under a contract with 
the National Aeronautics and Space Administration. DV acknowledges support from the grant 21-11058S of 
the Czech Science Foundation. We thank an anonymous reviewer for helpful comments.

\clearpage
\begin{table}
\centering
{
\begin{tabular}{lrrrrrrr}
\hline \hline
 \\   
source            & $\mu_e$   &  $\sigma_e$  & $\mu_1$  & $\sigma_1$ & $\mu_2$  & $\sigma_2$ & $w_1/w_2$  \\ 
                  & (or $\gamma_e$) &             & ($^\circ$)   &  ($^\circ$)     & ($^\circ$)   & ($^\circ$)     &   \\  
\hline
$\nu_6$           & 0.16      &  0.067      &  5.5    &  2.3      & 15.0     &  3.0       & 10  \\   
3:1               & 0.145     &  0.067      &  4.7    &  2.7      & 13.5     &  2.5       & 2.5 \\
5:2               & (0.1)     &   --        &  5.5    &  3.0      & 13.5     &  4.0       & 3.3  \\
7:3               & (0.085)   &   --        &  2.7    &  1.3      & 10.5     &  2.2       & 0.65 \\
8:3               & (0.1)     &   --        &  5.3    &  2.0      & 13.0     &  2.3       & 1.4  \\
9:4               & (0.09)    &   --        &  2.0    &  2.0      & 10.5     &  3.3       & 0.3  \\
11:5              & (0.11)    &   --        &  10.0   &  1.0      & 10.0     &  6.0       & 1.0  \\
2:1               & (0.12)    &    --       &  26.0   &  2.0      & 11.0     &  6.0       & 0.55  \\
\hline \hline
\end{tabular}
}
\caption{The eccentricity and inclination distributions adopted in this work for different sources.
The columns are: (1) source id., (2) the mean of the Gaussian distribution ($\mu_e$) or the scale
parameter of the Rayleigh distribution ($\gamma_e$, values in parentheses) in $e$, (3) the standard
deviation of the Gaussian distribution in $e$ ($\sigma_e$), (4-5) the mean and standard deviation
of the first Gaussian term in $i$ ($\mu_1$ and $\sigma_1$), (6-7) the mean and standard deviation
of the second Gaussian term in $i$ ($\mu_2$ and $\sigma_2$), and (8) the weight ratio of the two 
terms ($w_1/w_2$).}
\end{table}

\begin{table}
\centering
{
\begin{tabular}{lrrrr}
\hline \hline
 \\   
                  & min  & max  & $N_{\rm bin}$ & $\Delta$ \\  
\hline
$a$               & 0  & 4.2 au       & 42  & 0.1 au \\
$e$               & 0  & 1        & 20  & 0.05  \\ 
$i$               & 0  & 88$^\circ$  & 22  & 4$^\circ$ \\
$H$               & 15   & 25         & 40  & 0.25 \\
\hline \hline
\end{tabular}
}
\caption{The orbit and absolute magnitude binning used in this work. The columns are:
the (1) model variable, (2-3) minimum and maximum values considered here, (4) 
number of bins ($N_{\rm bin}$), and (5) bin size ($\Delta$).}
\end{table}

\begin{table}
\centering
{
\begin{tabular}{lrrrrr}
\hline \hline
label & parameter        & median & $-\sigma$ & $+\sigma$ & limit \\                      
\hline
\multicolumn{6}{c}{$\alpha$'s for $H=15$}                    \\
(1) & $\nu_6$          & 0.118   & 0.052    & 0.056 & --     \\ 
(2) & 3:1              & 0.219   & 0.040    & 0.041 & --     \\
(3) & 5:2              & 0.057   & 0.026    & 0.028 & --     \\ 
(4) & 7:3              & 0.008   & 0.005    & 0.009 & 0.012  \\ 
(5) & 8:3              & 0.093   & 0.020    & 0.021 & --     \\ 
(6) & 9:4              & 0.013   & 0.009    & 0.017 & 0.020  \\                 
(7) & 11:5             & 0.044   & 0.020    & 0.022 & --   \\   
(8) & 2:1              & 0.045   & 0.010    & 0.010 & --   \\   
(9) & inner weak       & 0.202   & 0.048    & 0.045 & --   \\  
(10) & Hungarias        & 0.082   & 0.022    & 0.022 & --   \\ 
(11) & Phocaeas         & 0.095   & 0.017    & 0.018 & --   \\ 
 -- & JFCs             & 0.012   & 0.008    & 0.013  & 0.017  \\ 
\hline
\multicolumn{6}{c}{$\alpha$'s for $H=25$}                \\
(12) & $\nu_6$          & 0.424   & 0.043    & 0.040 & --    \\ 
(13) & 3:1              & 0.338   & 0.034    & 0.035 & --    \\
(14) & 5:2              & 0.063   & 0.018    & 0.020 & --    \\ 
(15) & 7:3              & 0.004   & 0.003    & 0.006 & 0.007 \\ 
(16) & 8:3              & 0.010   & 0.008    & 0.014 & 0.016   \\ 
(17) & 9:4              & 0.007   & 0.005    & 0.010 & 0.012    \\                 
(18) & 11:5             & 0.009   & 0.007    & 0.013 & 0.014    \\   
(19) & 2:1              & 0.006   & 0.004    & 0.008 & 0.009    \\   
(20) & inner weak       & 0.033   & 0.023    & 0.036 & 0.049   \\  
(21) & Hungarias        & 0.056   & 0.027    & 0.030 & --     \\ 
(22) & Phocaeas         & 0.014   & 0.010    & 0.018 & 0.021  \\ 
-- & JFCs             & 0.008   & 0.006    & 0.011   & 0.014  \\ 
\hline
\multicolumn{6}{c}{$H$ {\it distribution}} \\
(23) & $N_{\rm ref}$      & 896     & 29       & 29    & --     \\  
(24) & $\gamma_2$        & 0.344   & 0.006    & 0.006 & --     \\
(25) & $\gamma_3$        & 0.328   & 0.004    & 0.004 & --      \\
(26) & $\gamma_4$        & 0.566   & 0.014    & 0.014 & --       \\
\hline
\multicolumn{6}{c}{{\it Disruption parameters}}      \\
(27) & $q^*_0$          &  0.144   & 0.004    & 0.007  & --     \\ 
(28) & $\delta q^*$     &  0.030  & 0.003    & 0.001   & --    \\  
\hline \hline
\end{tabular}
}
\caption{The median and uncertainities of our base model parameters.
The first column is the parameter/plot label in Fig. \ref{triangle} (JFCs do not appear in the 
figure). The uncertainties reported here were obtained from the posterior distribution 
produced by \texttt{MultiNest}. They do not account for uncertainties of the CSS detection 
efficiency. For parameters, for which the posterior distribution shown in Fig. \ref{triangle}
peaks near zero, the last column reports the upper limit (68.3\% of posteriors fall between
zero and that limit).}
\end{table}

\begin{table}
\centering
{
\begin{tabular}{rrrrrrrr}
\hline \hline
$H$ & $N_{\cal M}(H)$ & $H$ & $N_{\cal M}(H)$ & $H$ & $N_{\cal M}(H)$ & $H$ & $N_{\cal M}(H)$ \\                      
\hline
   15.1  &     55.8   &  17.6  &     783.3   &    20.1   &    5795  &     22.6   &    28970 \\   
   15.2  &     62.4   &  17.7  &     860.9   &    20.2   &    6180  &     22.7   &    31320 \\  
   15.3  &     69.7   &  17.8  &     945.1   &    20.3   &    6585  &     22.8   &    33930 \\    
   15.4  &     77.9   &  17.9  &     1036.3  &    20.4   &    7011  &     22.9   &    36850 \\    
   15.5  &     87.0   &  18.0  &     1134.9  &    20.5   &    7458  &     23.0   &    40120 \\   
   15.6  &     97.2   &  18.1  &     1241.4  &    20.6   &    7930  &     23.1   &    43780  \\   
   15.7  &     108.6  &  18.2  &     1356.3  &    20.7   &    8427  &     23.2   &    47910   \\   
   15.8  &     121.2   & 18.3  &     1480.1  &    20.8   &    8953  &     23.3   &    52580  \\   
   15.9  &     135.2   & 18.4  &     1613.2  &    20.9   &    9509  &     23.4   &    57870  \\  
   16.0  &     150.8   & 18.5  &     1756.1  &    21.0   &    10100 &     23.5   &    63880 \\    
   16.1  &     168.1   & 18.6  &     1909.4  &    21.1   &    10730 &     23.6   &    70760 \\    
   16.2  &     187.3   & 18.7  &     2073.6  &    21.2   &    11400 &     23.7   &    78630 \\    
   16.3  &     208.5   & 18.8  &     2249.2  &    21.3   &    12110 &     23.8   &    87670 \\    
   16.4  &     232.0   & 18.9  &     2436.8  &    21.4   &    12870 &     23.9   &    98100\\    
   16.5  &     258.0   & 19.0  &     2636.8  &    21.5   &    13690 &     24.0   &    110200 \\    
   16.6  &     286.7   & 19.1  &     2849.7  &    21.6   &    14570 &     24.1   &    124200 \\    
   16.7  &     318.4   & 19.2  &     3076.1  &    21.7   &    15520 &     24.2   &    140500 \\    
   16.8  &     353.2   & 19.3  &     3316.5  &    21.8   &    16540 &     24.3   &    159500\\   
   16.9  &     391.6   & 19.4  &     3571.3  &    21.9   &    17650 &     24.4   &    181400 \\    
   17.0  &     433.7   & 19.5  &     3841.0  &    22.0   &    18860 &     24.5   &    206900 \\    
   17.1  &     479.9   & 19.6  &     4126.1  &    22.1   &    20180 &     24.6   &    236400  \\    
   17.2  &     530.4   & 19.7  &     4426.9  &    22.2   &    21620 &     24.7   &    270500 \\    
   17.3  &     585.7   & 19.8  &     4743.9  &    22.3   &    23200 &     24.8   &    309800 \\ 
   17.4  &     646.0   & 19.9  &     5077.3  &    22.4   &    24940 &     24.9   &    355200 \\    
   17.5  &     711.8   & 20.0  &     5427.6  &    22.5   &    26850 &     25.0   &    407400 \\    
\hline \hline
\end{tabular}
}
\caption{The cumulative number of NEOs. For each absolute magnitude limit ($H$), the table reports 
the number of NEOs brighter than $H$ as estimated from our base model ($N_{\cal M}(H)$). The relative
1$\sigma$ uncertainty of the population estimates given here increases from $\simeq3$\% for $H<20$ 
to $\simeq 6$\% for $H<28$. The uncertainties were estimated from the posterior distribution 
produced by \texttt{MultiNest}. They do not account for uncertainties of the CSS detection efficiency.}
\end{table}

\begin{table}
\centering
{
\begin{tabular}{lrrr}
\hline \hline   
source          & $\tau$  & $p_{\rm imp}$  & $f_{\rm imp}$ \\  
                & (Myr)   &              & (Myr$^{-1}$) \\
\hline
$\nu_6$          & 6.64   & 0.02035   & 0.00306           \\ 
3:1              & 1.38   & 0.00273   & 0.00198          \\
5:2              & 0.29   & 0.00041   & 0.00141          \\ 
7:3              & 0.15   & 0.00006   & 0.00041          \\ 
8:3              & 1.20   & 0.00138   & 0.00149          \\ 
9:4              & 0.21   & 0.00028   & 0.00132          \\                 
11:5             & 0.29   & 0.00009   & 0.00029          \\   
2:1              & 0.41   & 0.00010   & 0.00025          \\   
inner weak       & 4.83   & 0.01262   & 0.00261          \\  
Hungarias        & 20.88  & 0.02802   & 0.00134          \\ 
Phocaeas         & 15.33  & 0.00666   & 0.00043          \\ 
\hline \hline
\end{tabular}
}
\caption{The average lifetime of NEOs ($\tau$), Earth impact probability ($p_{\rm imp}$)
  and impact flux ($f_{\rm imp}=p_{\rm imp}/\tau$). The values are given here for the fixed reference
  disruption distance $q^*=0.1$ au. }
\end{table}

\begin{table}
\centering
{
\begin{tabular}{lrrrr}
\hline \hline   
source          & Mercury  & Venus    & Earth  &  Mars  \\
                &   \%     &  \%      &  \%    &  \%    \\
\hline
$\nu_6$        & 0.273 & 2.042 & 2.035  & 0.317          \\ 
3:1            & 0.035 & 0.227 & 0.273  & 0.041         \\
5:2            & 0.006 & 0.023 & 0.041  & 0.004        \\ 
7:3            & 0.001 & 0.011 & 0.006  & 0.004        \\ 
8:3            & 0.015 & 0.099 & 0.138  & 0.067         \\ 
9:4            & 0.000 & 0.009 & 0.028  & 0.006         \\                 
11:5           & 0.000 & 0.017 & 0.009  & 0.017         \\   
2:1            & 0.001 & 0.013 & 0.010  & 0.006          \\   
inner weak     & 0.122 & 1.221 & 1.262  & 0.895          \\  
Hungarias      & 0.258 & 2.266 & 2.802  & 2.001          \\ 
Phocaeas       & 0.069 & 0.708 & 0.666  & 0.388          \\ 
\hline \hline
\end{tabular}
}
\caption{The impact probabilities ($p_{\rm imp}$) of NEOs from different sources. 
The values are given for the model with the fixed disruption distance $q^*=0.1$ au. The impact 
fluxes ($f_{\rm imp}$) for different planets can be computed by dividing the probabilities given 
here by the average NEO lifetimes given in Table 5.}
\end{table}

\clearpage
\begin{figure}
\epsscale{0.7}
\plotone{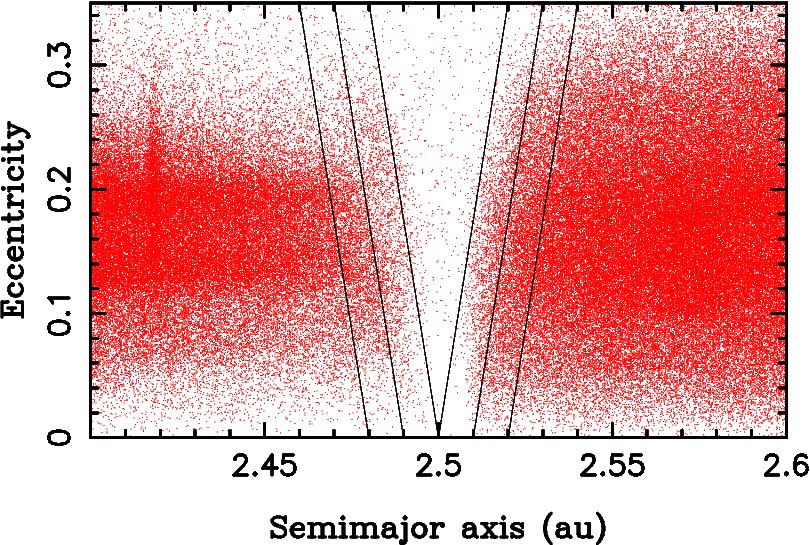}
\caption{The orbital distribution of main-belt asteroids (red dots) near the 3:1 resonance with Jupiter.
The inner V-shaped region approximates the dynamically unstable domain where test bodies representing the 
3:1 source were placed. The main belt asteroids with orbits in the two outer strips, $2.48<a<2.49$ au and 
$2.51<a<2.52$ au for $e=0$ and diagonally extending to $e>0$, were used to set up the eccentricity and 
inclination distributions of test bodies.}
\label{res31}
\end{figure}

\clearpage
\begin{figure}
\epsscale{0.4}
\plotone{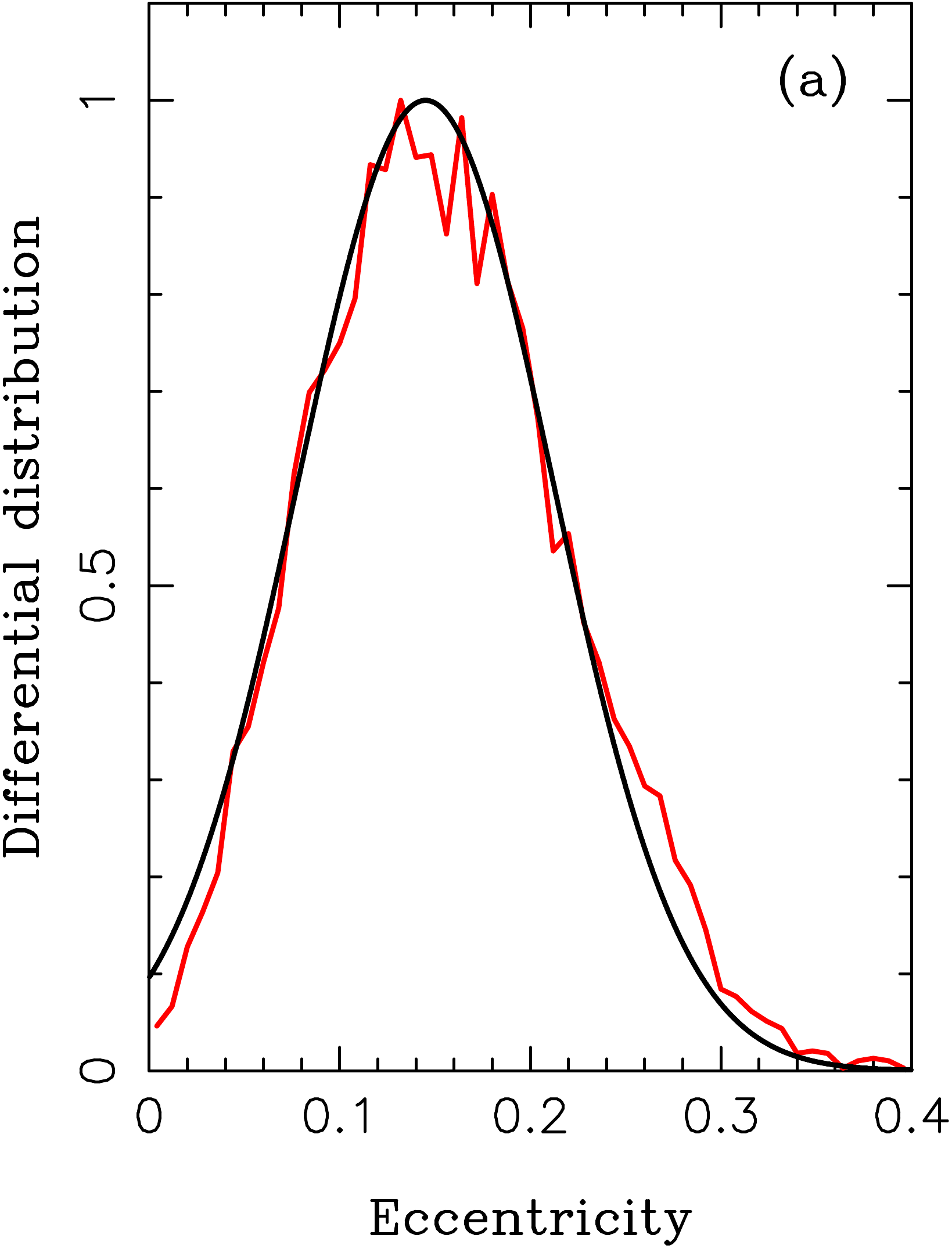}
\epsscale{0.364}
\plotone{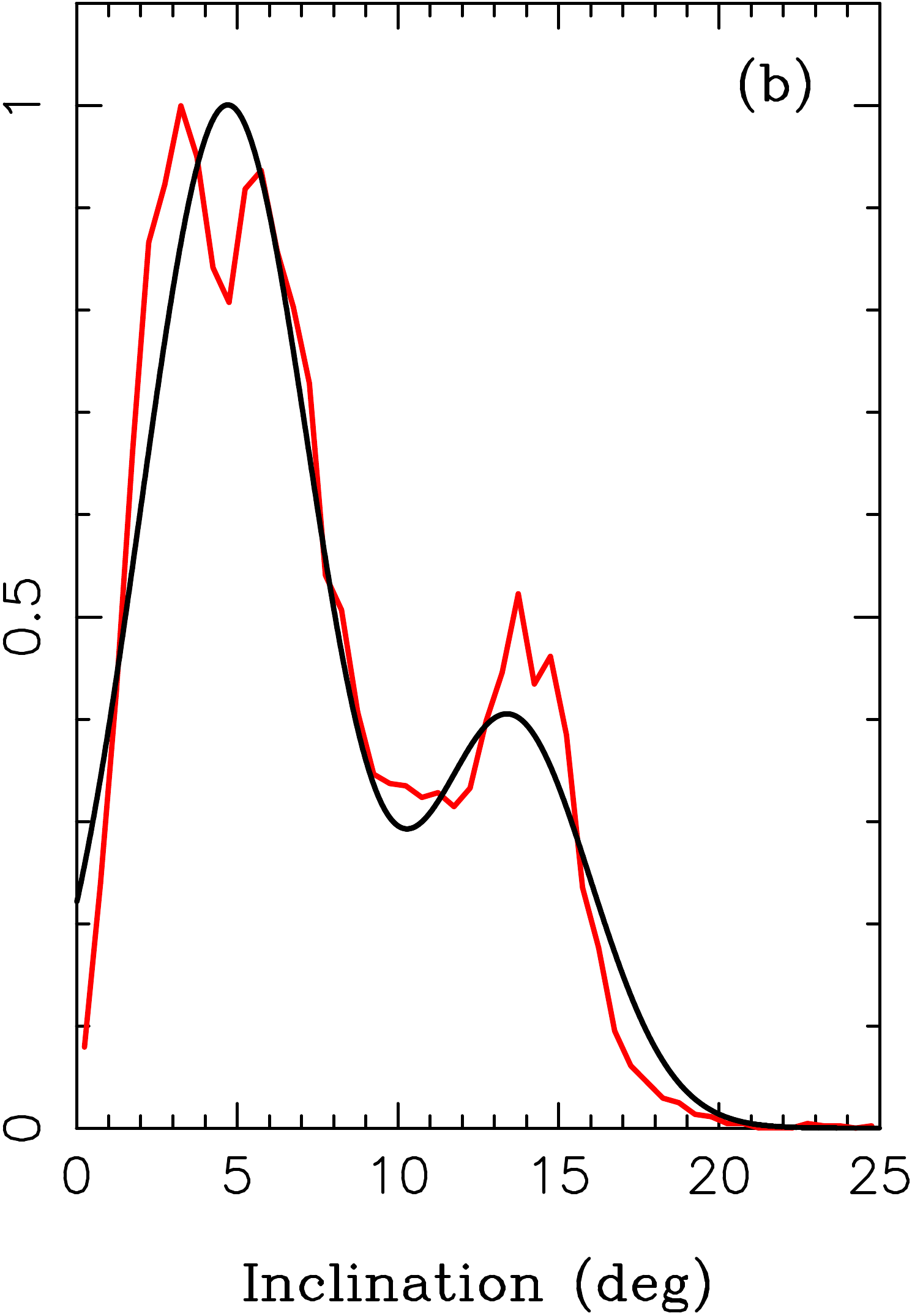}
\caption{The eccentricity (panel a) and inclination (panel b) distributions of bodies placed in the 
3:1 resonance with Jupiter. The red lines are the actual distributions of main belt asteroids near the
3:1 resonance. The black lines are the analytic approximation of these distributions described in the 
main text.}
\label{ei}
\end{figure}


\clearpage
\begin{figure}
\epsscale{0.3}
\plotone{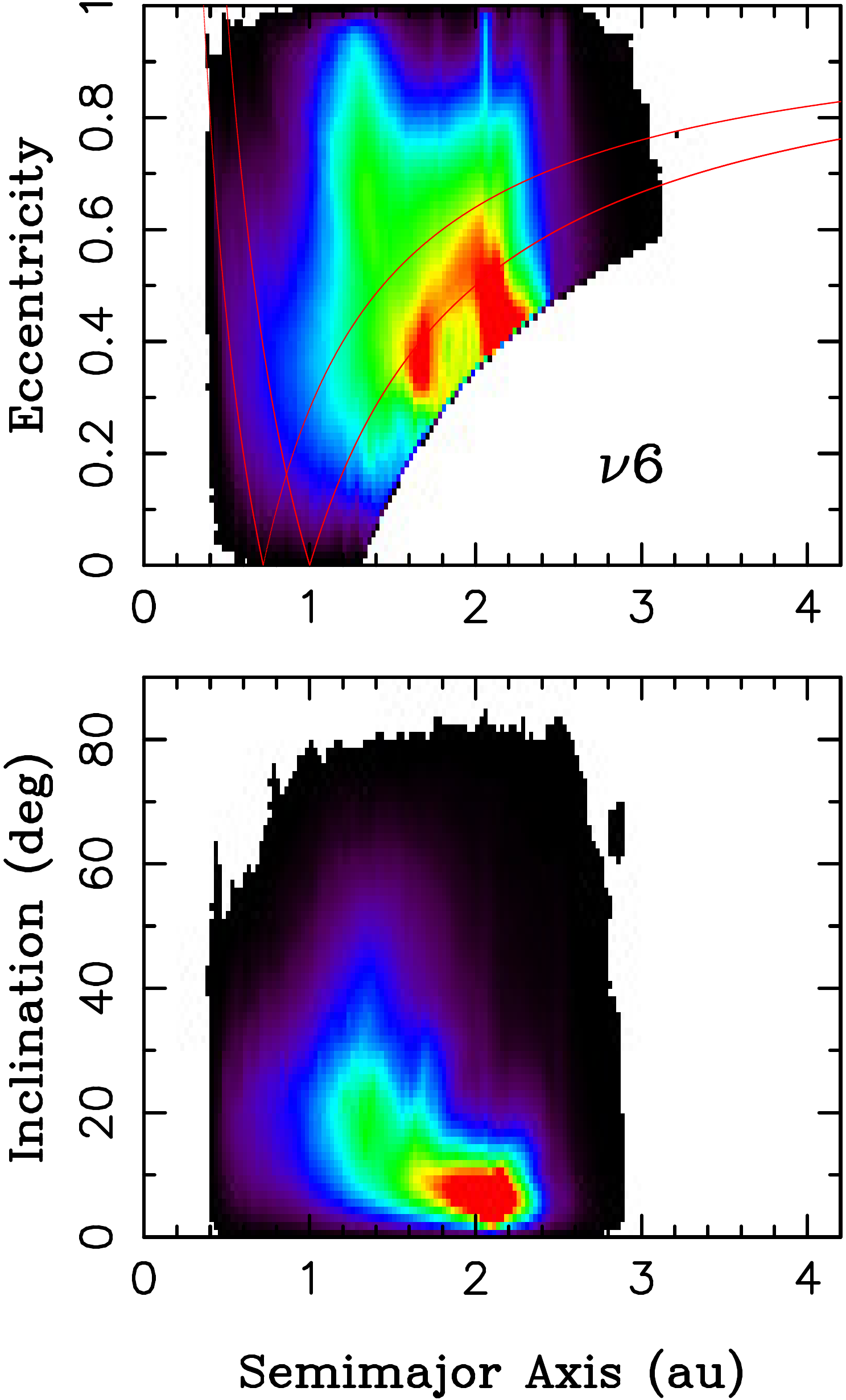}\hspace{2.mm}
\plotone{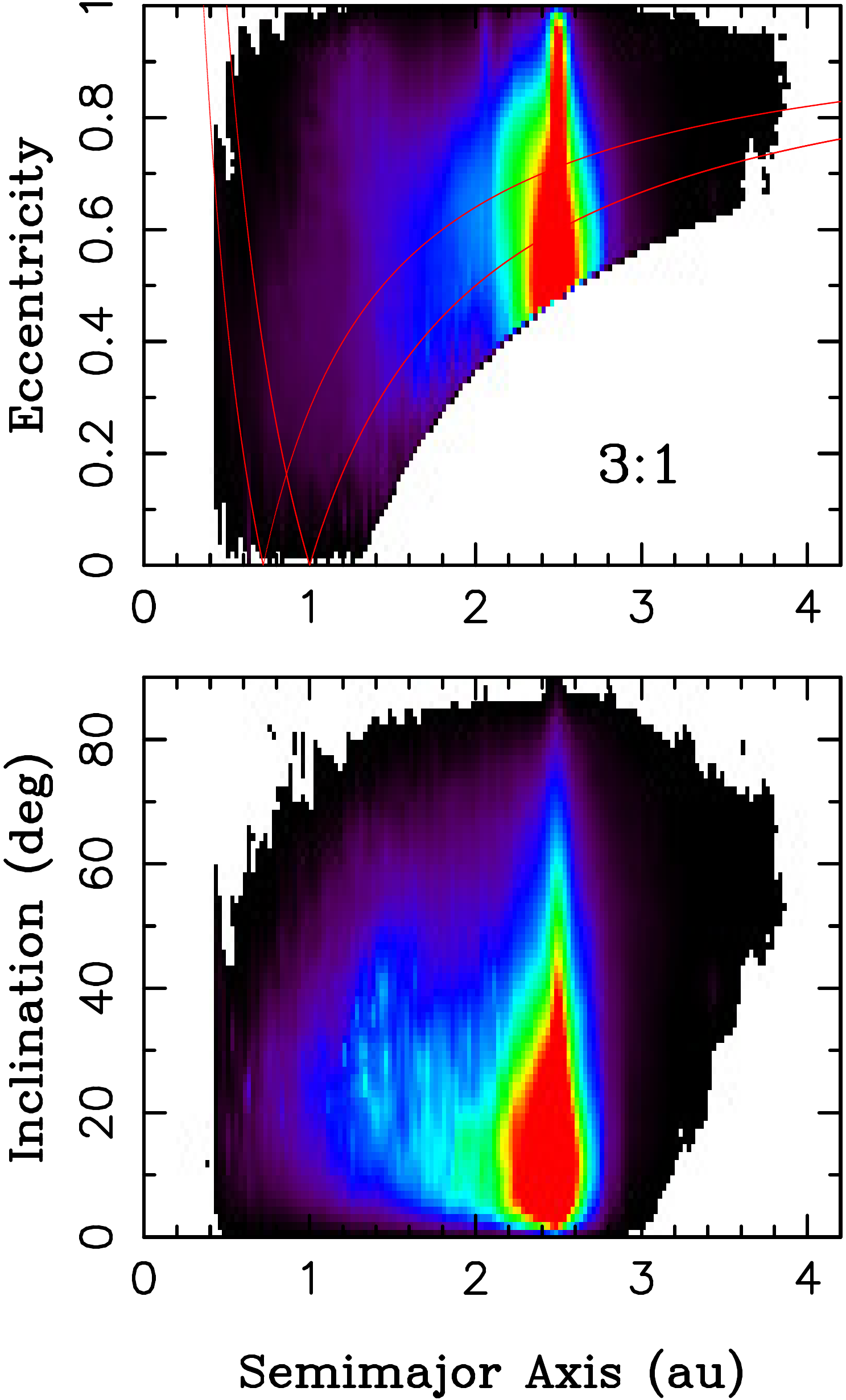}\hspace{2.mm}
\plotone{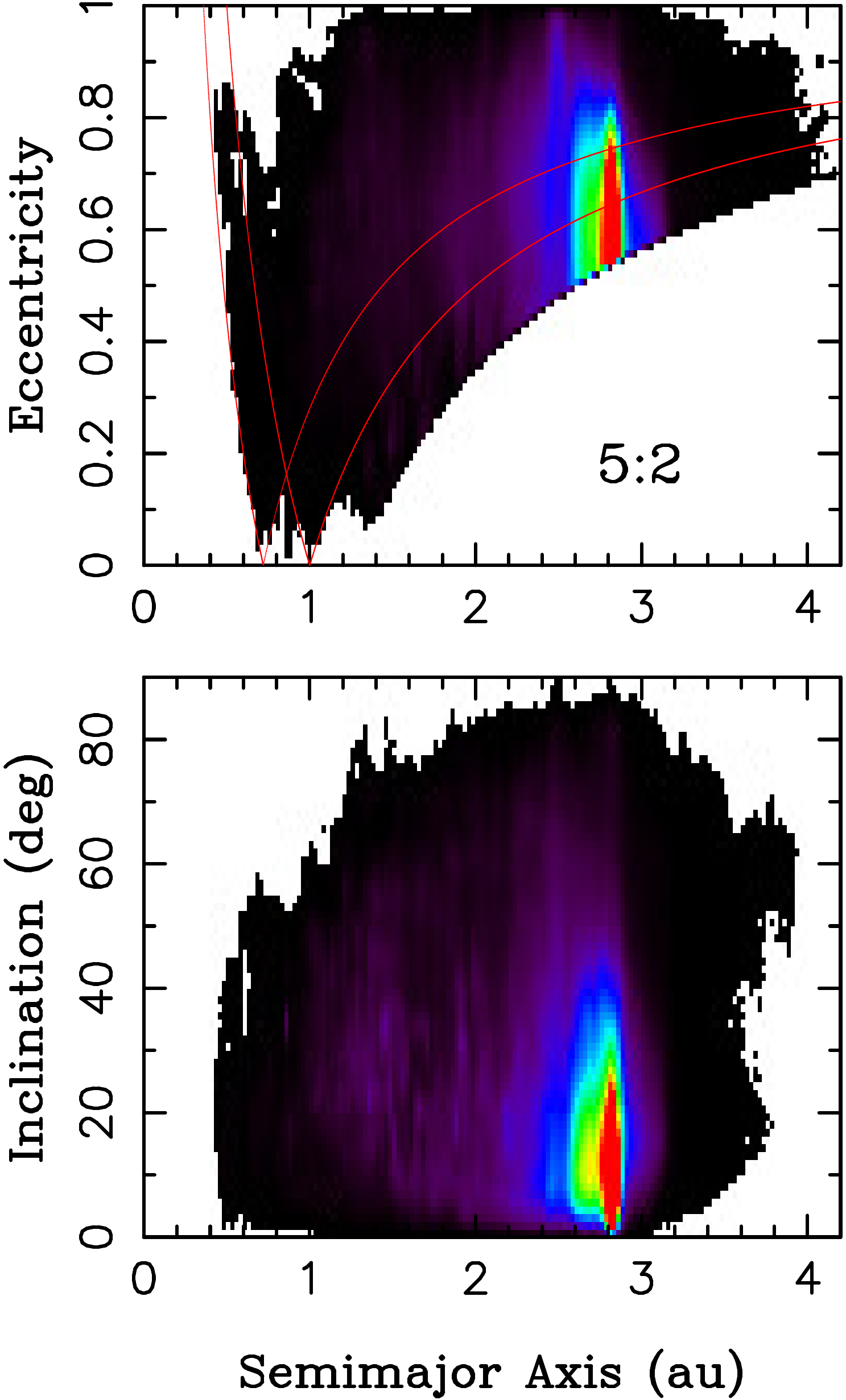}\vspace{3.mm}
\plotone{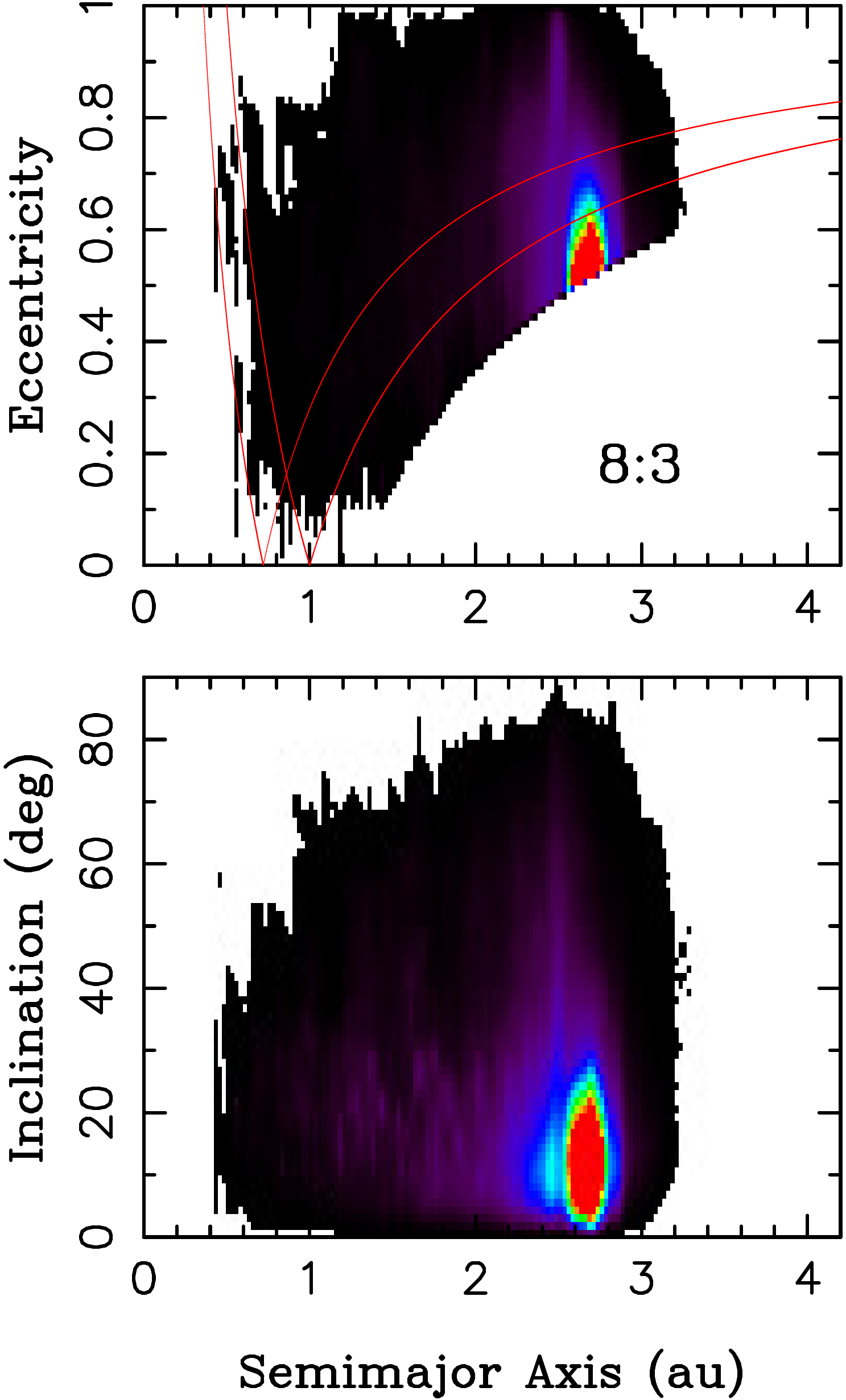}\hspace{2.mm}
\plotone{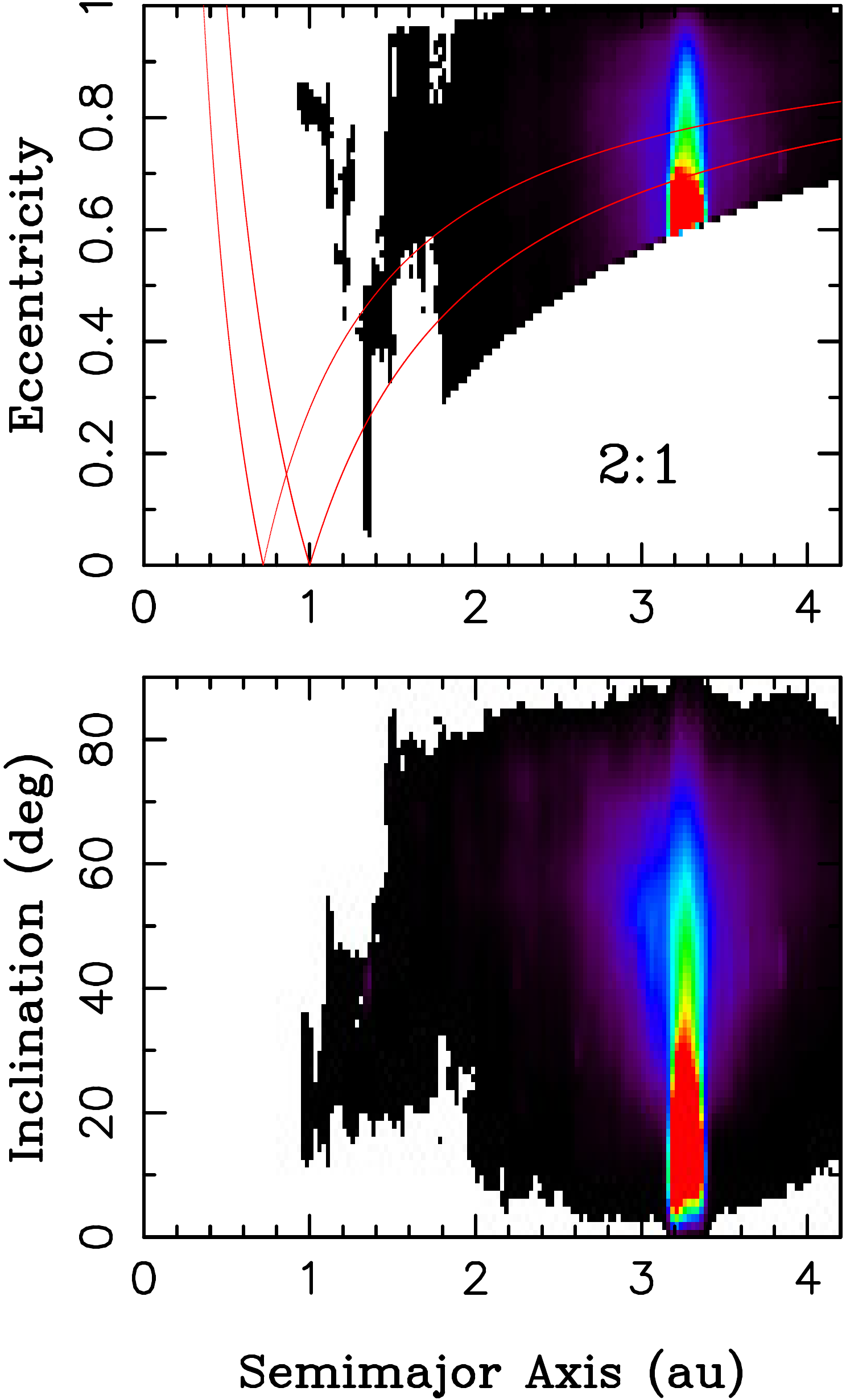}\hspace{2.mm}
\plotone{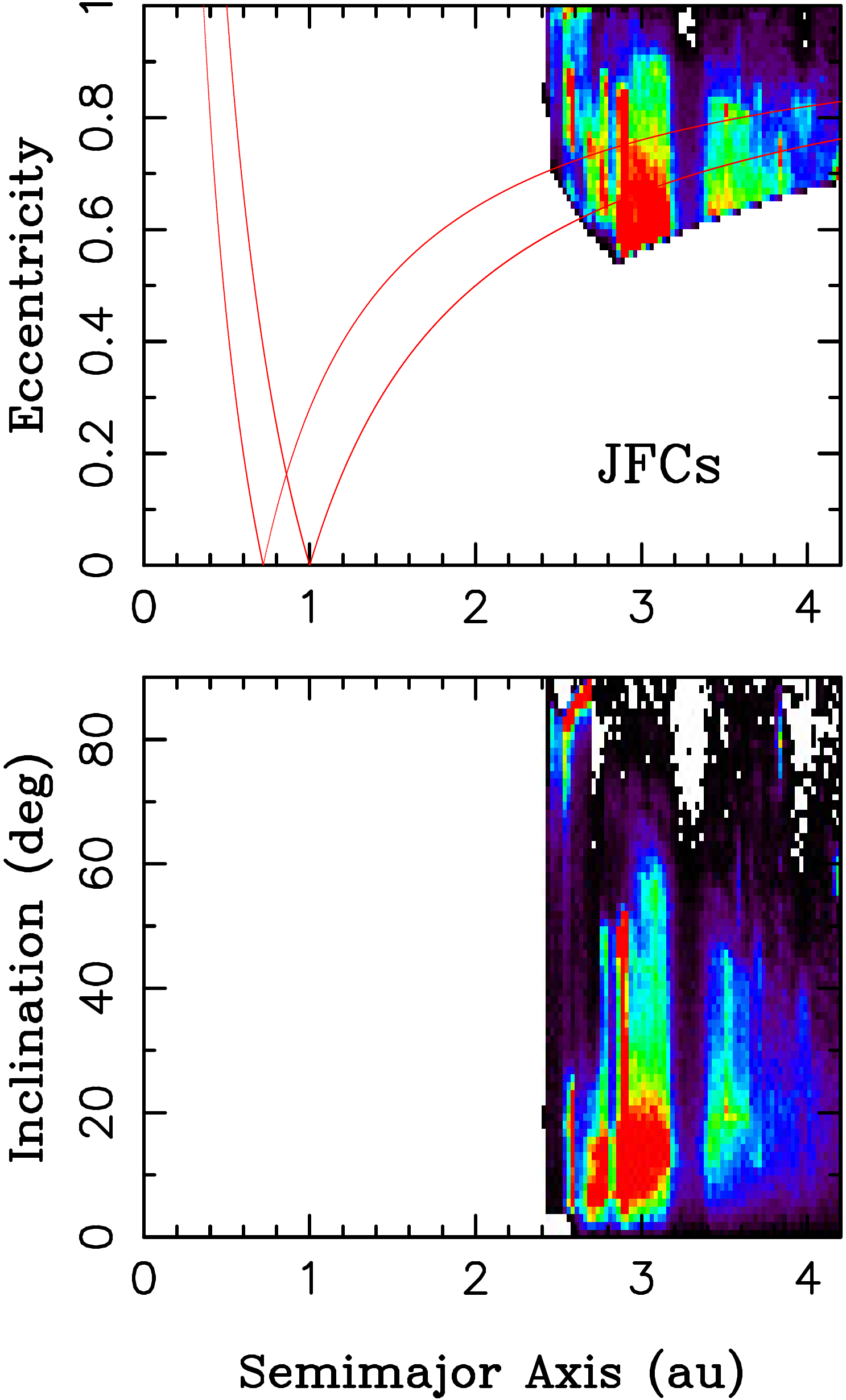}
\caption{The projected PDFs of model NEO orbits for different sources (projected $p_j(a,e,i)$): $\nu_6$, 3:1, 5:2, 8:3, 
2:1 and JFCs (from top left to bottom right). Higher values are shown by brighter 
colors. For reference, the red lines show orbits with $q=a_{\rm Earth}$, $Q=a_{\rm Earth}$, $q=a_{\rm Venus}$, 
and $Q=a_{\rm Venus}$, where $Q=a(1+e)$ is the aphelion distance, $a_{\rm Earth}=1.0$ au and $a_{\rm Venus}=0.72$ au.}
\label{reside}
\end{figure}

\clearpage
\begin{figure}
\epsscale{0.7}
\plotone{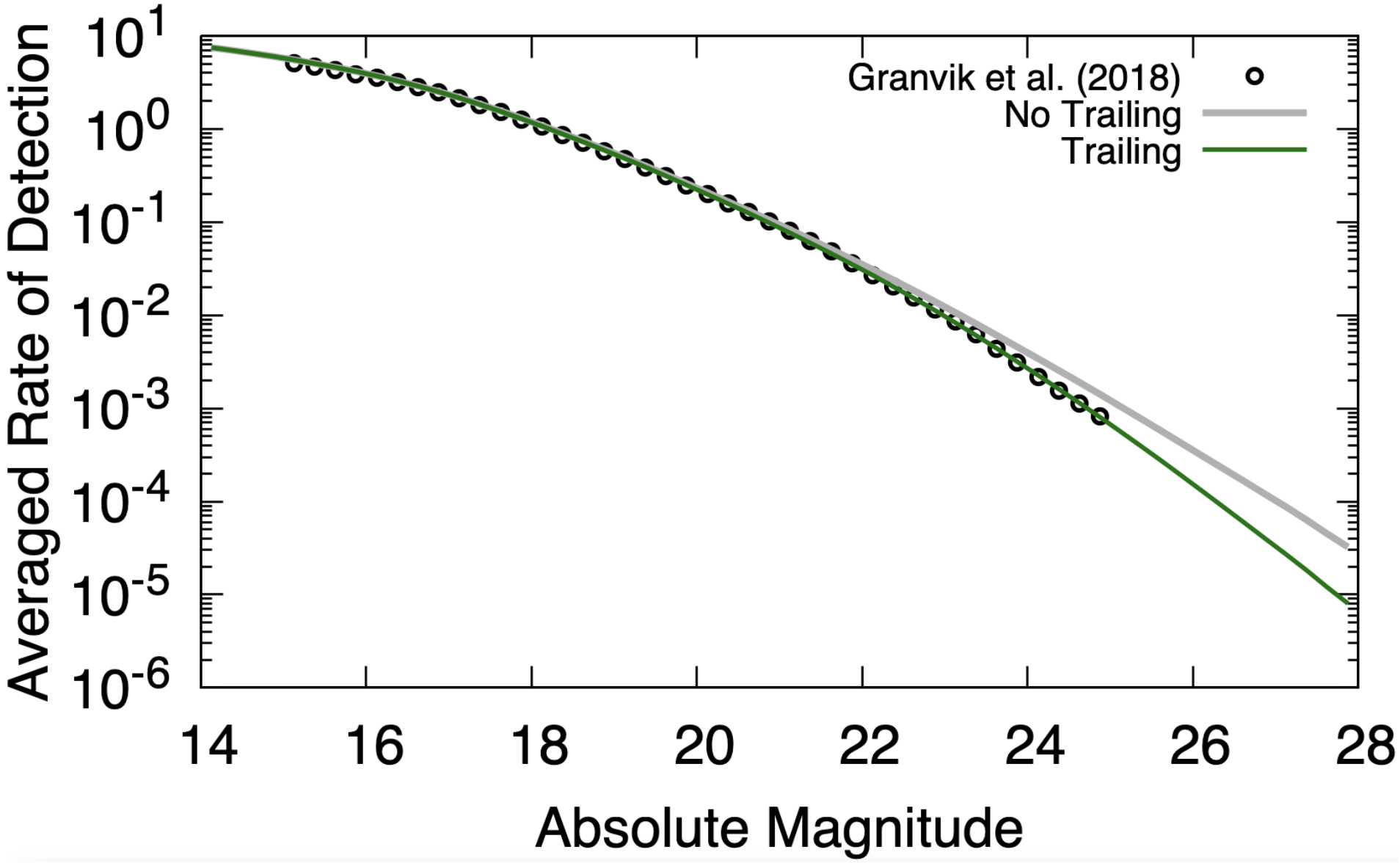}
\caption{The CSS's mean rate of detection -- the number of CSS FoVs in which a NEO with given orbital
  elements is expected to be detected -- is plotted as a function of the absolute magnitude (green
  line). The average of ${\cal R}(a,e,i,H)$, given in Eq. (\ref{calr}), was computed over the
  whole orbital domain. The original bias from Granvik et al. (2018) is shown by open circles.
  The gray line shows the detection rate when the trailing loss from Zavodny et al. (2008) is not
  accounted for.}
\label{bias1}
\end{figure}

\clearpage
\begin{figure}
\epsscale{1.0}
\plotone{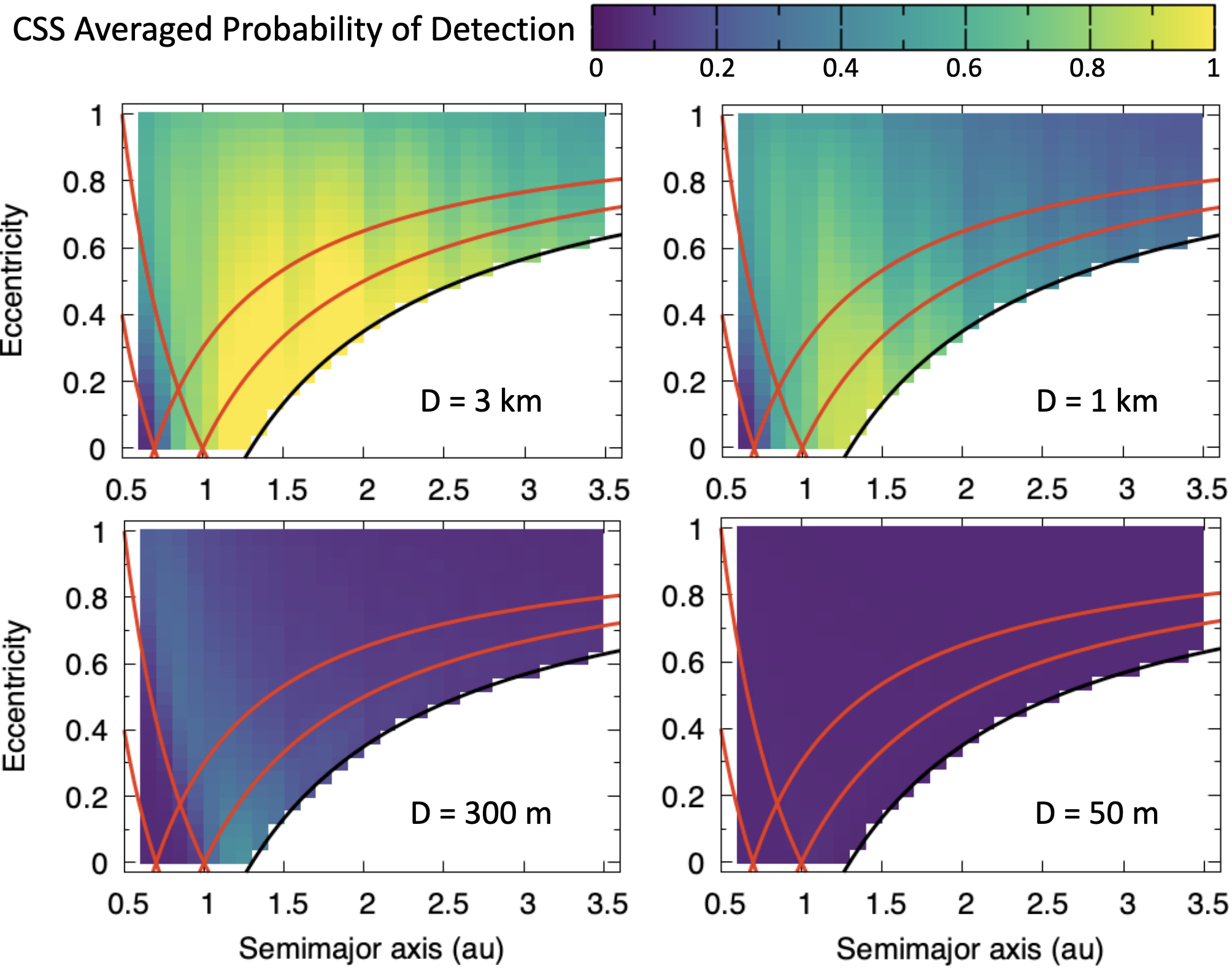}
\caption{The CSS detection probability (Eq. \ref{calp}) as a function of orbital elements for four different 
  absolute magnitude values.  From top-left to bottom-right, we plot ${\cal P}(a,e,i,H)$ for $H$ corresponding
  to objects with $D=3$ km, 1 km, 300 m and 50 m ($H=15.37$, 17.75, 20.37 and 24.26 for the reference albedo 
  $p_{\rm V}=0.14$). The detection probability was averaged over all inclinations bins. The vertical strips, 
  with ${\cal P}$ going up and down as a function of NEO's semimajor axis, are discussed in the main text.}
\label{bias2}
\end{figure}

\clearpage
\begin{figure}
\epsscale{0.8}
\plotone{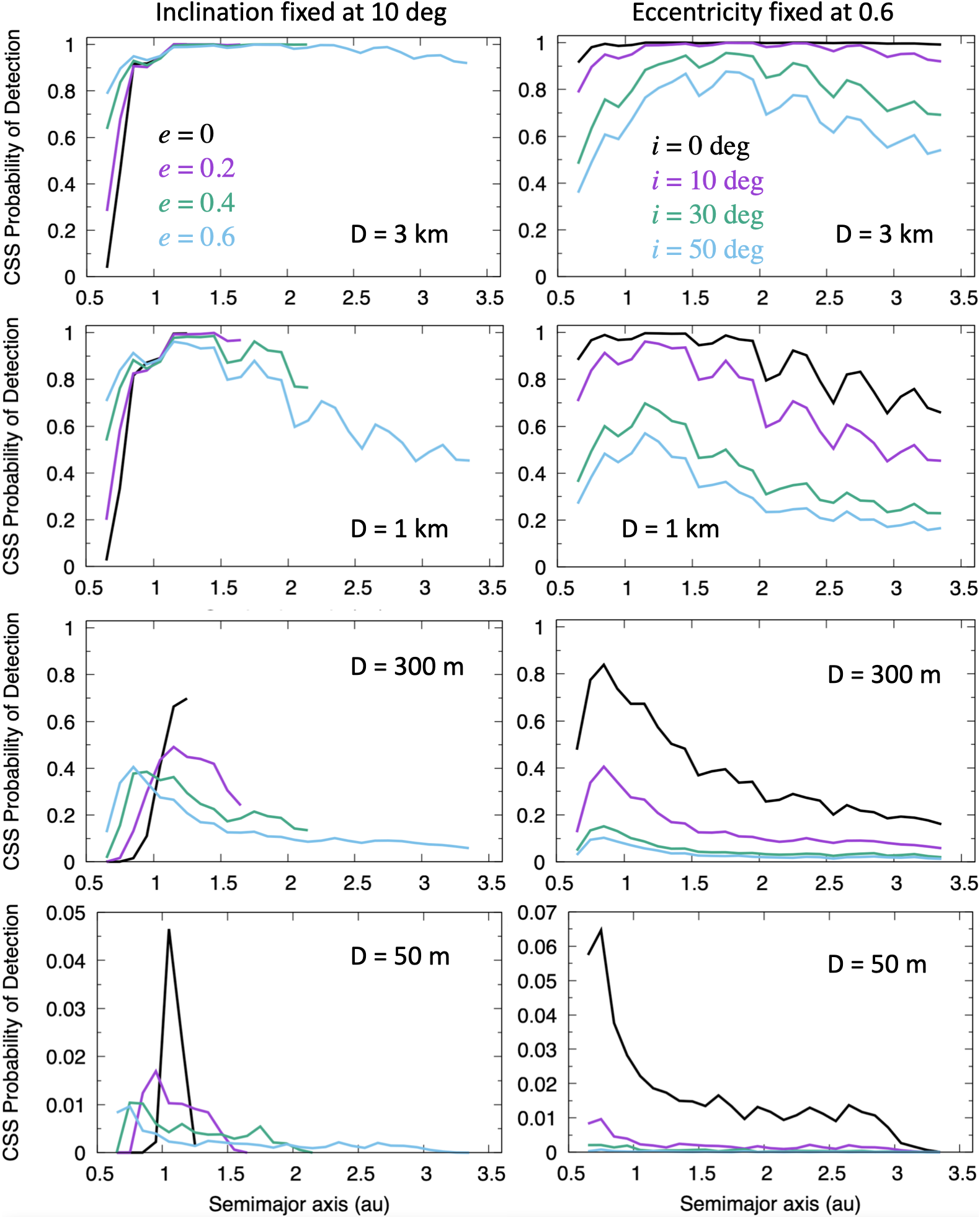}
\caption{The CSS detection probability (Eq. \ref{calp}) as a function of orbital elements for four 
different absolute magnitude values.  From top to bottom, we plot ${\cal P}(a,e,i,H)$ for 
$H$ corresponding to objects with $D=3$ km, 1 km, 300 m and 50 m ($H=15.37$, 17.75, 20.37 and 24.26 
for the reference albedo $p_{\rm V}=0.14$).
The plots in the left column show ${\cal P}$ for the fixed orbital inclination ($i=10^\circ$) and several
eccentricity values. The plots on the right show ${\cal P}$ for $e=0.6$ and several inclination values.
The detection probability was computed for orbits with $q<1.3$ au.}
\label{bias3}
\end{figure}



\clearpage
\begin{figure}
\epsscale{0.9}
\plotone{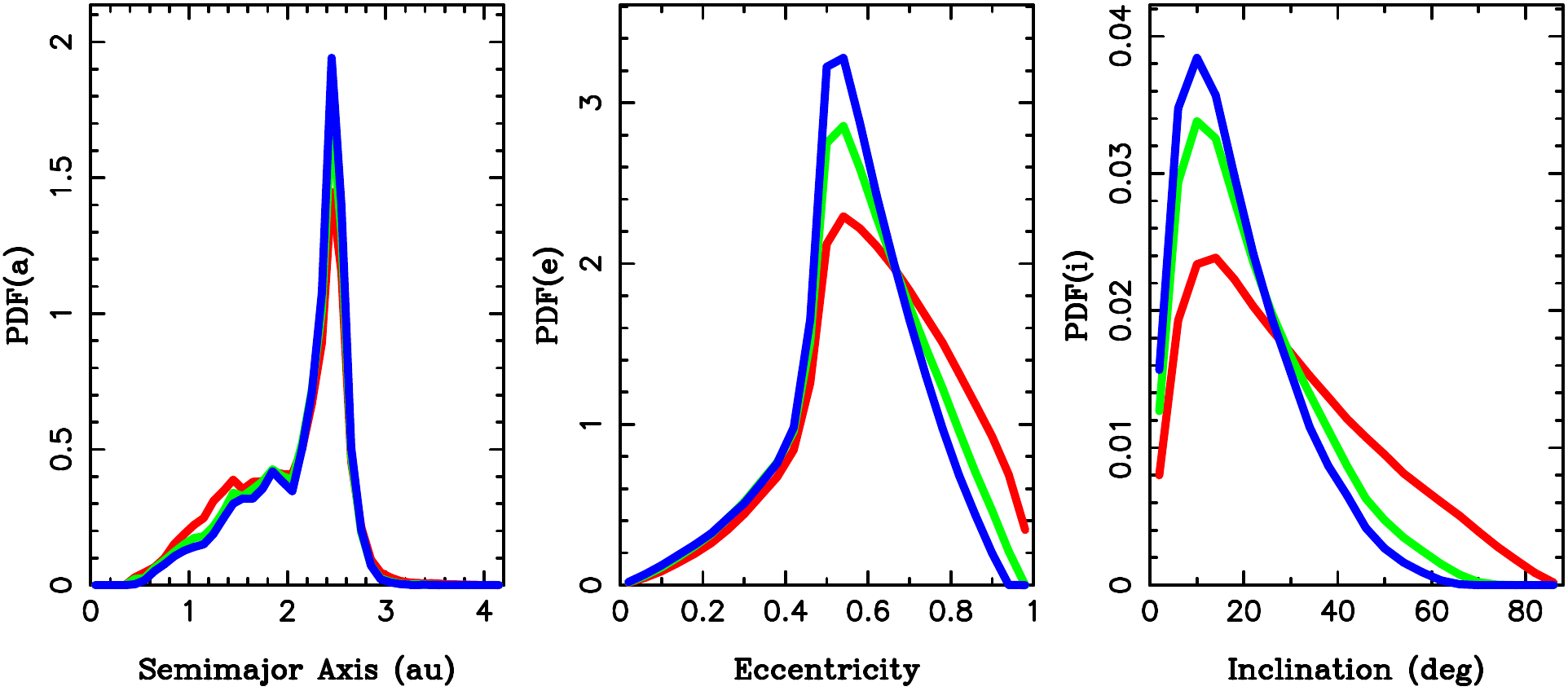}
\caption{The orbital distributions of NEOs from the 3:1 source for three disruption thresholds:
$q^*=0.005$ au (red line), $q^*=0.1$ au (green line), and $q^*=0.2$ au (blue line). By increasing the 
disruption distance in the model, we remove the orbits with high eccentricities, and the eccentricity
distribution becomes more peaked near $e=0.5$. At the same time, the inclination distribution becomes
narrower.}
\label{disrupt}
\end{figure}

\clearpage
\begin{figure}
\epsscale{1.0}
\plotone{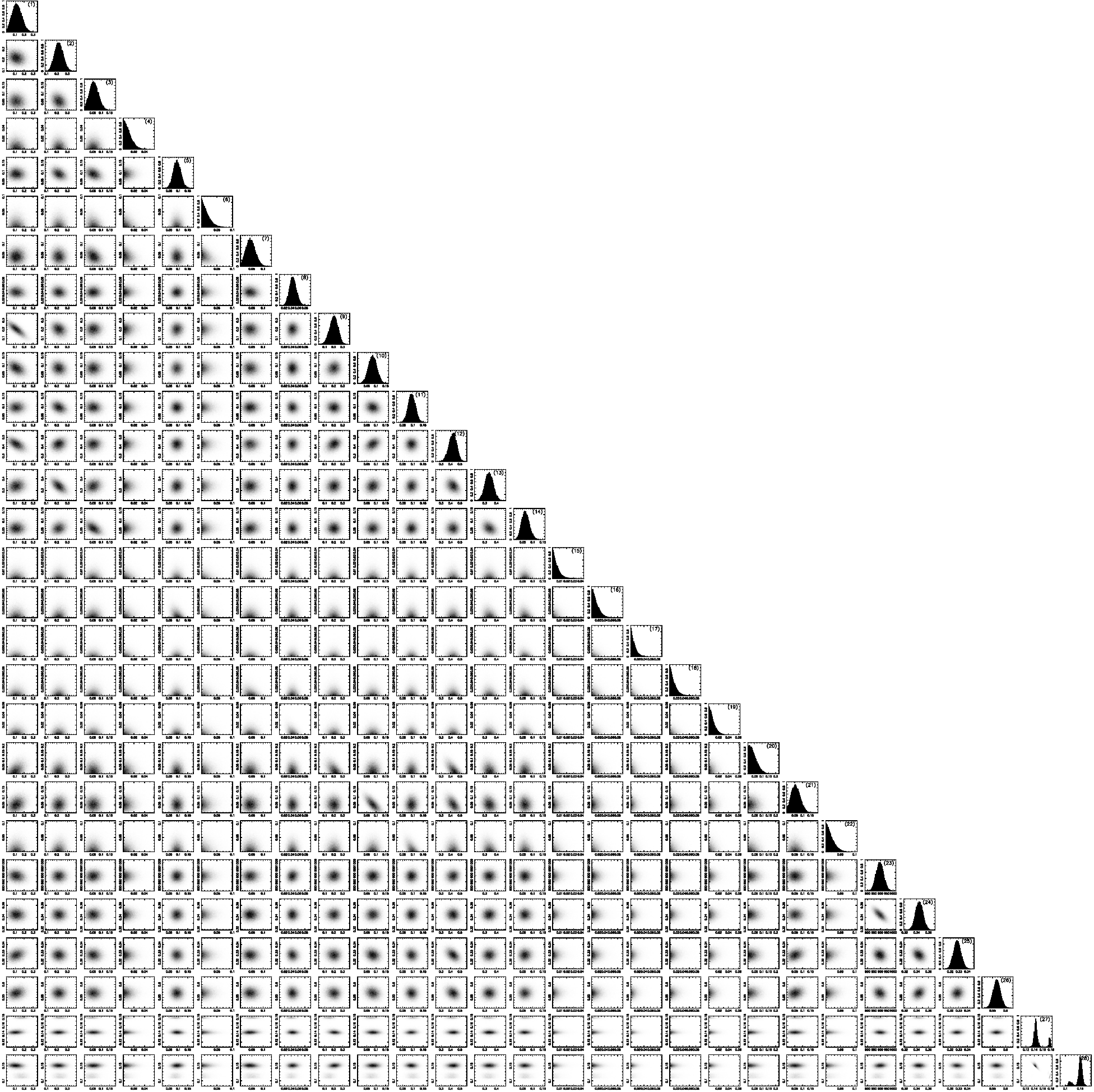}
\caption{The posterior distribution of 28 NEOMOD parameters from our base \texttt{MultiNest} 
fit to CSS. The individual plots are labeled (1) to (28) following the model parameter sequence 
given in Table 3.}
\label{triangle}
\end{figure}

\clearpage
\begin{figure}
\epsscale{0.9}
\plotone{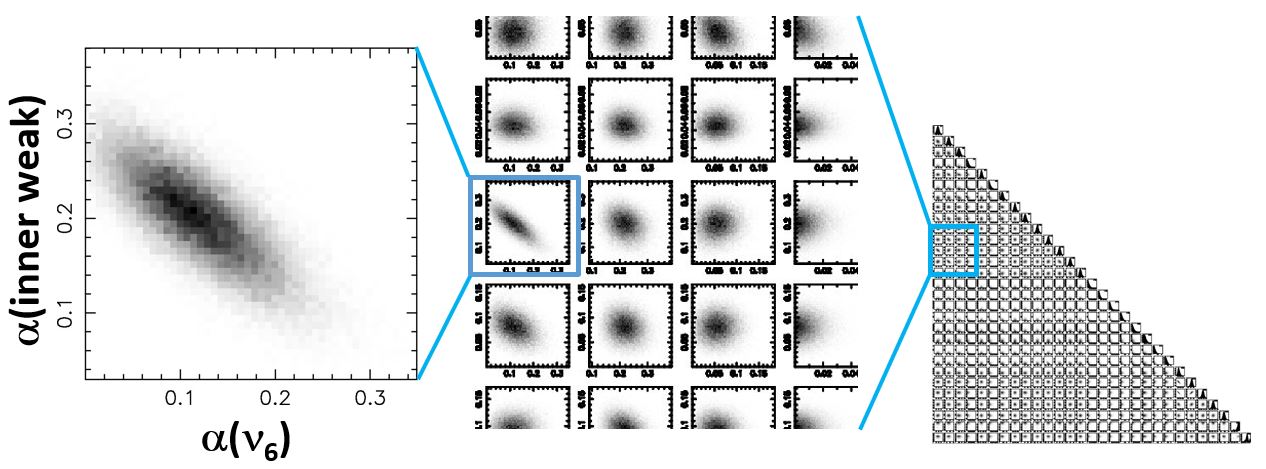}
\caption{The enlarged plot on the left illustrates the degeneracy between contributions of the 
$\nu_6$ resonance and weak resonances in the inner belt to bright NEOs ($H=15$). The two contributions are 
anti-correlated and sum up to $\simeq 30$\%.}
\label{degenerate}
\end{figure}


\clearpage
\begin{figure}
\epsscale{0.42}
\plotone{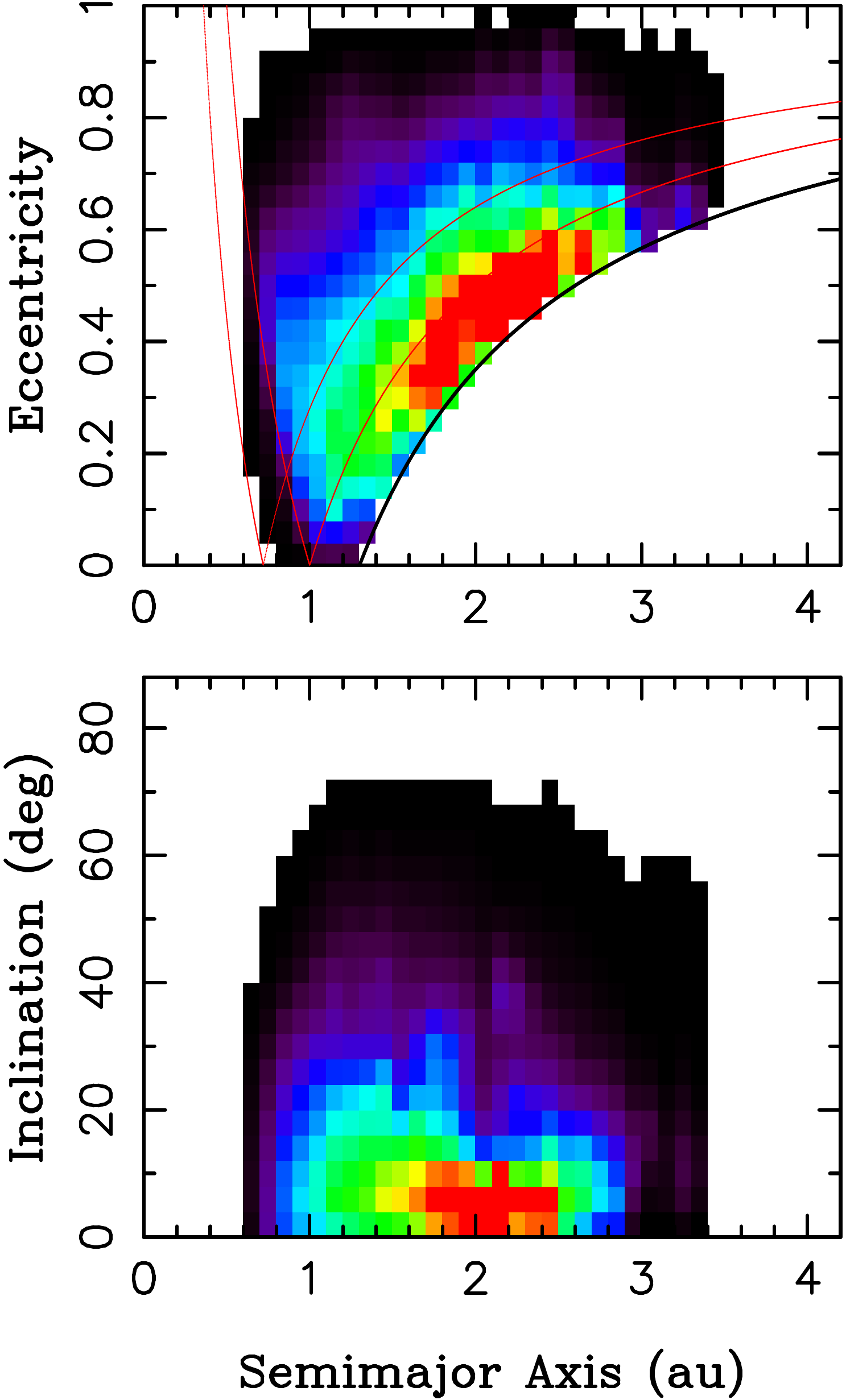}\hspace*{2.mm}
\epsscale{0.455}
\plotone{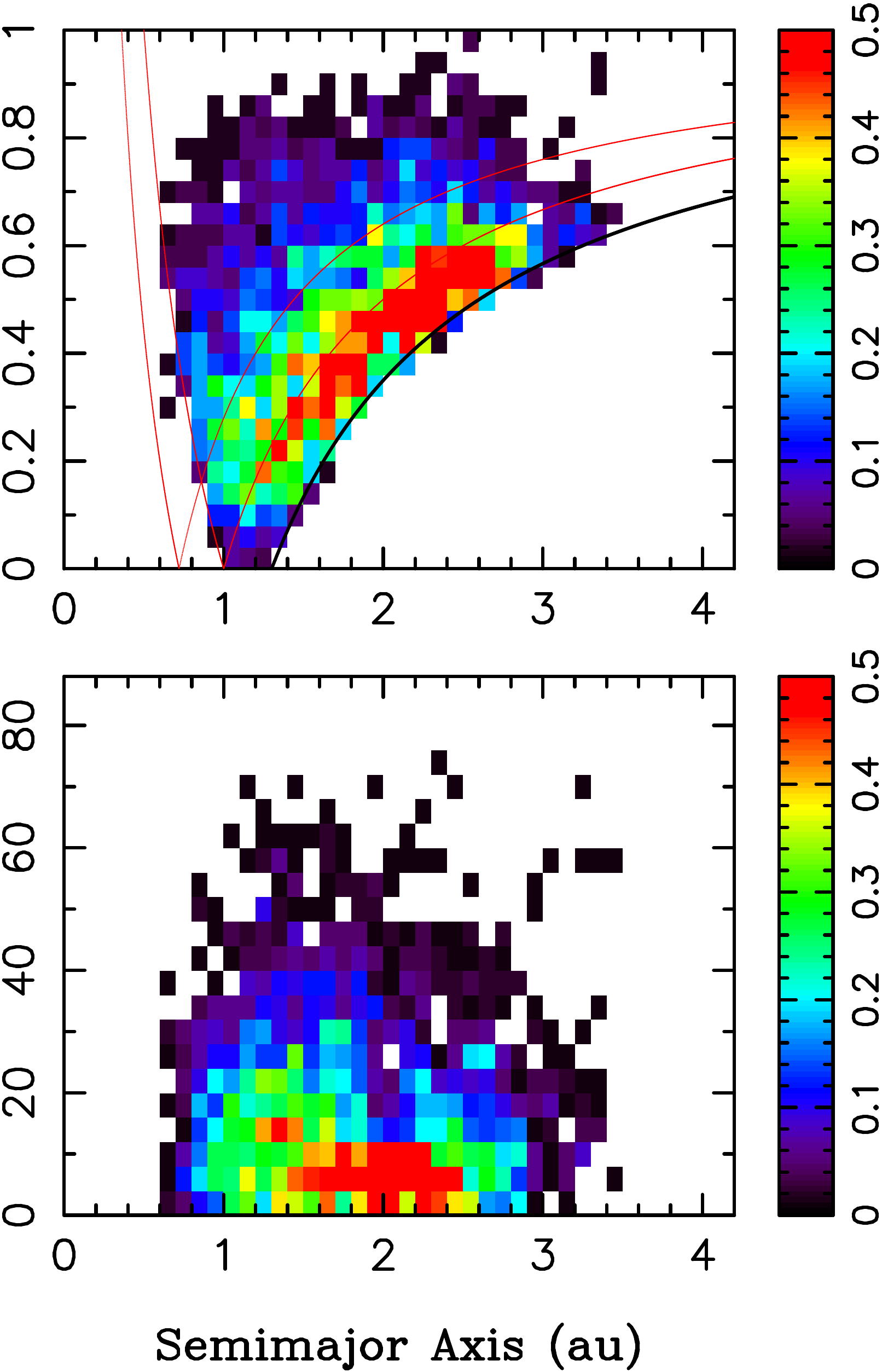}
\caption{The orbital distribution of NEOs from our biased based model (left panels) and the 
CSS NEO detections (right panels). The two distributions were binned with the same resolution 
and are shown here in the $(a,e)$ and $(a,i)$ projections. There are no NEOs with the aphelion
inside Venus orbit in CSS (and the biased model), because the pointing strategy of CSS 
had negligible low solar-elongation coverage.}
\label{bmodel}
\end{figure}

\clearpage
\begin{figure}
\epsscale{0.8}
\plotone{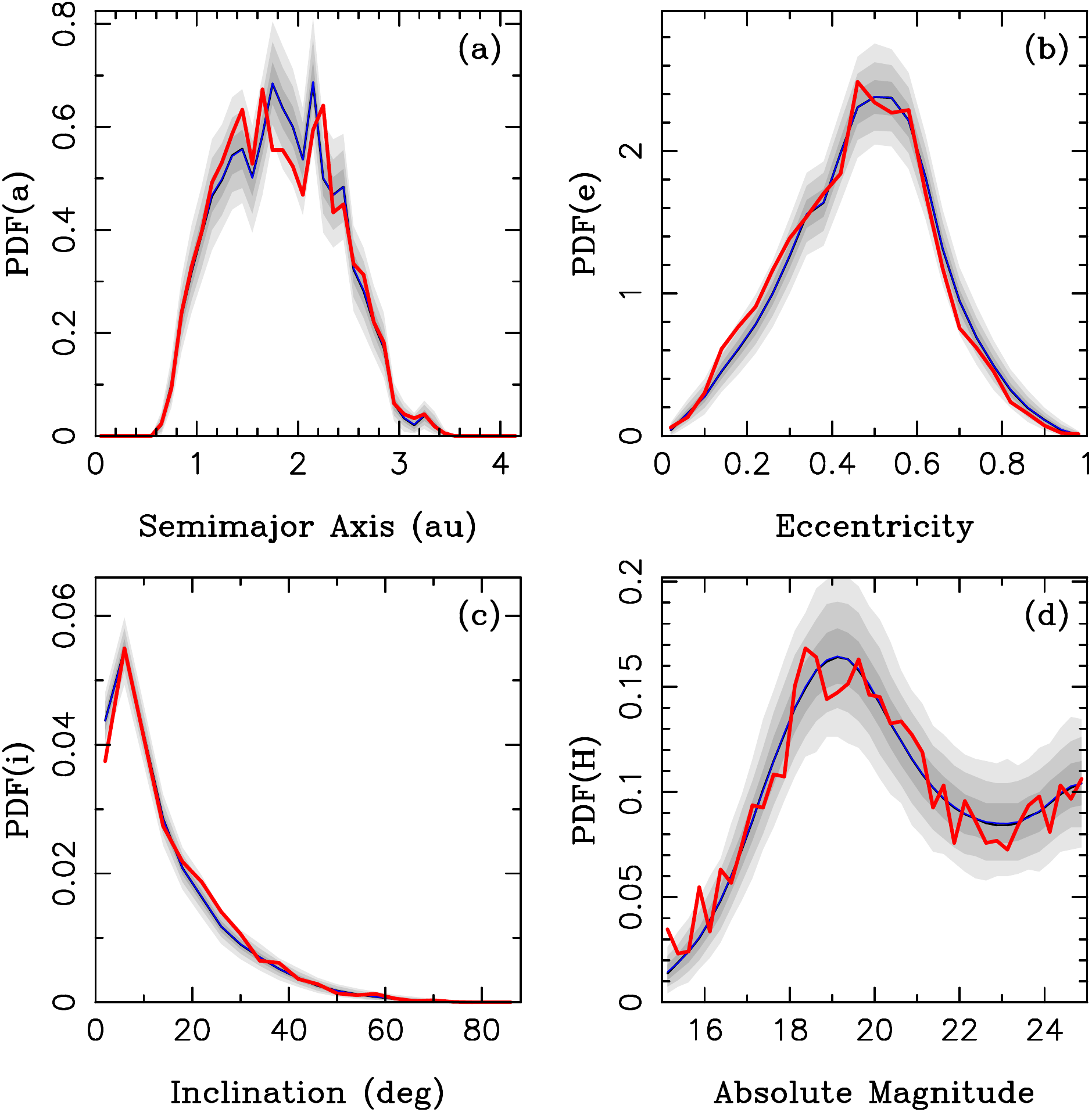}
\caption{The probability density functions (PDFs) of $a$, $e$, $i$, and $H$ from our 
biased base best-fit model (blue lines) and the CSS NEO detections (red lines). The shaded 
areas are 1$\sigma$ (bold gray), 2$\sigma$ (medium) and 3$\sigma$ (light gray) envelopes. We 
used the best-fit solution (i.e. the one with the maximum likelihood) from the base model and 
generated 30,000 random samples with 3,803 NEOs each (the sample size identical to the number 
of CSS's NEOs in the model domain; $15<H<25$). The samples were biased and binned with the 
standard binning (Table 2). We identified envelopes containing 68.3\% (1$\sigma$), 95.5\% 
(2$\sigma$) and 99.7\% (3$\sigma$) of samples and plotted them here. The K-S test 
probabilities are 9.7\%, 14\%, 32\% and 61\% for the $a$, $e$, $i$ and $H$ distributions,
respectively.}
\label{dif}
\end{figure}


\clearpage
\begin{figure}
\epsscale{0.48}
\plotone{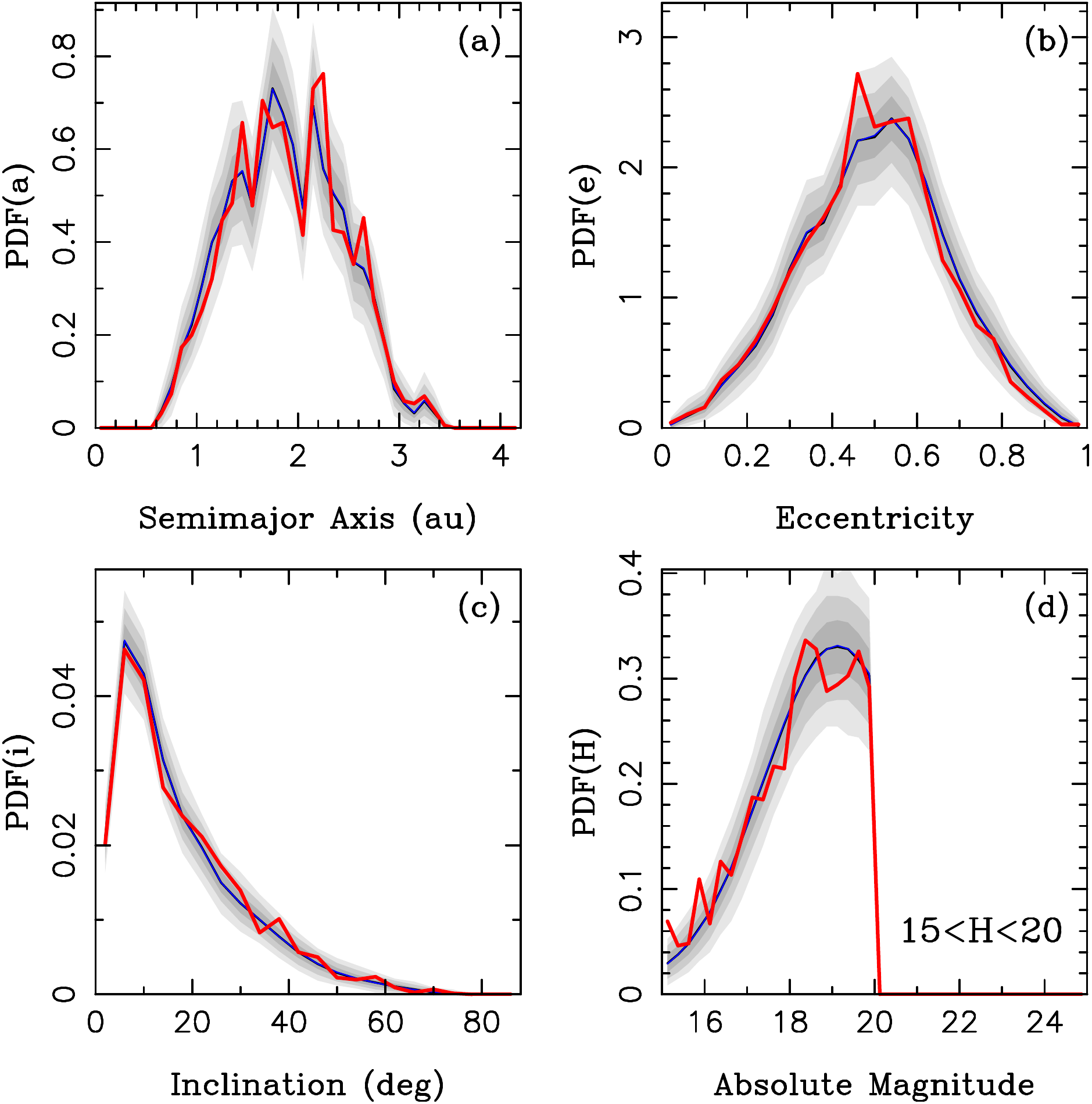}\hspace*{5.mm}
\plotone{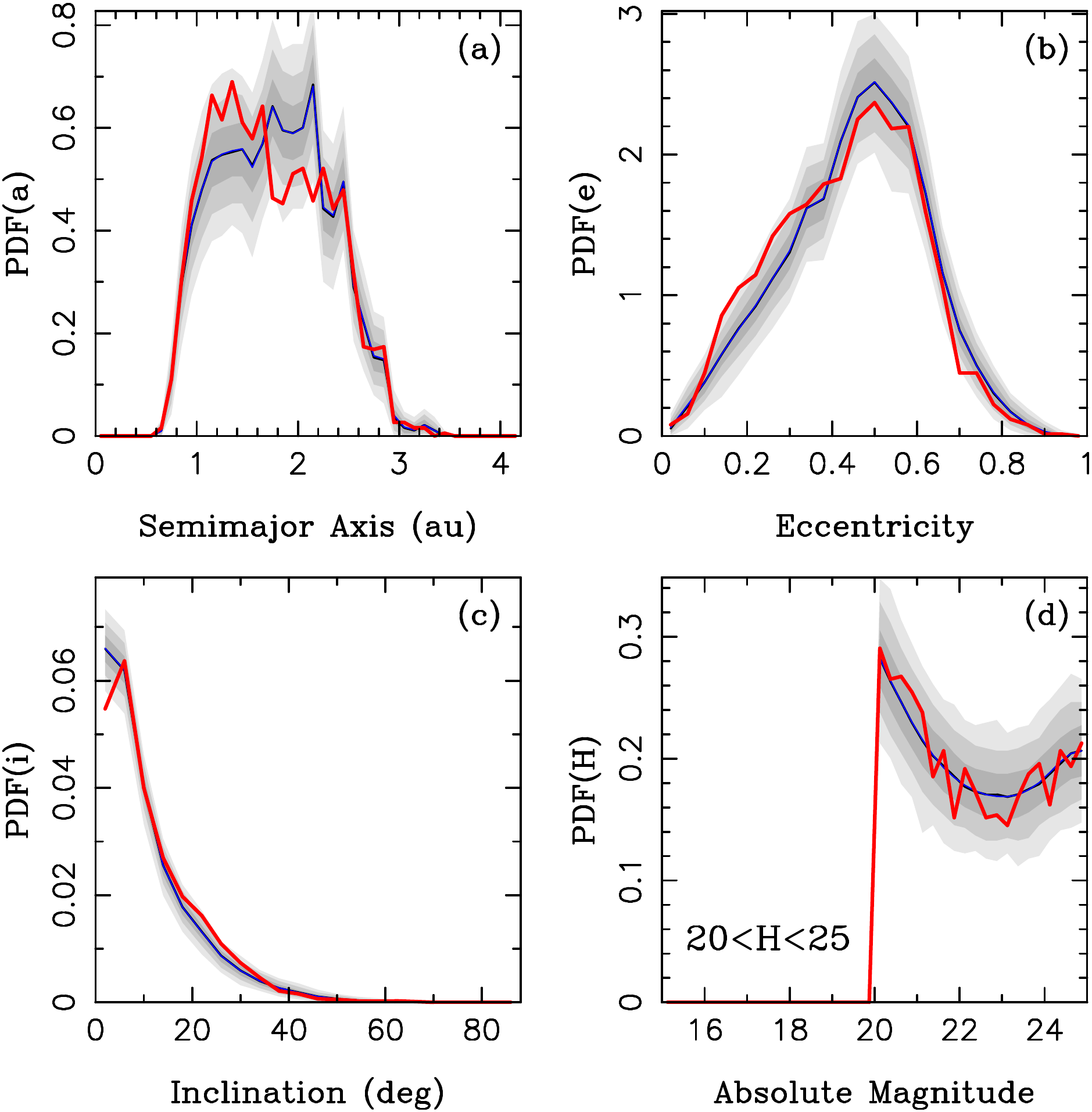}
\caption{The probability density functions (PDFs) of $a$, $e$, $i$, and $H$ from our biased base 
best-fit model (blue lines) are compared to the CSS NEO detections (red lines). The four panels on the left 
show the results for bright NEOs with $15<H<20$, and the four panels on the right show the results 
for faint NEOs with $20<H<25$. The shaded areas are 1$\sigma$ (bold gray), 2$\sigma$ (medium) and 
3$\sigma$ (light gray) envelopes. See caption of Fig. \ref{dif} for the method that we used to compute 
these envelopes. For $20<H<25$, the K-S test probabilities are $10^{-4}$ and 0.012 for the $a$ and $e$
distributions, respectively.}
\label{bright}
\end{figure}

\clearpage
\begin{figure}
\epsscale{0.6}
\plotone{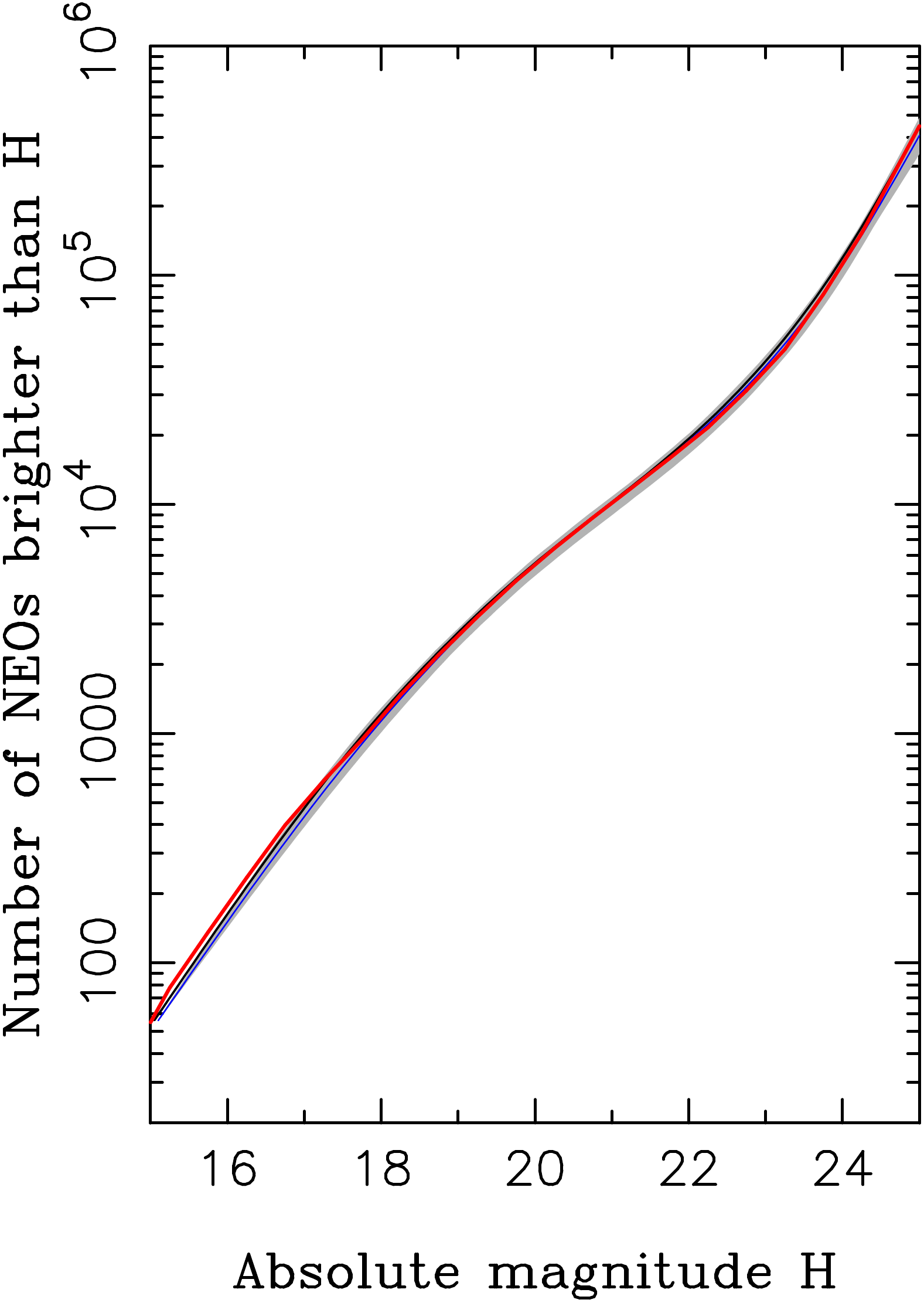}
\caption{The intrinsic (debiased) absolute magnitude distribution of NEOs from our base model (black line
is the median, blue line is the best fit) is compared to the magnitude distribution from Harris \& Chodas 
(2021) (red line). The gray area is the 3$\sigma$ envelope obtained from the posterior distribution
computed by \texttt{MultiNest}. It contains -- by definition -- 99.7\% of our base model posteriors.}
\label{harris}
\end{figure}

\clearpage
\begin{figure}
\epsscale{0.8}
\plotone{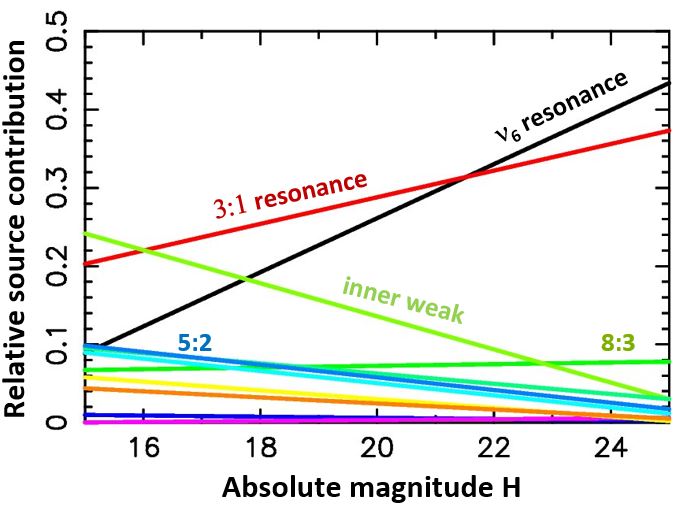}
\caption{The contribution of different NEO sources as a function of the absolute magnitude. The $\nu_6$
and 3:1 resonances are shown by the black and red lines. The light-green line is the contribution of 
weak resonances in the inner main belt. The plot shows the result for the maximum likelihood parameter
set from the base model. We simply plot $\alpha_j(15)$ and $\alpha_j(25)$ for each source and connect
them by a straight line (Sect 5.1). The uncertainties of $\alpha_j(15)$ and $\alpha_j(25)$ are listed 
in Table 3.}
\label{alphas}
\end{figure}

\clearpage
\begin{figure}
\epsscale{0.7}
\plotone{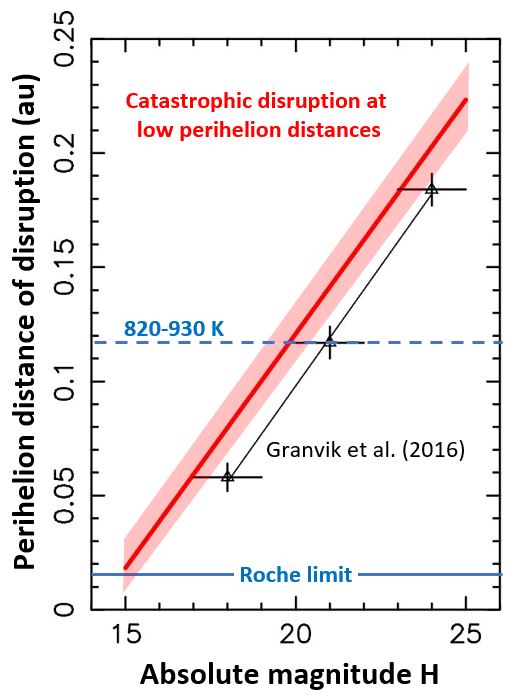}
\caption{Disruption models. Our base model ${\cal M}$, where the critical perihelion distance $q^*$
is assumed to be a linear function of absolute magnitude (the red line with the light red envelope containing 68\% of 
posteriors), is compared to Granvik et al. (2016) (triangles with error bars). The surface temperature
is estimated to be 820 K (average) and 930 K (subsolar) at 0.117 au (Granvik et al. 2016).}
\label{qstar}
\end{figure}

\clearpage
\begin{figure}
\epsscale{0.8}
\plotone{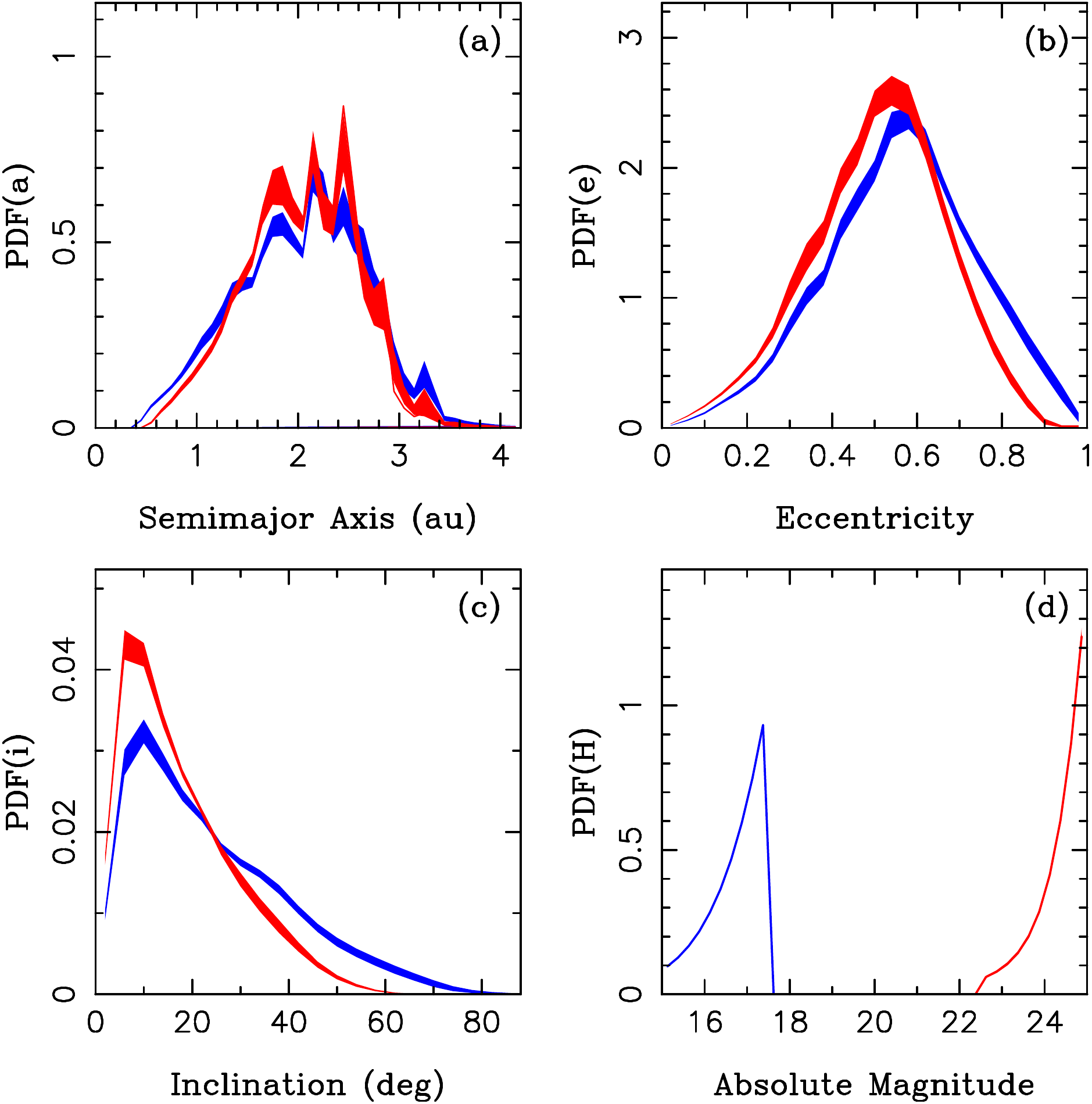}
\caption{The probability density functions (PDFs) of $a$, $e$, $i$, and $H$ from the intrinsic (debiased)
base model. The plot compares the distributions of bright NEOs with $15<H<17.5$ (blue) and faint 
NEOs with $22.5<H<25$ (red). The blue and red shaded areas are the 1$\sigma$ envelopes of our base model 
posteriors.}
\label{intr}
\end{figure}

\clearpage
\begin{figure}
\epsscale{0.7}
\plotone{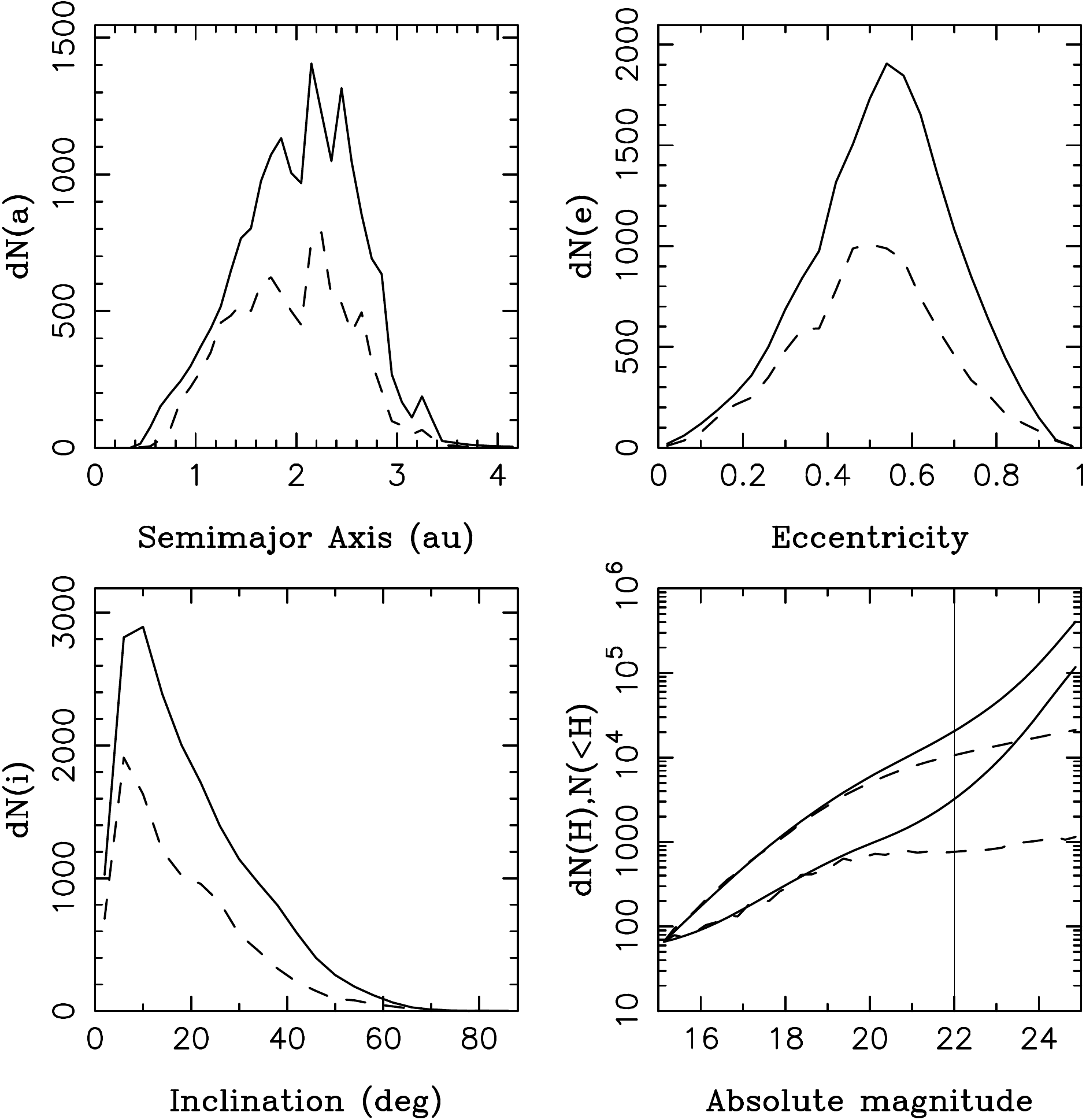}
\caption{The incompleteness of the known NEO population. For the $a$, $e$ and $i$ distributions, the dashed 
lines show the number of known NEOs with $H<22$, and the solid lines are the number of NEOs with $H<22$
inferred from our maximum likelihood base model (both given per bin interval; 0.1 au, 0.04 and 4$^\circ$). 
For the $H$ distribution in the bottom-right panel, we show both the cumulative and differential 
(per 0.25 mag) distributions (upper and lower lines, respectively; solid for model, dashed for known).
The uncertainty of the cumulative population estimates increases from $\simeq3$\% for $H<20$ to
$\simeq6$\% for $H<25$. The uncertainties were obtained from the posterior distribution produced by 
\texttt{MultiNest} and does not account for various uncertainties of the CSS detection efficiency.}
\label{incom}
\end{figure}

\clearpage
\begin{figure}
\epsscale{1.0}
\plotone{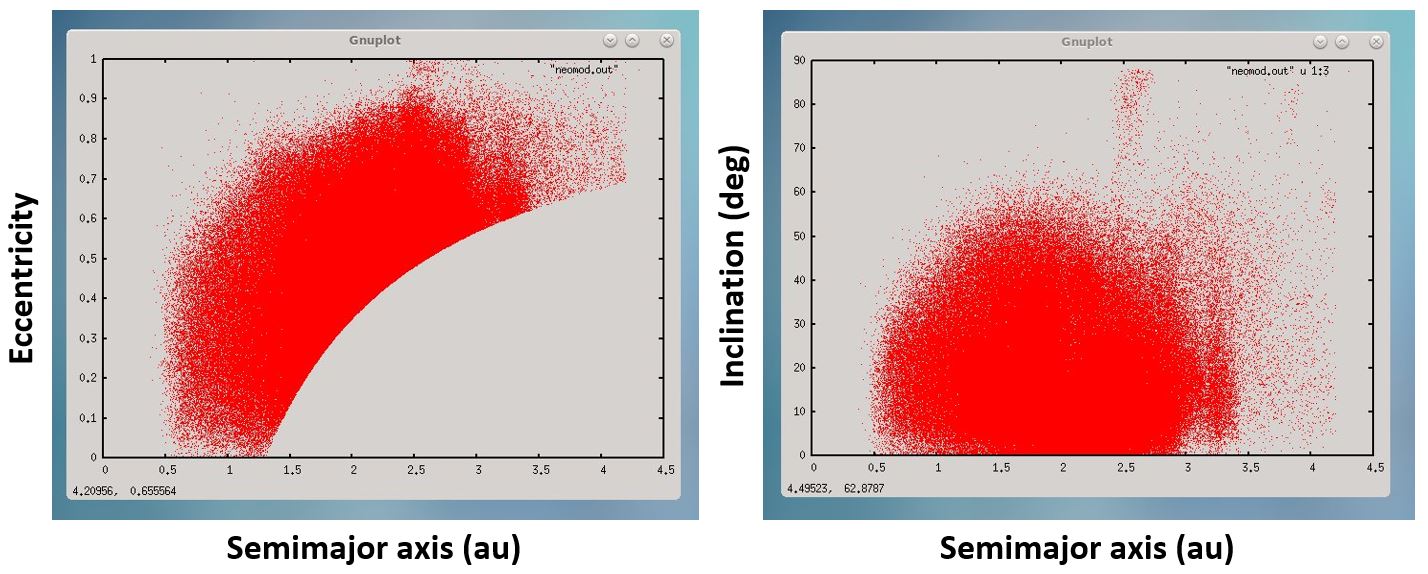}
\caption{A sample output from the NEOMOD Simulator that shows $\simeq 4.1 \times 10^5$ orbits of NEOs with
$H<25$.}
\label{neomod}
\end{figure}


\clearpage
\begin{figure}
\epsscale{0.7}
\plotone{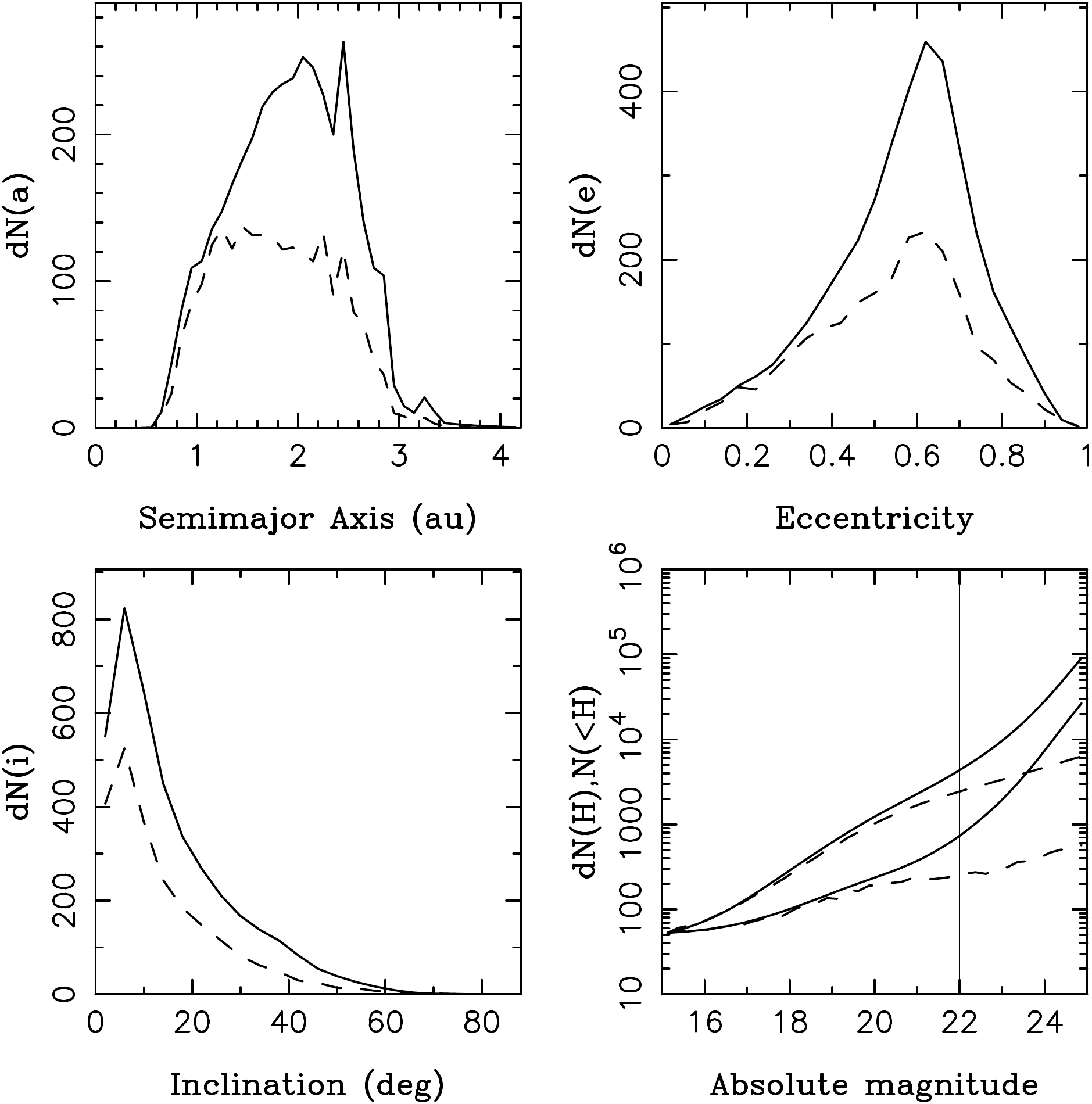}
\caption{The incompleteness of the known PHO population (MOID $<0.05$ au, $H<22$). For the $a$, $e$ and $i$ 
distributions, the dashed lines show the number of known PHOs with $H<22$, and the solid lines are the number 
of PHOs with $H<22$ inferred from our maximum likelihood base model (both given per bin interval; 0.1 au, 
0.04 and 4$^\circ$). For the $H$ distribution in the bottom-right panel, we show both the cumulative and differential 
(per 0.25 mag) distributions (upper and lower lines, respectively; solid for model, dashed for known).
The uncertainty of the cumulative population estimates increases from $\simeq3$\% for $H<20$ to
$\simeq6$\% for $H<25$. The uncertainties were obtained from the posterior distribution produced by 
\texttt{MultiNest} and does not account for various uncertainties of the CSS detection efficiency.}
\label{pho2}
\end{figure}



\begin{figure}
\epsscale{0.7}
\plotone{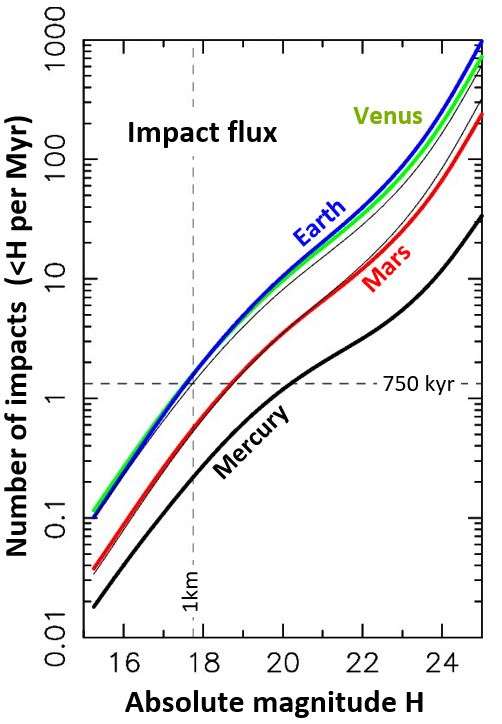}
\caption{The impact flux on the terrestrial planets as a function of NEO absolute magnitude. The black, green, blue 
and red lines show the impact flux for Mercury, Venus, Earth and Mars. The thin solid lines near the Earth flux 
is the base-model NEO magnitude distribution scaled with the fixed impact probability 
($1.5 \times 10^{-3}$ Myr$^{-1}$; see the main text). The thin solid line near the Mars flux is the Earth 
profile scaled down by a factor of 3. The horizontal dashed line shows the impact flux for $D>1$ km bodies 
from Morbidelli et al. (2021) (0.75-Myr average spacing between impacts). The vertical dashed line corresponds 
to $H=17.75$.}
\label{earth}
\end{figure}

\clearpage
\begin{figure}
\epsscale{0.8}
\plotone{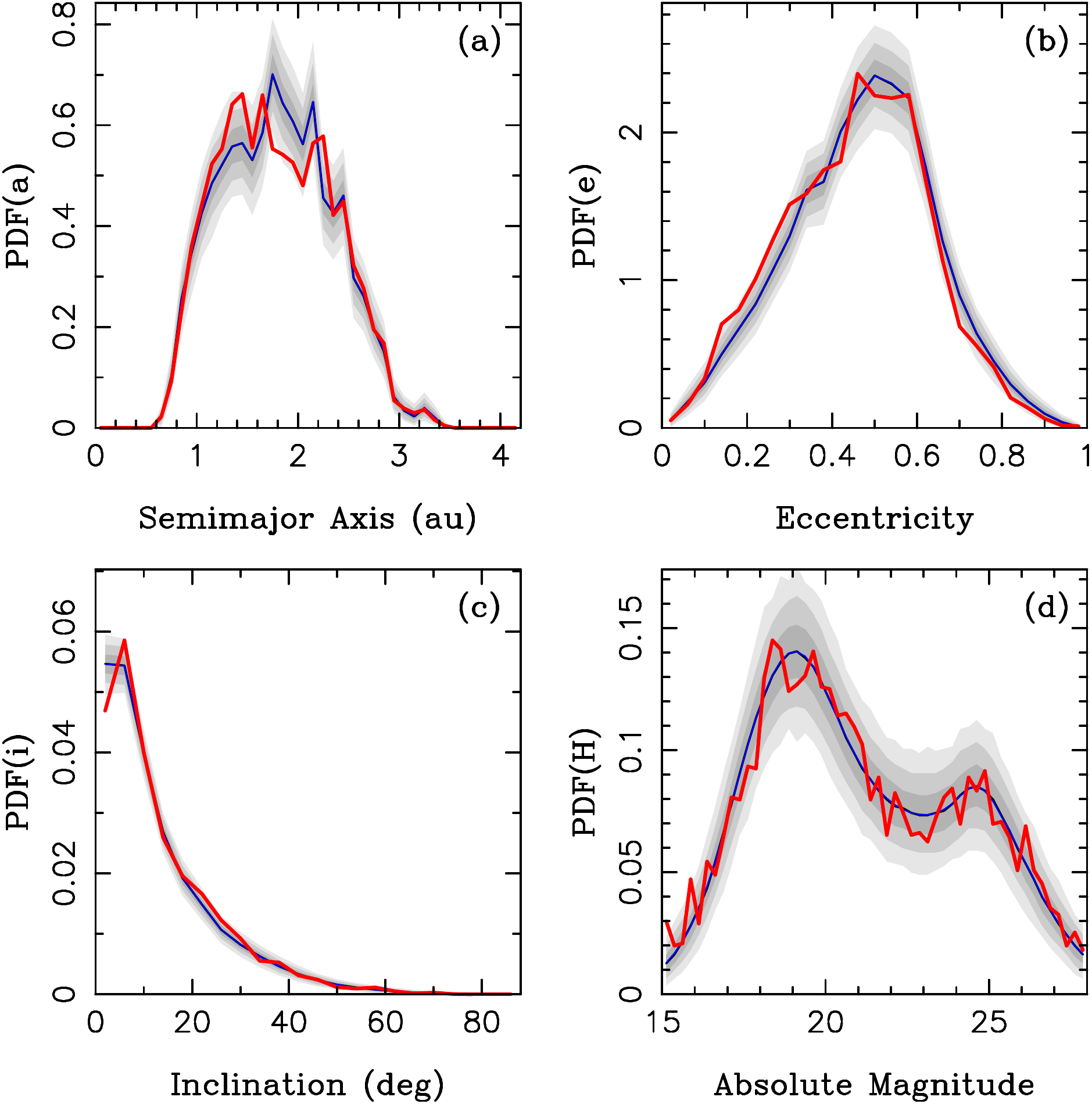}
\caption{The probability density functions (PDFs) of $a$, $e$, $i$, and $H$ from our modified base model 
with the extended magnitude range ($15<H<28$) (blue lines) is compared to the CSS NEO detections (red 
lines). The shaded areas are 1$\sigma$ (bold gray), 2$\sigma$ (medium) and 3$\sigma$ (light gray) envelopes. 
We used the best-fit solution (the one with the maximum likelihood) from the modified base model and 
generated 30,000 random samples with 4,412 NEOs each (the sample size identical to the number 
of CSS's NEOs in the model domain; $15<H<28$). The samples were biased and binned. We identified envelopes 
containing 68.3\% (1$\sigma$), 95.5\% (2$\sigma$) and 99.7\% (3$\sigma$) of samples and plotted them here.}
\label{aux1}
\end{figure}

\clearpage
\begin{figure}
\epsscale{0.7}
\plotone{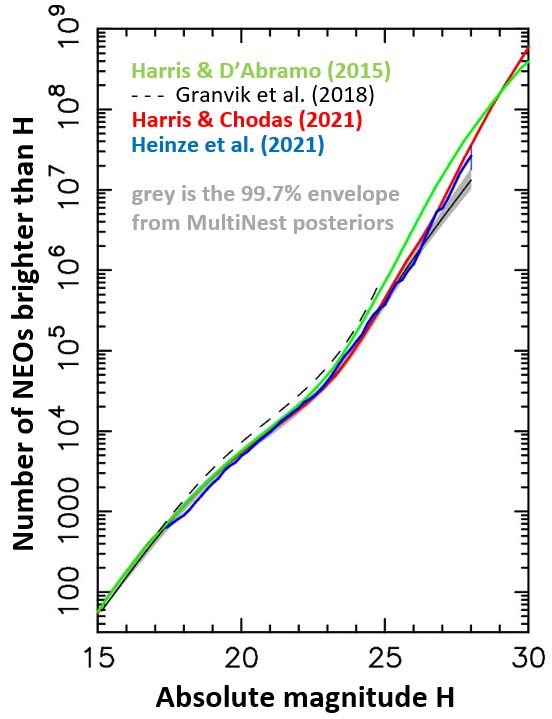}
\caption{The (intrinsic) absolute magnitude distribution from the modified base model where we extended 
the model domain to $15<H<28$ (solid black line). The gray area shows the 99.7\% envelope of posteriors 
from the \texttt{MultiNest} fit. For reference, we also plot the magnitude distributions from
Harris \& D'Abramo (2015) (green line), Granvik et al. (2018) (dashed black line for $17<H<25$), 
Harris \& Chodas (2021) (red line), and Heinze et al. (2021) (blue line; the vertical blue bar 
at $H=28$ shows the 1$\sigma$ uncertainty reported in Heinze et al. for the number of NEO with 
$H<28$).}
\label{aux2}
\end{figure}

\begin{figure}
\epsscale{0.7}
\plotone{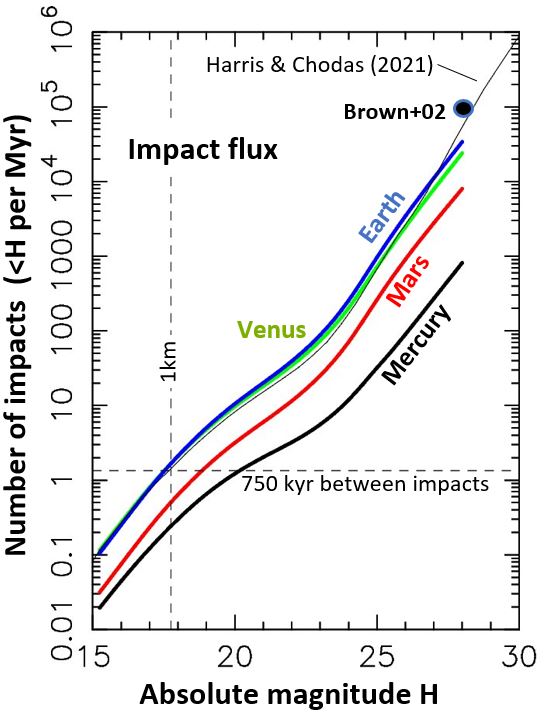}
\caption{The impact flux on the terrestrial planets as function of NEOs absolute magnitude. The black, green, blue 
and red lines show the impact flux for Mercury, Venus, Earth and Mars from Eq. (\ref{fimp}). The thin solid line 
near the Earth flux is the NEO magnitude distribution from Harris \& Chodas (2021) scaled with the fixed 
impact probability ($1.5 \times 10^{-3}$ Myr$^{-1}$; see the main text).
The black dot approximately marks the constraint from bolide detonations in the Earth atmosphere
(Brown et al. 2002). The horizontal dashed line shows the impact flux for $D>1$ km bodies from Morbidelli et 
al. (2021) (0.75-Myr average spacing between impacts). The vertical dashed line corresponds to $H=17.75$.}
\label{impacts}
\end{figure}


\end{document}